\documentclass[aps,prd,twocolumn,superscriptaddress,showpacs,floatfix]{revtex4}
\usepackage{bm}
\usepackage[dvips]{graphicx}
\usepackage{amsmath,amssymb,times}

\newcommand{\bequ}{\begin{equation}}
\newcommand{\eequ}{\end{equation}}
\newcommand{\bea}{\begin{eqnarray}}
\newcommand{\eea}{\end{eqnarray}}

\def\lsim{~\,\makebox(1,1){$\stackrel{<}{\widetilde{}}$}\,~}

\DeclareSymbolFont{boldletters}{OML}{cmm} {b}{it}
\DeclareSymbolFontAlphabet{\mathbit}{boldletters}
\DeclareMathSymbol{\alpha}{\mathalpha}{letters}{"0B}
\DeclareMathSymbol{\beta}{\mathalpha}{letters}{"0C}
\DeclareMathSymbol{\gamma}{\mathalpha}{letters}{"0D}
\DeclareMathSymbol{\delta}{\mathalpha}{letters}{"0E}
\DeclareMathSymbol{\epsilon}{\mathalpha}{letters}{"0F}
\DeclareMathSymbol{\zeta}{\mathalpha}{letters}{"10}
\DeclareMathSymbol{\eta}{\mathalpha}{letters}{"11}
\DeclareMathSymbol{\theta}{\mathalpha}{letters}{"12}
\DeclareMathSymbol{\iota}{\mathalpha}{letters}{"13}
\DeclareMathSymbol{\kappa}{\mathalpha}{letters}{"14}
\DeclareMathSymbol{\lambda}{\mathalpha}{letters}{"15}
\DeclareMathSymbol{\mu}{\mathalpha}{letters}{"16}
\DeclareMathSymbol{\nu}{\mathalpha}{letters}{"17}
\DeclareMathSymbol{\xi}{\mathalpha}{letters}{"18}
\DeclareMathSymbol{\pi}{\mathalpha}{letters}{"19}
\DeclareMathSymbol{\rho}{\mathalpha}{letters}{"1A}
\DeclareMathSymbol{\sigma}{\mathalpha}{letters}{"1B}
\DeclareMathSymbol{\tau}{\mathalpha}{letters}{"1C}
\DeclareMathSymbol{\upsilon}{\mathalpha}{letters}{"1D}
\DeclareMathSymbol{\phi}{\mathalpha}{letters}{"1E}
\DeclareMathSymbol{\chi}{\mathalpha}{letters}{"1F}
\DeclareMathSymbol{\psi}{\mathalpha}{letters}{"20}
\DeclareMathSymbol{\omega}{\mathalpha}{letters}{"21}
\DeclareMathSymbol{\varepsilon}{\mathalpha}{letters}{"22}
\DeclareMathSymbol{\vartheta}{\mathalpha}{letters}{"23}
\DeclareMathSymbol{\varpi}{\mathalpha}{letters}{"24}
\DeclareMathSymbol{\varrho}{\mathalpha}{letters}{"25}
\DeclareMathSymbol{\varsigma}{\mathalpha}{letters}{"26}
\DeclareMathSymbol{\varphi}{\mathalpha}{letters}{"27}
\DeclareMathSymbol{\Gamma}{\mathalpha}{letters}{"00}
\DeclareMathSymbol{\Delta}{\mathalpha}{letters}{"01}
\DeclareMathSymbol{\Theta}{\mathalpha}{letters}{"02}
\DeclareMathSymbol{\Lambda}{\mathalpha}{letters}{"03}
\DeclareMathSymbol{\Xi}{\mathalpha}{letters}{"04}
\DeclareMathSymbol{\Pi}{\mathalpha}{letters}{"05}
\DeclareMathSymbol{\Sigma}{\mathalpha}{letters}{"06}
\DeclareMathSymbol{\Upsilon}{\mathalpha}{letters}{"07}
\DeclareMathSymbol{\Phi}{\mathalpha}{letters}{"08}
\DeclareMathSymbol{\Psi}{\mathalpha}{letters}{"09}
\DeclareMathSymbol{\Omega}{\mathalpha}{letters}{"0A}

\begin{document}
%\preprint{SAGA-HE-265}
\title{ Color screening potential at finite density in two-flavor lattice QCD with Wilson fermions }

\author{Junichi Takahashi}
\email[]{takahashi@phys.kyushu-u.ac.jp}
\affiliation{Department of Physics, Graduate School of Sciences, Kyushu University,
             Fukuoka 812-8581, Japan}

\author{Keitaro Nagata}
\email[]{knagata@post.kek.jp}
\affiliation{KEK Theory Center, High Energy Accelerator Research Organization (KEK),
             Tsukuba 305-0801, Japan}

\author{Takuya Saito}
\email[]{tsaitou@kochi-u.ac.jp}
\affiliation{Integrated Information Center, Kochi University,
             Kochi 780-8520, Japan}

\author{Atsushi Nakamura}
\email[]{nakamura@riise.hiroshima-u.ac.jp}
\affiliation{Research Institute for Information Science and Education, Hiroshima University,
             Higashi-Hiroshima 739-8527, Japan}

\author{Takahiro Sasaki}
\email[]{sasaki@phys.kyushu-u.ac.jp}
\affiliation{Department of Physics, Graduate School of Sciences, Kyushu University,
             Fukuoka 812-8581, Japan}

\author{Hiroaki Kouno}
\email[]{kounoh@cc.saga-u.ac.jp}
\affiliation{Department of Physics, Saga University,
             Saga 840-8502, Japan}

\author{Masanobu Yahiro}
\email[]{yahiro@phys.kyushu-u.ac.jp}
\affiliation{Department of Physics, Graduate School of Sciences, Kyushu University,
             Fukuoka 812-8581, Japan}

\date{\today}

%%%%%%%%%%%%%%%%%%%%%%%%%%%%%%%%%%%%%%%%%%%%%%%%%%%%%%%%%%%%%%%%%%%%%%%%%%%
%%%%%  Abstruct 
%%%%%%%%%%%%%%%%%%%%%%%%%%%%%%%%%%%%%%%%%%%%%%%%%%%%%%%%%%%%%%%%%%%%%%%%%%%
\begin{abstract}
We investigate the chemical-potential ($\mu$) dependence of static-quark free energies in both the real and imaginary $\mu$ regions, performing lattice QCD simulations at imaginary $\mu$ and extrapolating the results  to the real $\mu$ region with analytic continuation. 
Lattice QCD calculations are done on a $16^{3}\times 4$ lattice with the clover-improved two-flavor Wilson fermion action and the renormalization-group improved Iwasaki gauge action. 
Static-quark potentials are evaluated from the Polyakov-loop correlation functions in the deconfinement phase.   
As the analytic continuation, the potential calculated at imaginary $\mu=i\mu_{\rm I}$ is expanded into a Taylor-expansion series of  $i\mu_{\rm I}/T$ up to 4th order and the pure imaginary variable $i\mu_{\rm I}/T$ is replaced by the real one $\mu_{\rm R}/T$. 
At real $\mu$, the 4th-order term weakens $\mu$ dependence of the potential sizably. 
At long distance, all of the color singlet and non-singlet potentials tend to twice the single-quark free energy, indicating that the interactions between static quarks are fully color-screened for finite $\mu$. 
For both real and imaginary $\mu$, the color-singlet $q{\bar q}$ and the color-antitriplet $qq$ interaction are attractive, whereas the color-octet $q{\bar q}$ and the color-sextet $qq$ interaction are repulsive.
The attractive interactions have stronger $\mu/T$ dependence than the repulsive interactions.
The color-Debye screening mass is extracted from the color-singlet potential at imaginary $\mu$, and the mass is extrapolated to real $\mu$ by analytic continuation. 
The screening mass thus obtained has stronger $\mu$ dependence than the prediction of the hard thermal loop perturbation theory at both real and imaginary $\mu$.
\end{abstract}

\maketitle
%%%%%%%%%%%%%%%%%%%%%%%%%%%%%%%%%%%%%%%%%%%%%%%%%%%%%%%%%%%%%%%%%%%%%%%%%%%
%%%%%  Introduction 
%%%%%%%%%%%%%%%%%%%%%%%%%%%%%%%%%%%%%%%%%%%%%%%%%%%%%%%%%%%%%%%%%%%%%%%%%%%
\section{Introduction}
Recent relativistic heavy-ion collision experiments have revealed various properties of QCD, suggesting the realization of the QCD phase transition from the hadronic phase to the quark gluon plasma (QGP) phase~\cite{QGP}, although no clear evidence of the transition is presented yet.  
Meanwhile, lattice QCD (LQCD) simulations provide very precise information on QCD particularly at finite temperature ($T$) and small quark-number chemical potential ($\mu$), although LQCD simulations have the serious sign problem at large $\mu/T$; see for  example Ref.~\cite{Muroya}.   
The LQCD simulations are complementary to the experiments, since the former is  suitable to understand  static properties of QCD and the latter is to investigate dynamical properties of QCD.\\
\indent 
Finite-density calculations in LQCD are affected by the well-known sign problem, that is, the fermion determinant  $\det M(\mu)$ becomes complex for finite $\mu$ and thus prohibits the use of conventional numerical algorithms.
In order to avoid the sign problem, several approaches have been proposed so far~\cite{Muroya,Forcrand1}. One is the  imaginary chemical potential approach. For pure imaginary chemical potential ($\mu=i\mu_{\rm I}$), the fermion determinant is real, so that LQCD simulations become feasible. 
Observables at real $\mu$ are extracted from those at imaginary $\mu$ with analytic continuation.\\
\indent 
In the imaginary $\mu$ region, QCD has two characteristic properties, the Roberge-Weiss (RW) periodicity and the RW phase transition~\cite{RW,Sakai:2008py}. The QCD grand partition function has a periodicity of $2\pi/N_{c}$ in $\mu_{\rm I}/T$:  
\bea
Z\left( \frac{\mu_{\rm I}}{T} \right)=Z\left( \frac{\mu_{\rm I}}{T} + \frac{2\pi k}{N_{c}} \right)
\eea
for integer $k$ and the number of color $N_{c}=3$. 
This is called the RW periodicity. 
Roberge and Weiss also showed that a first-order phase transition occurs at $T \ge T_{\rm RW}$ and $\mu/T=i\pi/N_{c}$. 
This is named the RW phase transition, and $T_{\rm RW}$ is slightly larger than the pseudo-critical temperature $T_{\rm pc}$ of the deconfinement transition at zero $\mu$. 
These features are remnants of ${\mathrm Z}_{N_{c}}$ symmetry in the pure gauge limit. The order parameter of the RW phase transition is a ${\cal C}$-odd quantity such as the phase of the Polyakov loop~\cite{Kouno:2009bm}, where ${\cal C}$ means charge conjugation. 
These properties are confirmed by LQCD simulations~\cite{Forcrand2,Wu,D'Elia1,D'Elia2,Forcrand3,Nagata}.\\
\indent
The free energies between two static quarks are fundamental quantities to understand medium effects in QGP.
For example, the color-Debye screening mass is the inverse of the range of the color-singlet potential determined from the free energies.
The potential largely affects the behavior of heavy-quark bound states such as $J/\Psi$ and $\Upsilon$ in QGP created at the center of heavy-ion collisions~\cite{Matsui}. 
In LQCD, the static-quark potential is evaluated from the Polyakov-loop correlation function. 
For zero chemical potential, $T$ dependence of the static-quark potential was investigated by quenched QCD~\cite{Kaczmarek1,Saito1,Saito2} and full QCD with staggered-type~\cite{Kaczmarek2} and Wilson-type quark actions~\cite{Bornyakov,Maezawa1,Maezawa2}. 
For small $\mu/T$, it was analyzed by the Taylor-expansion method with staggered-type~\cite{Doering} and Wilson-type quark actions~\cite{Ejiri}. 
In the analysis~\cite{Ejiri}, the expansion coefficients are taken up to 2nd order of $\mu/T$.\\
\indent
In this paper, we investigate $\mu$ dependence of static-quark free energies and the color-Debye screening mass in both the imaginary and real $\mu$ regions, performing LQCD simulations at imaginary $\mu$ with standard numerical algorithms and extrapolating the result to the real  $\mu=\mu_{\rm R}$ region with  analytic continuation. 
LQCD simulations are done on a $16^{3}\times 4$ lattice with the clover-improved two-flavor Wilson fermion action and the renormalization-group (RG) improved Iwasaki gauge action. 
We consider two temperatures above $T_{\rm pc}$, i.e., $T/T_{\rm pc}=1.20$ and 1.35. 
Following the previous LQCD simulation~\cite{Ejiri} at small $\mu/T$, we compute static-quark free energies along the line of constant physics at $m_{\rm PS}/m_{\rm V}=0.80$. 
This corresponds to considering an intermediate quark mass. 
As the analytic continuation, the static-quark potential at imaginary $\mu=i\mu_{\rm I}$ is expanded into a Taylor-expansion series of $i\mu_{\rm I}/T$ and pure imaginary variable $i\mu_{\rm I}/T$ is replaced by real one $\mu_{\rm R}/T$.\\
\indent
In the present work the Taylor-expansion coefficients of the static-quark potential are evaluated up to 4th order, whereas the coefficients were computed up to 2nd order in Ref.~\cite{Ejiri}. 
It is found that the 4th-order term yields non-negligible contributions to $\mu$ dependence of the static-quark potentials at real $\mu$. At long distance, all of the color singlet and non-singlet potentials tend to twice the single-quark free energy, indicating that the interactions between static quarks are fully color-screened.
Although this property is known for finite $T$ and zero $\mu$~\cite{Maezawa1}, the present work shows that the property persists also for finite $\mu$.
For both real and imaginary $\mu$, the color-singlet $q{\bar q}$ and the color-antitriplet $qq$ interaction are attractive, whereas the color-octet $q{\bar q}$ and the color-sextet $qq$ interaction are repulsive. 
The attractive interactions become weak as $(\mu/T)^2$ increases, whereas the repulsive interactions little depend on $\mu$.
The color-Debye screening mass at imaginary $\mu$ is extracted from the color-singlet potential there. The mass at real $\mu$ is extrapolated from the mass at imaginary $\mu$ by analytic continuation, i.e., by expanding the mass at imaginary $\mu$ into a power series of $i\mu_{\rm I}/T$ up to 2nd order and replacing $i\mu_{\rm I}$ by $\mu_{\rm R}$. The $(\mu /T)$ dependence of the screening mass is found to be stronger than the prediction of the hard thermal loop perturbation theory (HTLpt).\\
\indent 
This paper is organized as follows. Section~II presents the lattice action and the definition of static-quark free energies. 
In Sec.~III, we show simulation parameters and numerical results for the Polyakov loop and the static-quark free energies in the color-singlet, -octet, -antitriplet and -sextet channels. 
We also extract the color-Debye screening mass from the color-singlet potential and compare it with the results of the hard thermal loop perturbation theory. 
Section IV is devoted to a summary.

%%%%%%%%%%%%%%%%%%%%%%%%%%%%%%%%%%%%%%%%%%%%%%%%%%%
%%%  Lattice formulation
%%%%%%%%%%%%%%%%%%%%%%%%%%%%%%%%%%%%%%%%%%%%%%%%%%%
\section{Lattice Formulation}
\label{Lattice Formulation}

\subsection{Lattice action}
We use the RG-improved Iwasaki gauge action $S_{g}$~\cite{Iwasaki} and the clover-improved two-flavor Wilson quark action $S_{q}$~\cite{clover-Wilson} defined by
\bea
S&=&S_{g}+S_{q}, \\
S_{g}&=&-\beta \sum_{x} \left(c_{0} \sum^{4}_{\mu<\nu;\mu,\nu=1}W^{1\times 1}_{\mu\nu}(x)\right. \nonumber \\
&&\left. +c_{1} \sum^{4}_{\mu\neq\nu;\mu,\nu=1}W^{1\times 2}_{\mu\nu}(x)\right), \\
S_{q}&=&\sum_{f=u,d}\sum_{x,y}\bar{\psi}^{f}_{x}M_{x,y}\psi^{f}_{y},
\eea
where $\beta=6/g^{2}$, $c_{1}=-0.331$, $c_{0}=1-8c_{1}$, and
\bea
M_{x,y}=&&\delta_{xy}-\kappa\sum^{3}_{i=1}\{(1-\gamma_{i})U_{x,i}\delta_{x+\hat{i},y}+(1+\gamma_{i})U^{\dag}_{y,i}\delta_{x,y+\hat{i}}\} \nonumber \\
&&-\kappa\{e^{\mu}(1-\gamma_{4})U_{x,4}\delta_{x+\hat{4},y}+e^{-\mu}(1+\gamma_{4})U^{\dag}_{y,4}\delta_{x,y+\hat{4}}\} \nonumber \\
&&-\delta_{xy}c_{\mathrm{SW}}\kappa\sum_{\mu<\nu}\sigma_{\mu\nu}F_{\mu\nu}.
\eea
Here $\kappa$ is the hopping parameter, $\mu$ is the quark chemical potential in lattice unit, and $F_{\mu\nu}$ is the lattice field strength,  $F_{\mu\nu}=(f_{\mu\nu}-f^{\dag}_{\mu\nu})/(8i)$ with $f_{\mu\nu}$ the standard clover-shaped combination of gauge links. For the clover coefficient $c_{\mathrm{SW}}$, we adopt a mean field value using $W^{1\times 1}$ calculated in the one-loop perturbation theory~\cite{Iwasaki} : $c_{\mathrm{SW}}=(W^{1\times 1})^{-3/4}=(1-0.8412\beta^{-1})^{-3/4}$. We denote the spatial and temporal lattice size as $N_{s}$ and $N_{t}$, respectively. At $\mu=0$, a value of $\kappa$ is determined for each $\beta$ along the line of constant physics with $m_{\mathrm{PS}}/m_{\mathrm{V}} = 0.80$ obtained in Ref.~\cite{Khan1,Khan2,Maezawa2}.

\subsection{Static-quark free energies}
The Polyakov loop is defined as 
\bea
L(\bm{x})=\prod^{N_{t}}_{t=1}U_{4}(\bm{x},t)
\eea
with link variables $U_{\mu} \in \mathrm{SU}(3)$. 
At imaginary $\mu$, the ensemble average of the Polyakov loop becomes a complex number, 
$\langle \mathrm{Tr}L(0) \rangle \equiv \Phi e^{i\theta}$. 
The modulus is related to the single-quark free energy $F_q$ as 
\bea
 \Phi=e^{-F_q/T}.
\eea
The modulus and phase, $\Phi$ and $\theta$, are the order parameters of the confinement/deconfinement and RW phase transitions~\cite{Kouno:2009bm}, respectively. 
After taking an appropriate gauge fixing, one can derive the static-quark free energies (potentials) $V_M$ of color channel $M$ 
from the Polyakov-loop correlators~\cite{Nadkarni1,Nadkarni2}: 
\bea
e^{-V_{1}(r,T,\mu)/T}&=&\frac{1}{3}\langle \mathrm{Tr}L^{\dag}(\bm{x})L(\bm{y}) \rangle, \\
e^{-V_{8}(r,T,\mu)/T}&=&\frac{1}{8}\langle \mathrm{Tr}L^{\dag}(\bm{x})\mathrm{Tr}L(\bm{y}) \rangle \nonumber \\
&& \qquad \qquad -\frac{1}{24}\langle \mathrm{Tr}L^{\dag}(\bm{x})L(\bm{y}) \rangle, \\
e^{-V_{6}(r,T,\mu)/T}&=&\frac{1}{12}\langle \mathrm{Tr}L(\bm{x})\mathrm{Tr}L(\bm{y}) \rangle \nonumber \\
&& \qquad \qquad +\frac{1}{12}\langle \mathrm{Tr}L(\bm{x})L(\bm{y}) \rangle, \\
e^{-V_{3^{\ast}}(r,T,\mu)/T}&=&\frac{1}{6}\langle \mathrm{Tr}L(\bm{x})\mathrm{Tr}L(\bm{y}) \rangle \nonumber \\
&& \qquad \qquad -\frac{1}{6}\langle \mathrm{Tr}L(\bm{x})L(\bm{y}) \rangle,
\eea
where $r=|\bm{x}-\bm{y}|$ and the subscripts $M=(1, 8, \mathrm{3}^{\ast}, 6)$ mean the color-singlet, -octet, -antitriplet and -sextet channels, respectively. 
We adopt both the Coulomb gauge fixing and the Landau gauge fixing. 
As shown in Sec. III, however, the latter breaks the RW periodicity of $V_M$, whereas the former preserves it. We then mainly use the Coulomb gauge fixing in this paper.\\
\indent 
For the color-singlet potential $V_{1}(r)$, the corresponding Polyakov-loop correlator is 1 at $r=0$, and hence $V_{1}(0)=0$. For the color non-singlet potentials, the corresponding Polyakov-loop correlators are not 1 at $r=0$.
Therefore, the color-singlet potential tends to a common value independent of $T$ and $\mu$ in the short-distance limit, but the color non-singlet potentials do not. These properties will be shown explicitly in Sec. \ref{Numerical results}.\\
\indent 
In general, the $V_{M}$  ($M=1, 8, 3^{\ast}, 6$) are complex at finite imaginary $\mu$. 
The real part of $V_{M}$ is ${\cal C}$-even and the imaginary part is ${\cal C}$-odd. 
This can be easily understood by expanding $V_{M}$ into a power series of $i\mu_{\rm I}/T$:
\bea
\frac{V_{M}(r,T,\mu_{\rm I})}{T}&=&v_{0}(r)+iv_{1}(r)\left(\frac{\mu_{\rm I}}{T}\right)+v_{2}(r)\left(\frac{\mu_{\rm I}}{T}\right)^{2} \nonumber \\
&&+iv_{3}(r)\left(\frac{\mu_{\rm I}}{T}\right)^{3}+v_{4}(r)\left(\frac{\mu_{\rm I}}{T}\right)^{4},  
\label{Eq:expansion-1}
\eea
where we consider terms up to 4th order. 
The potential $V_{M}$ at real $\mu$ is obtained from that at imaginary  $\mu$ by analytic continuation, i.e., by replacing $i\mu_{\rm I}/T$ by $\mu_{\rm R}/T$:
\bea
\frac{V_{M}(r,T,\mu_{\rm R})}{T}&=&v_{0}(r)+v_{1}(r)\left(\frac{\mu_{\rm R}}{T}\right)-v_{2}(r)\left(\frac{\mu_{\rm R}}{T}\right)^{2} \nonumber \\
&&-v_{3}(r)\left(\frac{\mu_{\rm R}}{T}\right)^{3}+v_{4}(r)\left(\frac{\mu_{\rm R}}{T}\right)^{4}.  
\eea
\indent
The WHOT-QCD Collaboration calculated the Taylor-expansion coefficients of $V_{M}$ up to 2nd order by using the Taylor-expansion method and the reweighting technique with the Gaussian approximation for the distribution of the complex phase of the quark determinant~\cite{Ejiri}. 
In this work, meanwhile, we obtain the coefficients up to 4th order from $V_{M}$ at imaginary $\mu$ by expanding it as in \eqref{Eq:expansion-1}. 
As shown in Sec. III, the $v_{2}(r)$ thus obtained is consistent with the results (ref.~\cite{Ejiri}) calculated directly with the Taylor-expansion method, and the $v_{4}(r)$ yields non-negligible contributions to $\mu_{\rm R}$ dependence of $V_{M}$.
We do not adopt the definition of the renormalized Polyakov-loop correlators that are adjusted the absolute value as discussed in Ref.~\cite{Kaczmarek3}.
However, our approach is adequate to investigate the dependence of the static-quark potentials on $\mu$, since the lattice spacing $a$ is common for each $\mu$.
The renormalization procedure thus changes only $v_0$, but not $v_2$ and $v_4$.

%%%%%%%%%%%%%%%%%%%%%%%%%%%%%%%%%%%%%%%%%%%%%%
%%%%%%%%  Numerical results 
%%%%%%%%%%%%%%%%%%%%%%%%%%%%%%%%%%%%%%%%%%%%%%
\section{Numerical results}
\label{Numerical results}

The Hybrid Monte-Carlo algorithm is used to generate full QCD configurations with two-flavor dynamical quarks. The simulations are performed on a lattice of either $N_{s}^{3}\times N_{t} = 12^3 \times 4$ or $16^3 \times 4$.  The step size of the molecular dynamics is $\delta \tau = 0.01$ and the step number of the dynamics is $N_{\tau} = 100$. The acceptance ratio is more than 95\%. We generated 16,000 trajectories and removed the first 1,000 trajectories as thermalization for all the parameter set. We measured the static-quark potential at every 100 trajectories.
The relation of parameters $\kappa$ and $\beta$ to the corresponding $T/T_{\rm pc}$ is determined in Ref.~\cite{Khan1,Khan2,Maezawa2}; see Table \ref{table-para} for the relation.

%%%%%%%%%%%%%%%%
%%% Table I
%%%%%%%%%%%%%%%%
\begin{table}[h]
\begin{center}
\begin{tabular}{|c|c|c|c|c|}
\hline 
$N_{s}$ & $\kappa$ & $\beta$ & $T/T_{\mathrm{pc}}$ & $\mu_{\rm I}/T$
\\
\hline
~~$12$~~ & ~~$0.140070$~~ & ~~$1.85$~~ & ~~$0.99(5)$~~ & ~~$0\sim \pi/3$~~\\ \cline{2-5}
             & ~~$0.138817$~~ & ~~$1.90$~~ & ~~$1.08(5)$~~ & ~~$0\sim \pi/3$~~\\ \cline{2-5}
             & ~~$0.137716$~~ & ~~$1.95$~~ & ~~$1.20(6)$~~ & ~~$0\sim 1.10$~~\\ \cline{2-5}
             & ~~$0.136931$~~ & ~~$2.00$~~ & ~~$1.35(7)$~~ & ~~$0\sim 1.10$~~\\
\hline 
~~$16$~~ & ~~$0.137716$~~ & ~~$1.95$~~ & ~~$1.20(6)$~~ & ~~$0\sim 1.20$~~\\ \cline{2-5}
             & ~~$0.136931$~~ & ~~$2.00$~~ & ~~$1.35(7)$~~ & ~~$0\sim 1.20$~~\\
\hline 
\end{tabular}
\caption{
Summary of the simulation parameter sets determined in Ref.~\cite{Khan1,Khan2,Maezawa2}. $T_{\mathrm{pc}}$ is the pseudocritical temperature at $\mu=0$. 
In this parameter setting, the lattice spacing $a$ is about $0.14 \sim 0.2$~fm. }
\label{table-para}
\end{center}
\end{table}
%%%%%%%%%%%

%%%%%%%%%%%%%%%%%%%%%%%%%%%%%%%%%%%%%%%%%%%%%%
%%%%%%%%  Polyakov loop
%%%%%%%%%%%%%%%%%%%%%%%%%%%%%%%%%%%%%%%%%%%%%%

\subsection{Polyakov loop}
\label{Sec:Polyakov loop}
\begin{figure}[htbp]
\begin{center}
\hspace{10pt}
 \includegraphics[width=0.445\textwidth]{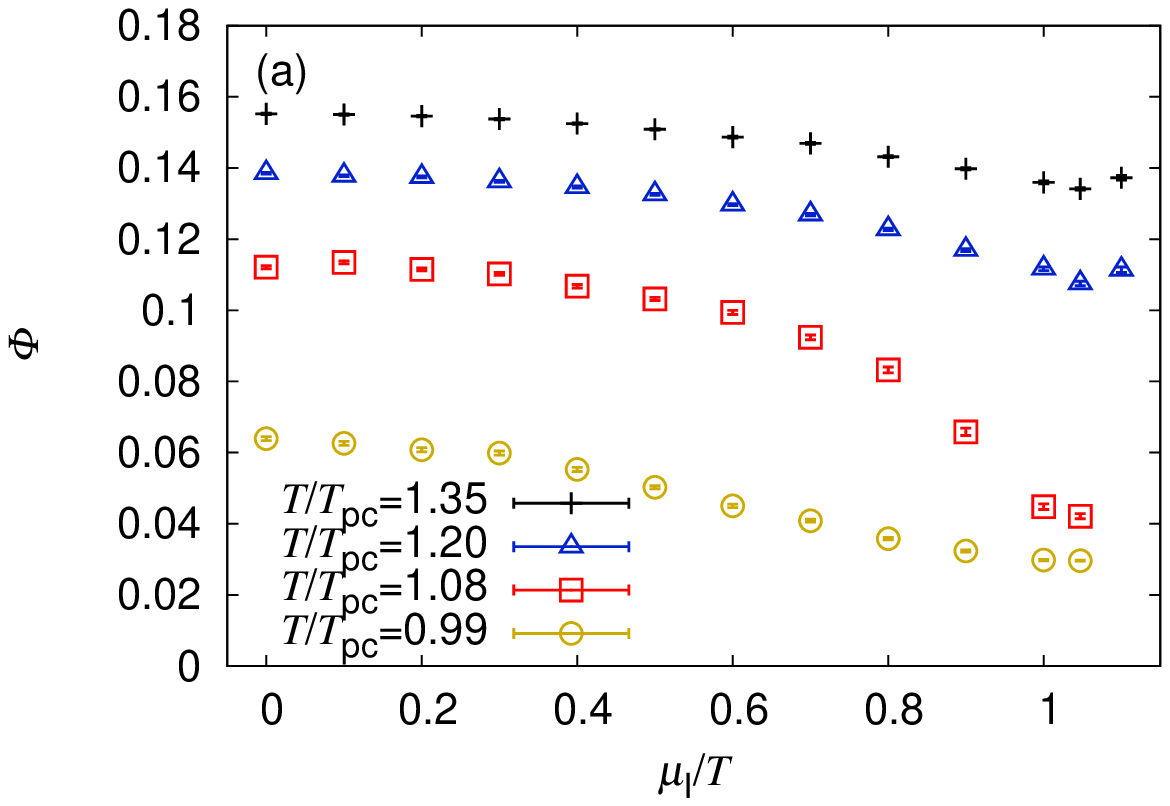}
 \includegraphics[width=0.445\textwidth]{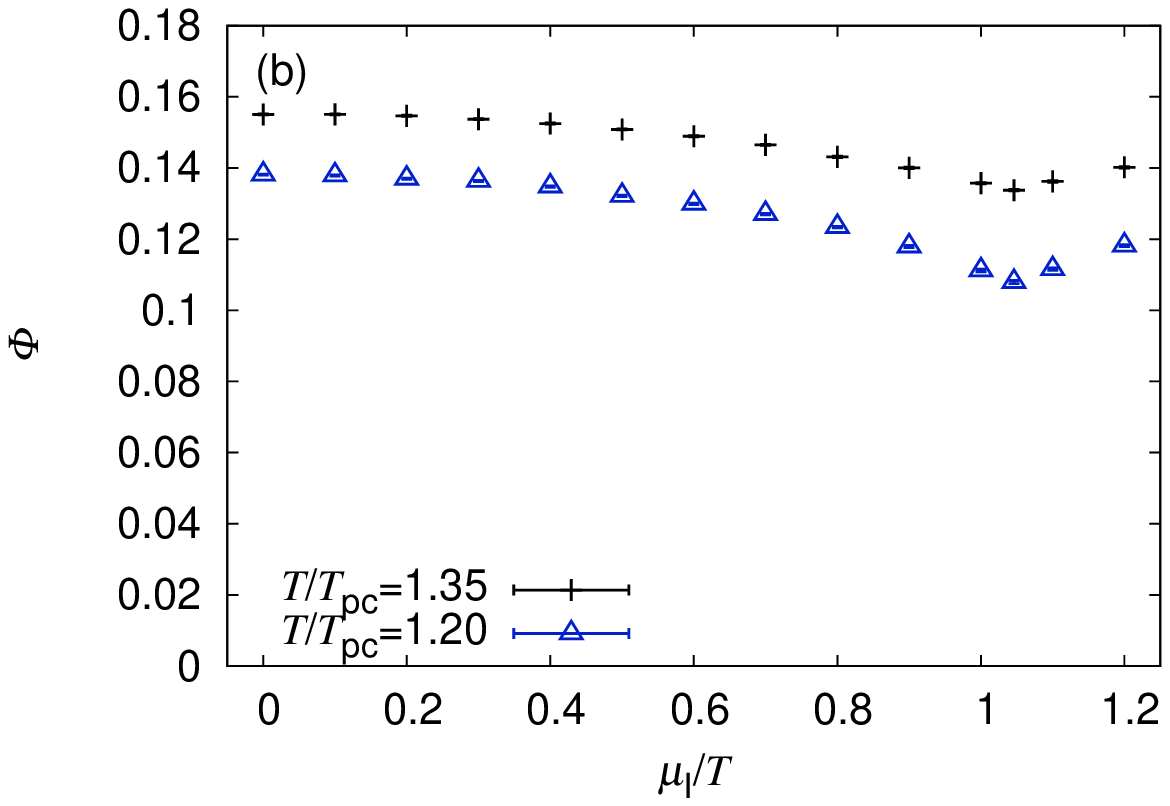}
\end{center}
\vspace{10pt}
\caption{
$\mu_{\rm I}/T$ dependence of $\Phi$ at various values of $T$ for  
(a) a $12^3\times 4$ lattice and (b) a $16^3\times 4$ lattice.
}
\label{absPol}
\end{figure}

\begin{figure}[htbp]
\begin{center}
\hspace{10pt}
 \includegraphics[width=0.445\textwidth]{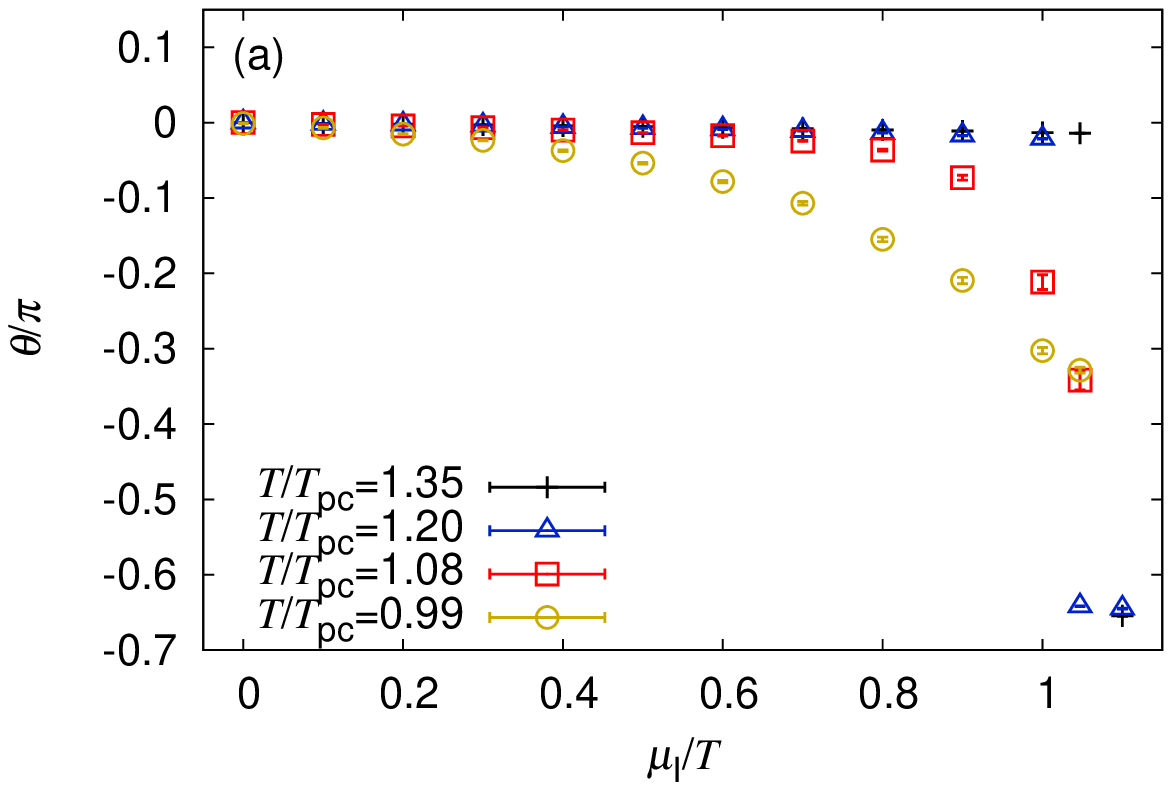}
 \includegraphics[width=0.445\textwidth]{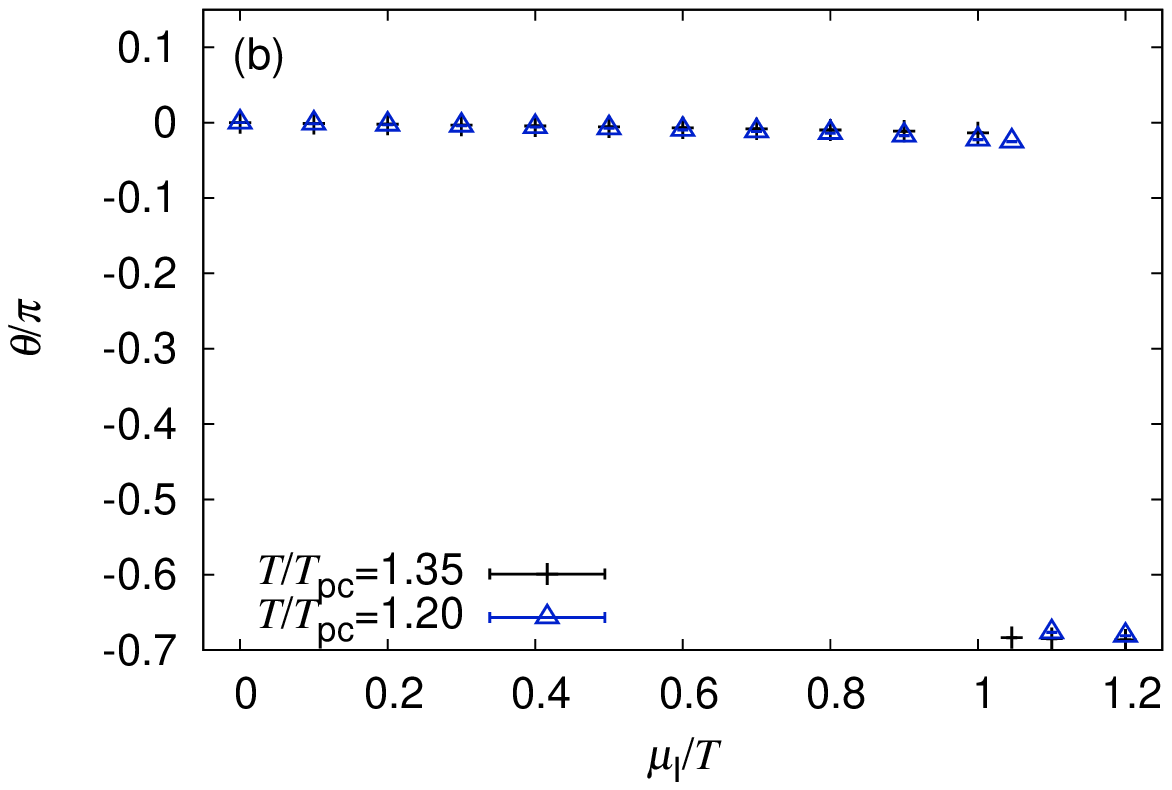}
\end{center}
\vspace{10pt}
\caption{
$\mu_{\rm I}/T$ dependence of $\theta$ at various values of $T$ 
for (a) a $12^3\times 4$ lattice and (b) a $16^3 \times 4$ lattice.
}
\label{phasePol}
\end{figure}

First we investigate  the behavior of the expectation value of the Polyakov loop, $\langle \mathrm{Tr}L(0) \rangle \equiv \Phi e^{i\theta}$, at finite imaginary $\mu$.
Figures \ref{absPol} and \ref{phasePol} show $\mu_{\rm I}/T$ dependence of $\Phi$ and $\theta$, respectively, and panels (a) and (b) correspond to results of  $12^3\times 4$ and  $16^3\times 4$ lattices, respectively. 
For both the lattice sizes, $\Phi$ as a ${\cal C}$-even quantity has a cusp and $\theta$ as a ${\cal C}$-odd quantity has a jump from 0 to $-2\pi/3$ at $\mu_{\rm I}/T=\pi/3$, when $T/T_{\rm pc}=1.2$ and 1.35. 
These properties are characteristic of the RW phase transition~\cite{Sakai:2008py}, and the critical endpoint $T_{\rm RW}$ of the RW phase transition is located at $1.08T_{\rm pc} < T_{\rm RW} < 1.2T_{\rm pc}$.
Obviously, an order parameter of the transition is a  ${\cal C}$-odd quantity~\cite{Kouno:2009bm}. 
The modulus $\Phi$ as a ${\cal C}$-even quantity is mirror symmetric with respect to the line of $\mu_{\rm I}/T=\pi/3$.
This is a result of the RW periodicity $\Phi(\mu_{\rm I}/T)=\Phi(\mu_{\rm I}/T+2\pi/3)$ and ${\cal C}$-evenness $\Phi(\mu_{\rm I}/T)=\Phi(-\mu_{\rm I}/T)$.\\
\indent
For $T/T_{\mathrm{pc}}=1.08$, $\Phi$ rapidly decreases as $\mu_{\rm I}/T$ increases, indicating that the system is in the deconfinement phase at small $\mu_{\rm I}/T$ but in the confinement phase near $\mu_{\rm I}/T=\pi/3$. 
At $T/T_{\mathrm{pc}}=1.20$ and $1.35$, the system is always in the deconfinement phase, and calculated results agree with each other between 
$12^3 \times 4$ and $16^3 \times 4$ lattice sizes.
This indicates that the $16^3 \times 4$ lattice is large enough. 
We then calculate the static-quark potential at $T/T_{\mathrm{pc}}=1.20$ and $1.35$ on a $16^3 \times 4$ lattice.

\onecolumngrid

%%%%%%%%%%%%%%%
%    Fig. 1
%%%%%%%%%%%%%%%
%%%%%%%%%%%%%%

%%%%%%%%%%%%%%%
%    Fig. 2
%%%%%%%%%%%%%%%
%%%%%%%%%%%%%%

%%%%%%%%%%%%%%%
%    Fig. 3
%%%%%%%%%%%%%%%
\begin{figure}[h]
\begin{center}
\hspace{10pt}
\includegraphics[width=0.445\textwidth]{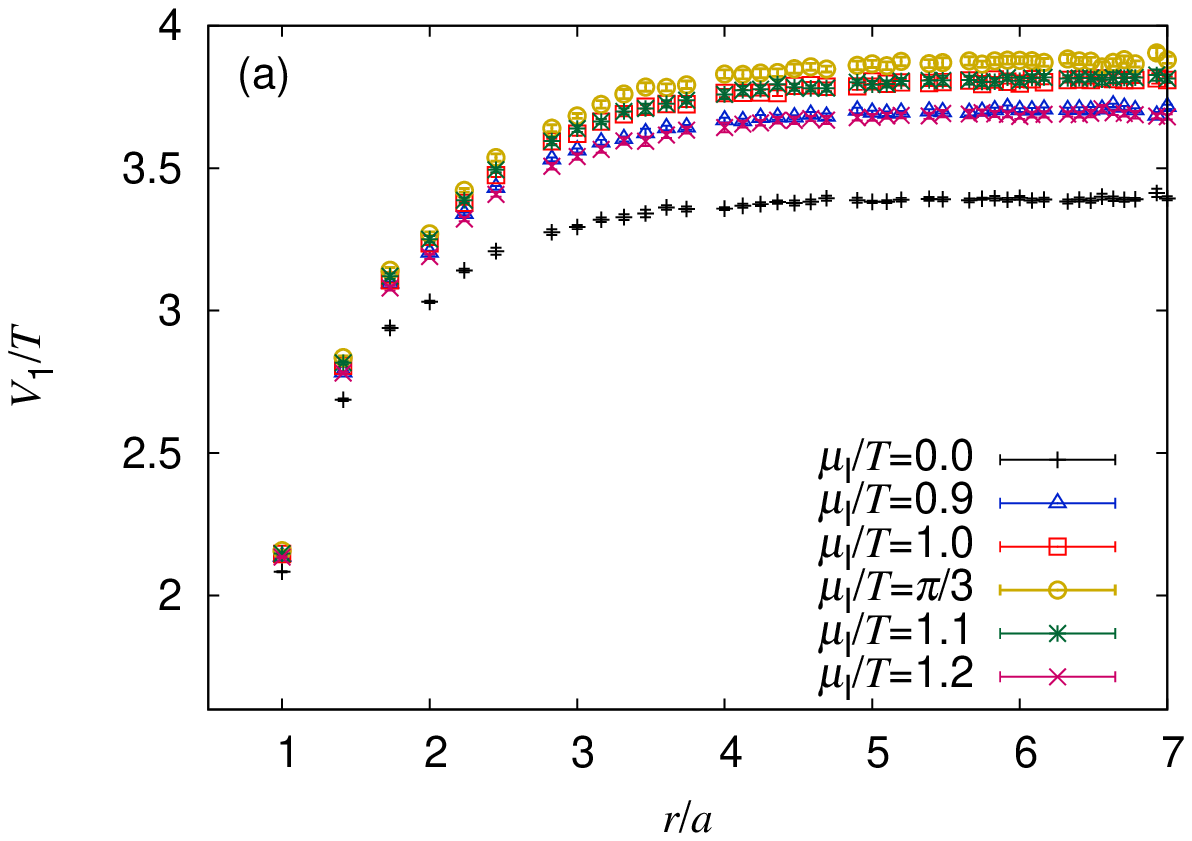}
\includegraphics[width=0.445\textwidth]{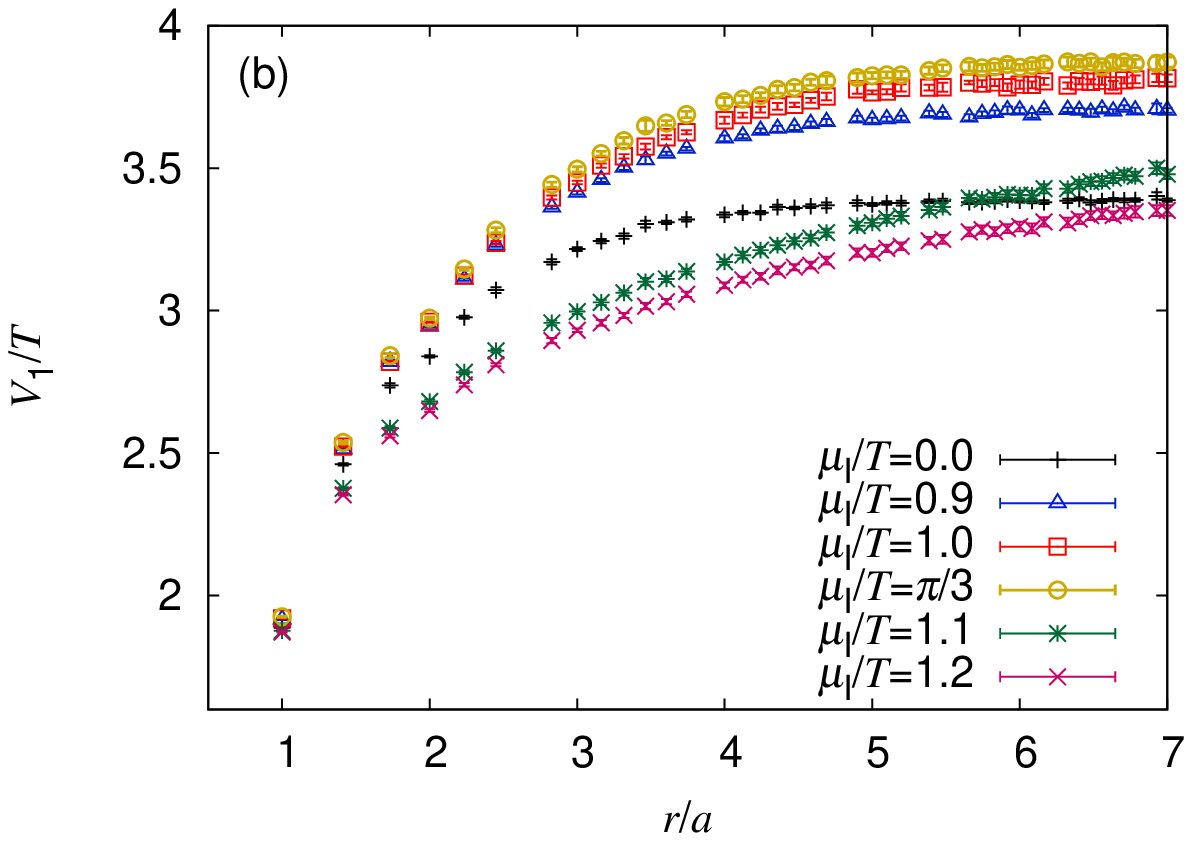}
\end{center}
\vspace{10pt}
\caption{
$\mu_{\rm I}/T$ dependence of the color-singlet $q\bar{q}$ potential  at $T/T_{\mathrm{pc}}=1.20$ 
based on (a) the Coulomb 
and (b) the Landau gauge fixing.
}
\label{oRWt-1pot_CL}
\end{figure}
%%%%%%%%%%%%%%

%%%%%%%%%%%%%%%
%    Fig. 4
%%%%%%%%%%%%%%%
\begin{figure}[h]
\begin{center}
\hspace{10pt}
  \includegraphics[width=0.445\textwidth]{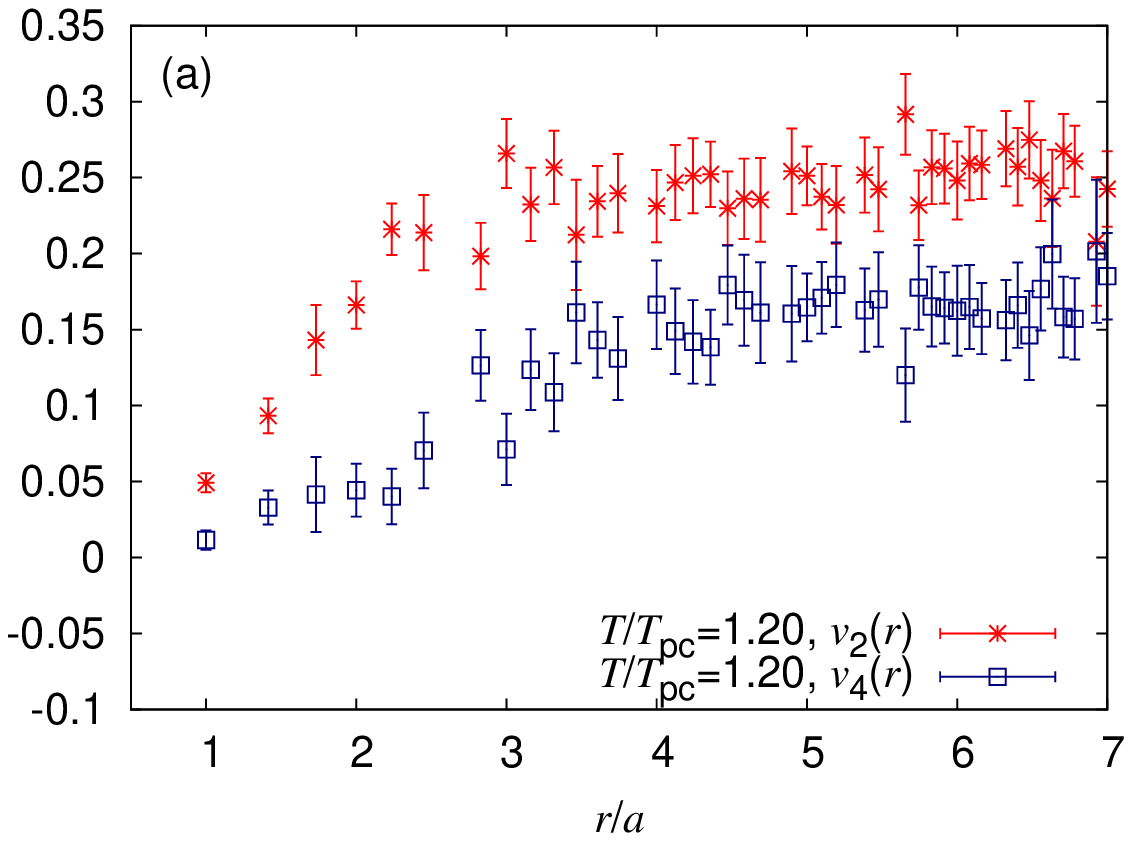}
  \includegraphics[width=0.445\textwidth]{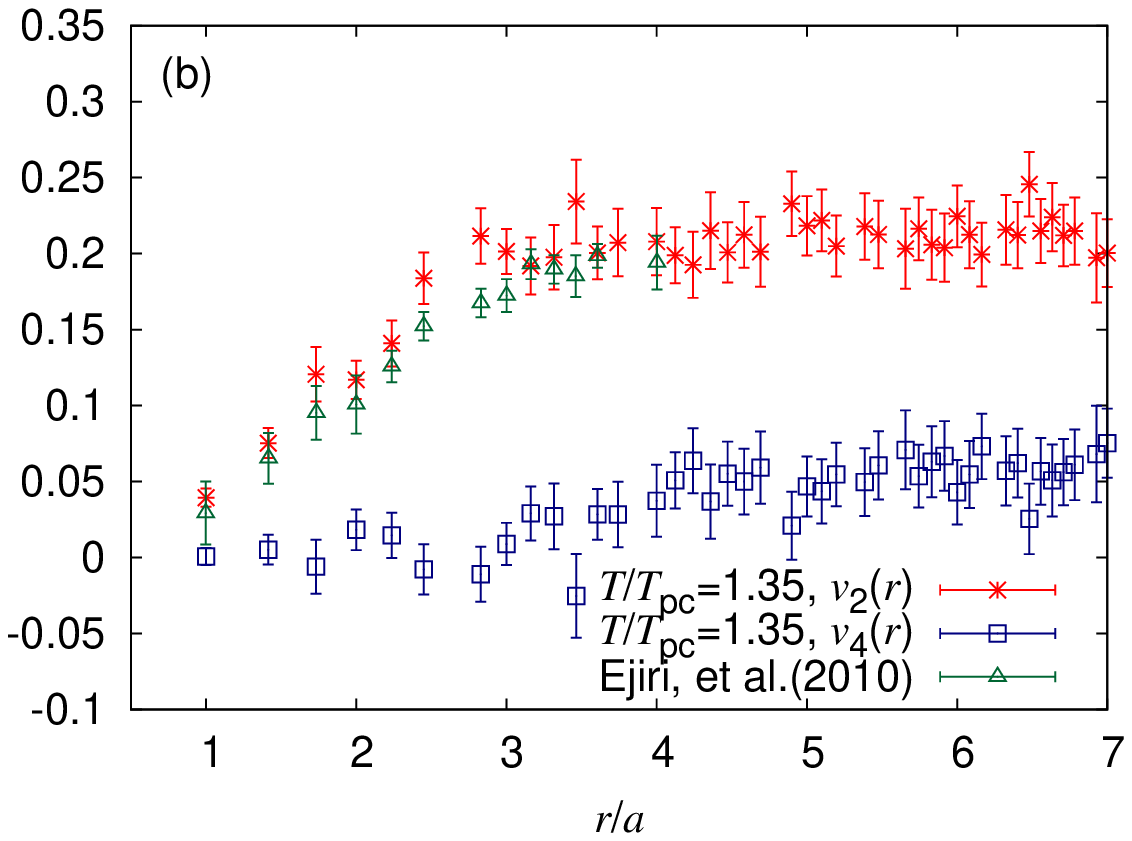}
\end{center}
\vspace{10pt}
\caption{Taylor-expansion coefficients, $v_2(r)$ and $v_4(r)$, of $V_{1}(r)$ at (a) $T/T_{\mathrm{pc}}=1.20$ and (b) $T/T_{\mathrm{pc}}=1.35$.
See Tables ~\ref{table-V1V8b1950} and \ref{table-V1V8b2000} in Appendix A for the numerical results.
In panel (b), the triangles denote the results of the Taylor-expansion method in Ref. \cite{Ejiri}. 
}
\label{Taylor-expansion-coefficient}
\end{figure}
%%%%%%%%%%%%%%

%%%%%%%%%%%%%%%
%    Fig. 5
%%%%%%%%%%%%%%%
\begin{figure}[h]
\begin{center}
\hspace{10pt}
 \includegraphics[width=0.445\textwidth]{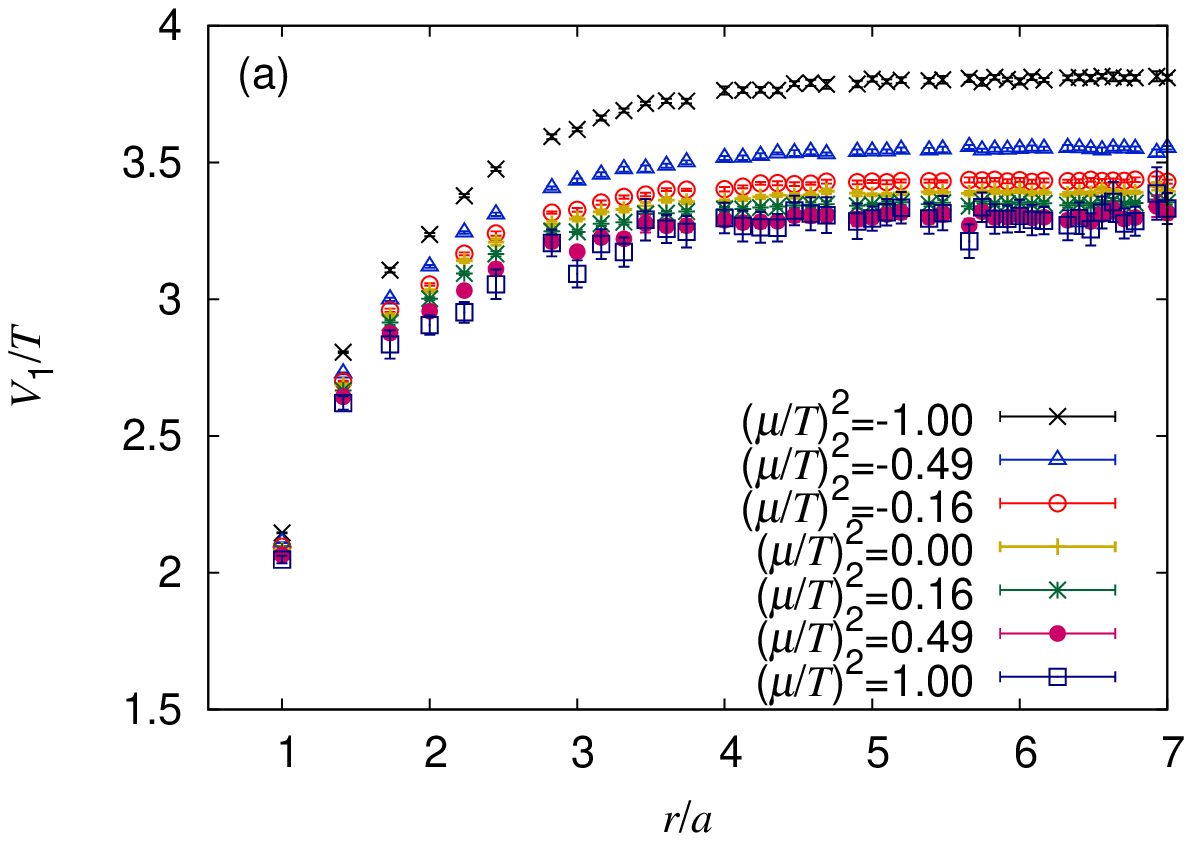}
 \includegraphics[width=0.445\textwidth]{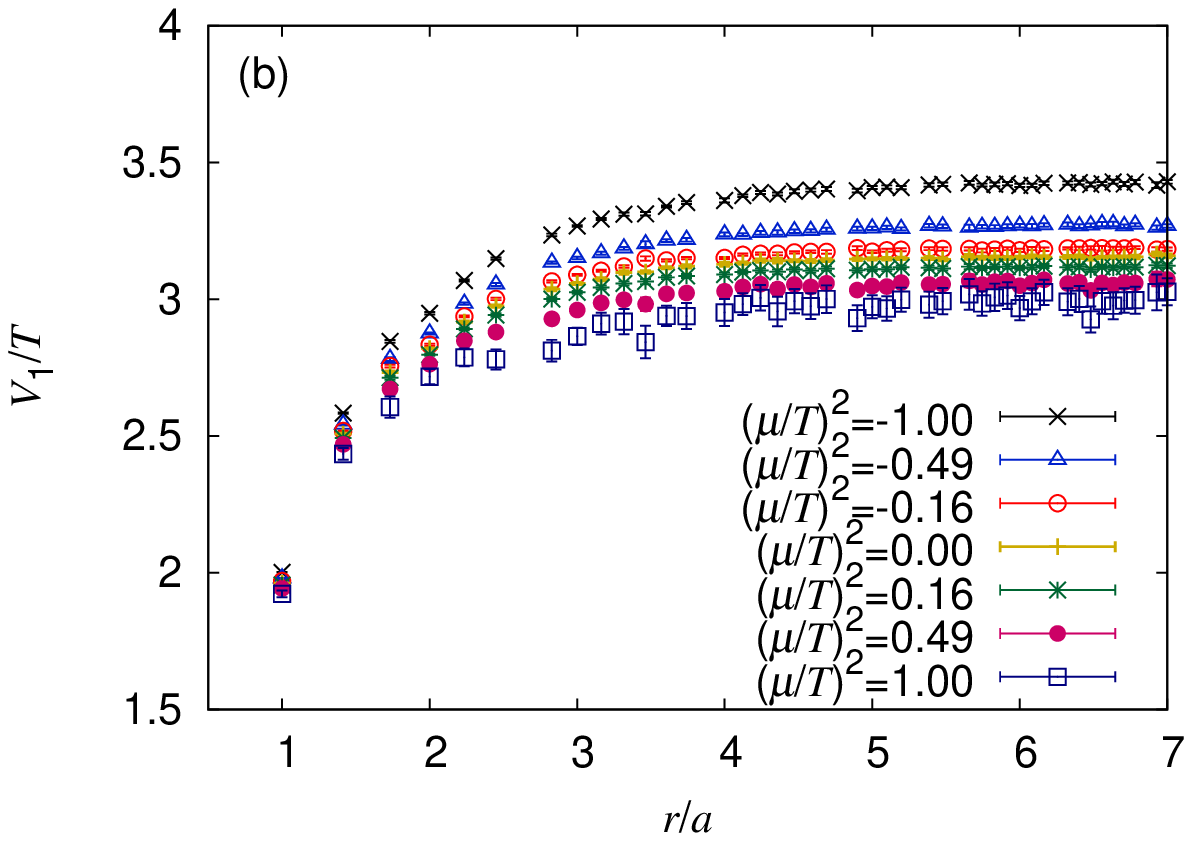}
\end{center}
\vspace{10pt}
\caption{
$\mu/T$ dependence of the color-singlet $q\bar{q}$ potential for (a) $T/T_{\mathrm{pc}}=1.20$ and (b) $T/T_{\mathrm{pc}}=1.35$.
}
\label{1pot_mu2-dep}
\end{figure}
%%%%%%%%%%%%%%

%%%%%%%%%%%%%%%
%    Fig. 6
%%%%%%%%%%%%%%%
\begin{figure}[h]
\begin{center}
\hspace{10pt}
 \includegraphics[width=0.445\textwidth]{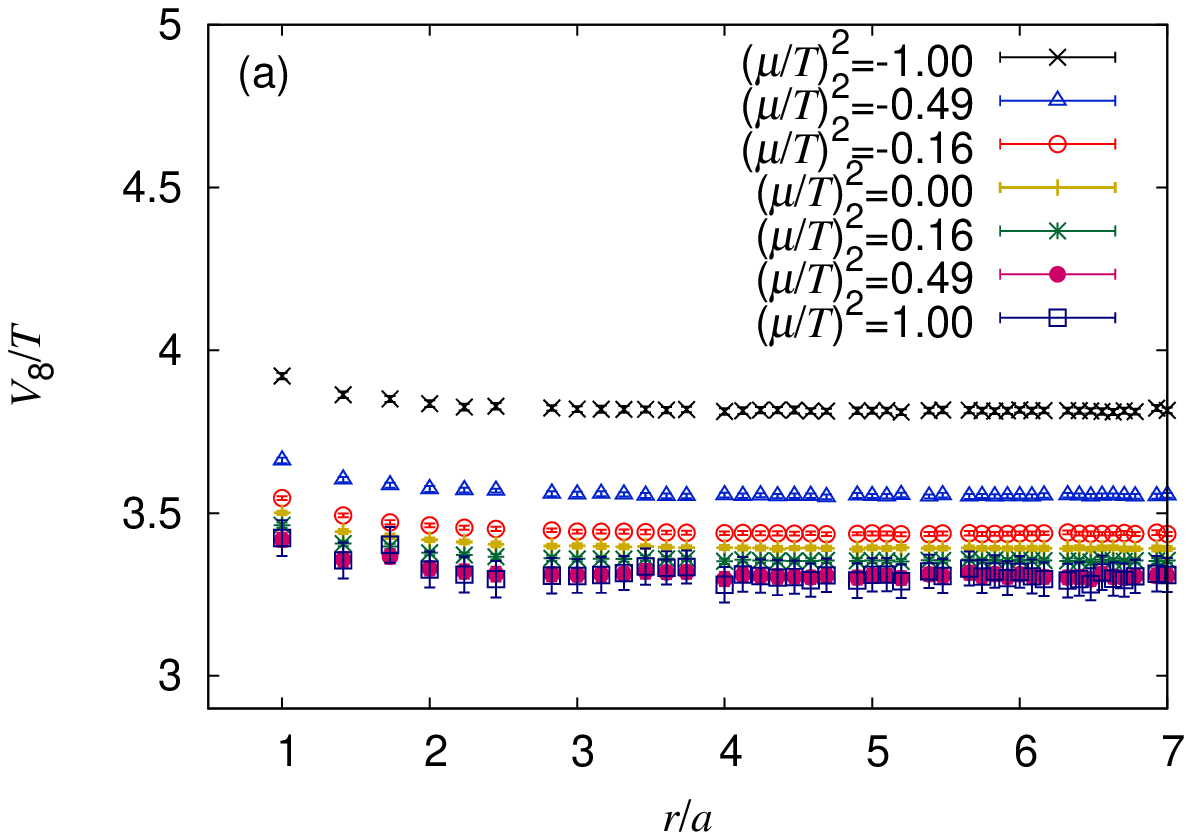}
 \includegraphics[width=0.445\textwidth]{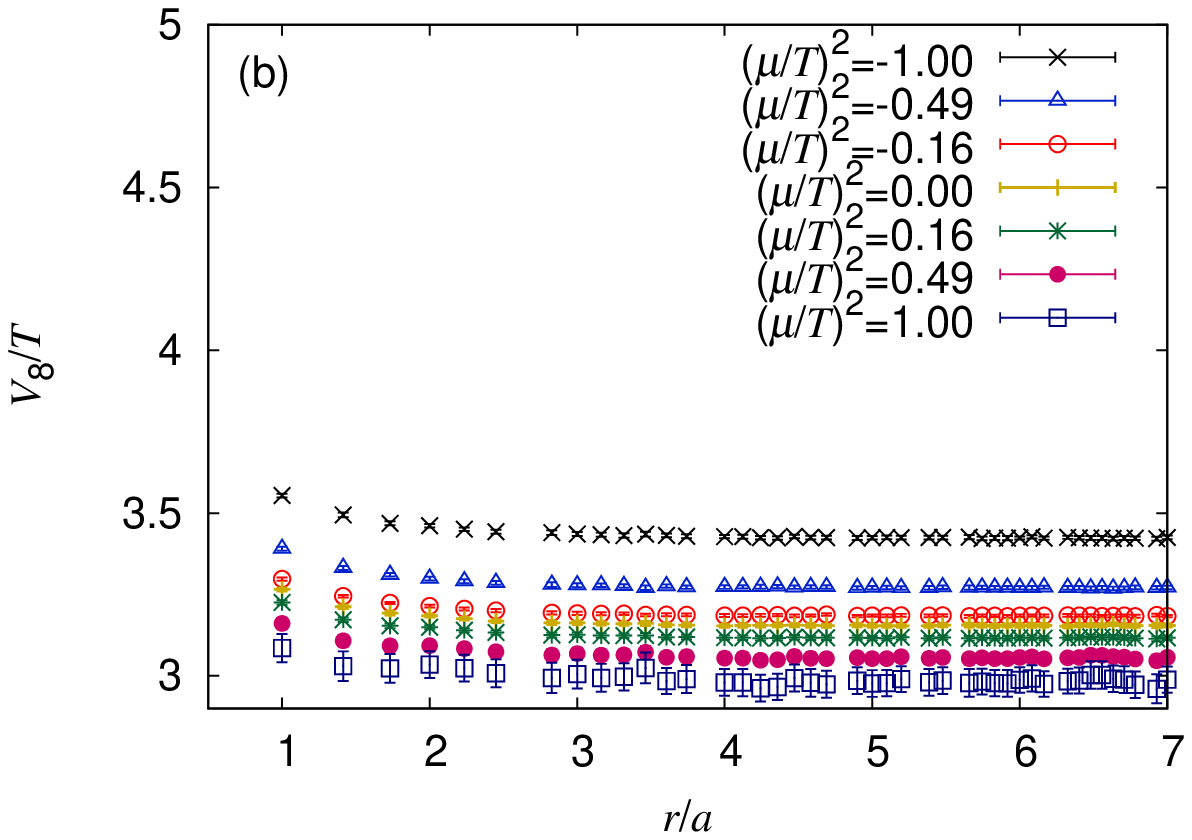}
\end{center}
\vspace{10pt}
\caption{
$\mu/T$ dependence of the color-octet $q\bar{q}$ potential 
for (a) $T/T_{\mathrm{pc}}=1.20$ and (b) $T/T_{\mathrm{pc}}=1.35$ .
}
\label{octet-pot_C}
\end{figure}
%%%%%%%%%%%%%%

%%%%%%%%%%%%%%%
%    Fig. 7
%%%%%%%%%%%%%%%
\begin{figure}[h]
\begin{center}
\hspace{10pt}
 \includegraphics[width=0.445\textwidth]{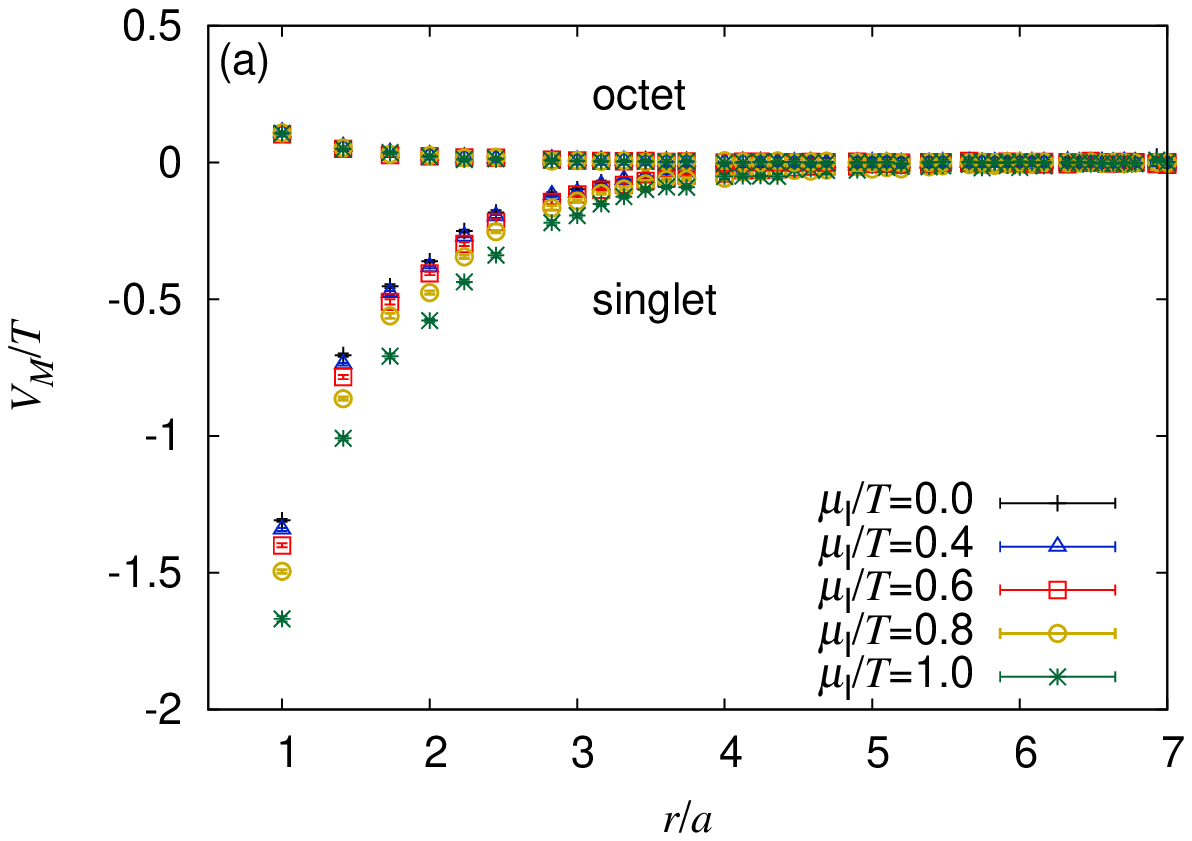}
 \includegraphics[width=0.445\textwidth]{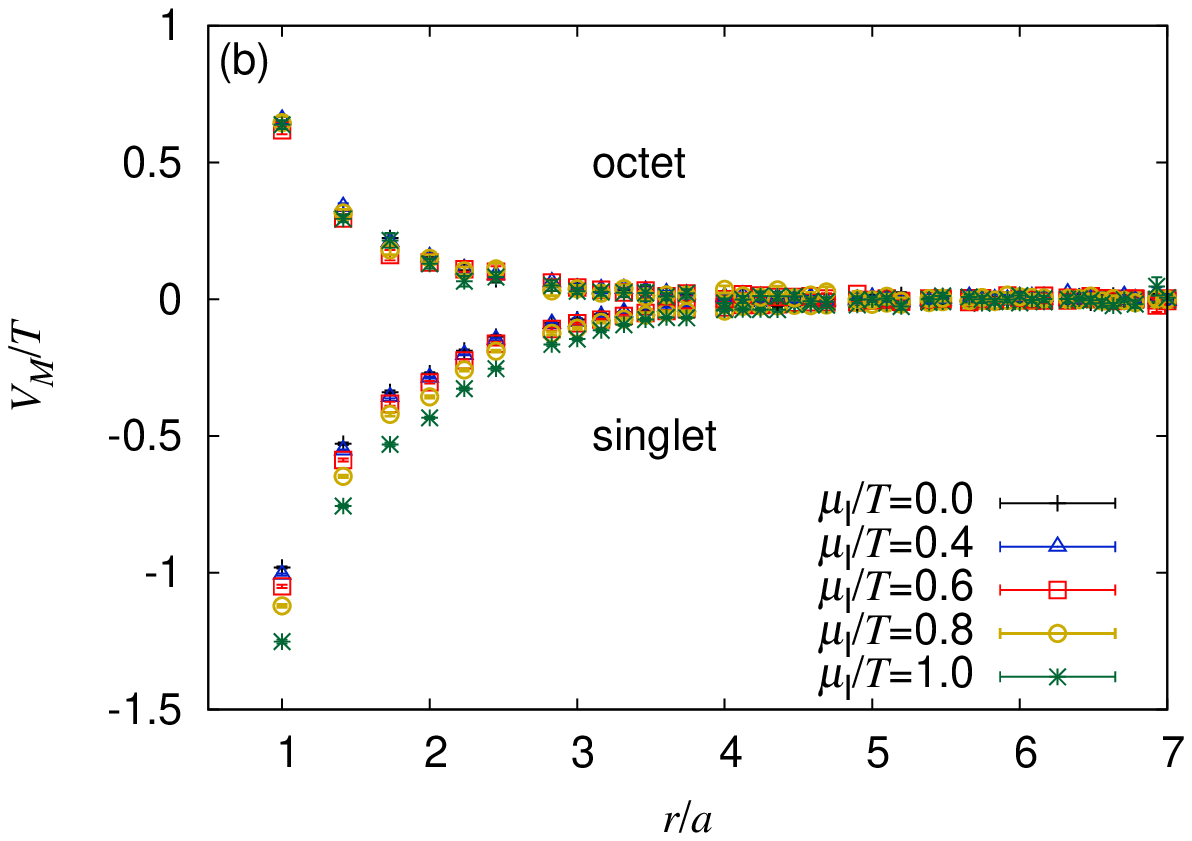}
\end{center}
\vspace{10pt}
\caption{
$\mu_{\rm I}/T$ dependence of the subtracted $q\bar{q}$ potentials in the color-singlet and -octet channels at $T/T_{\mathrm{pc}}=1.20$.
The potentials are divided by the absolute values of the corresponding 
Casimir factors in panel (b). Such a normalization is not taken in panel (a).
}
\label{1-8pot_mu-dep_b1950_re}
\end{figure}
%%%%%%%%%%%%%%

%%%%%%%%%%%%%%%
%    Fig. 8
%%%%%%%%%%%%%%%
\begin{figure}[h]
\begin{center}
\hspace{10pt}
 \includegraphics[width=0.445\textwidth]{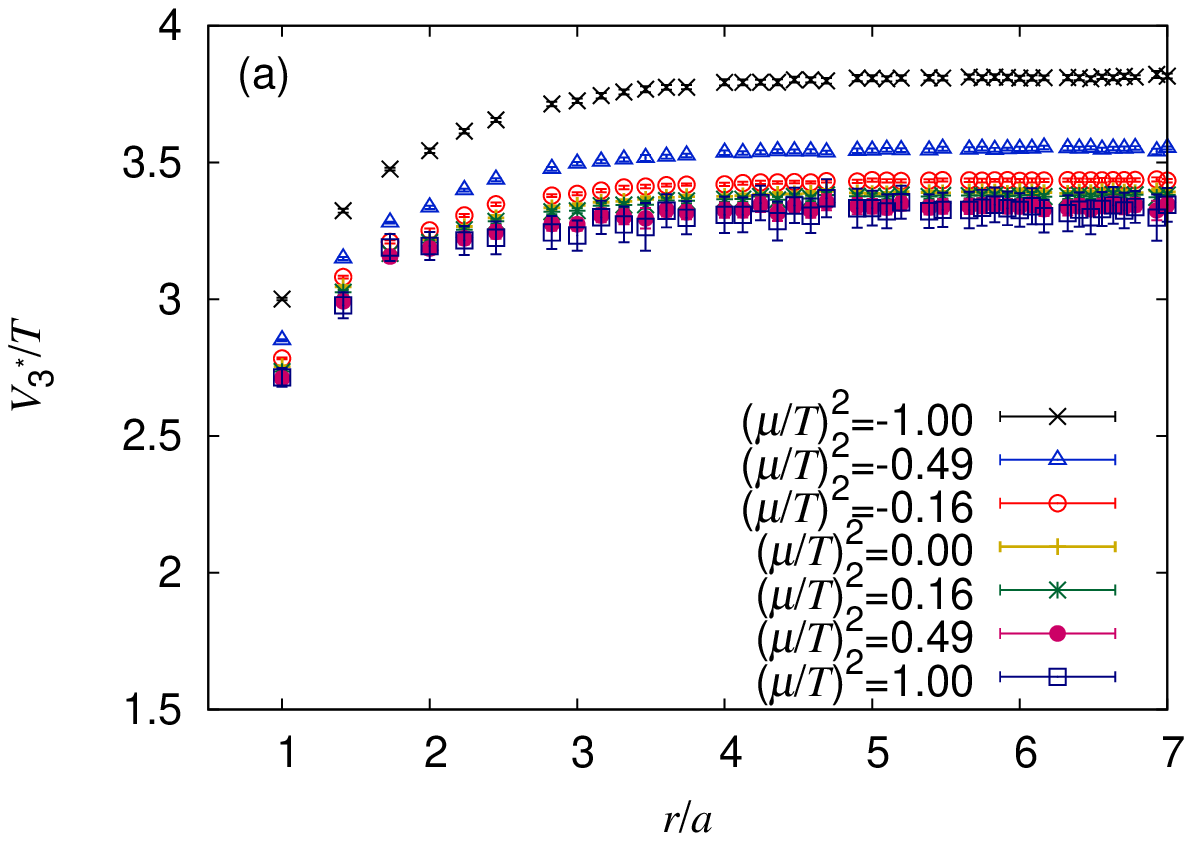}
 \includegraphics[width=0.445\textwidth]{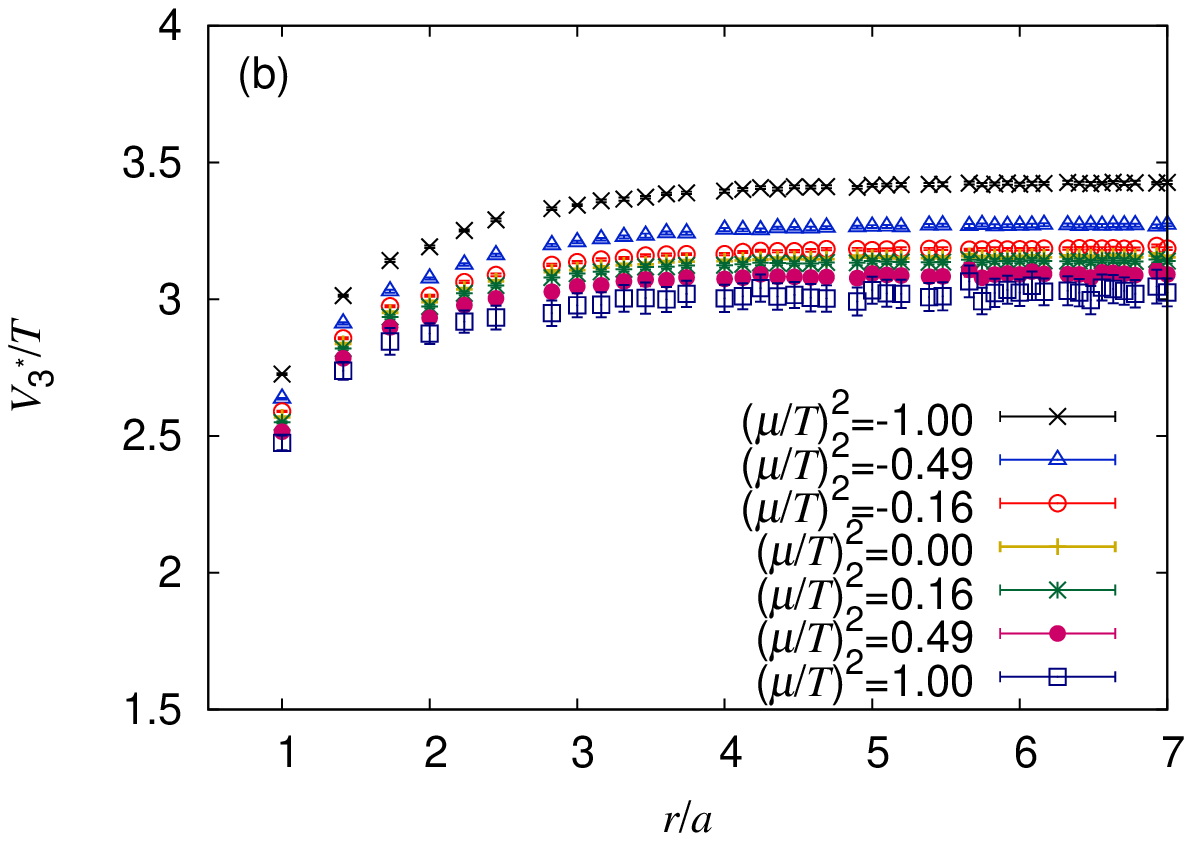}
\end{center}
\vspace{10pt}
\caption{
$\mu/T$ dependence of the real part of the color-antitriplet $qq$ potential 
for (a) $T/T_{\mathrm{pc}}=1.20$ and (b) $T/T_{\mathrm{pc}}=1.35$ .
}
\label{antitriplet-pot_C}
\end{figure}
%%%%%%%%%%%%%%

%%%%%%%%%%%%%%%
%    Fig. 9
%%%%%%%%%%%%%%%
\begin{figure}[h]
\begin{center}
\hspace{10pt}
 \includegraphics[width=0.445\textwidth]{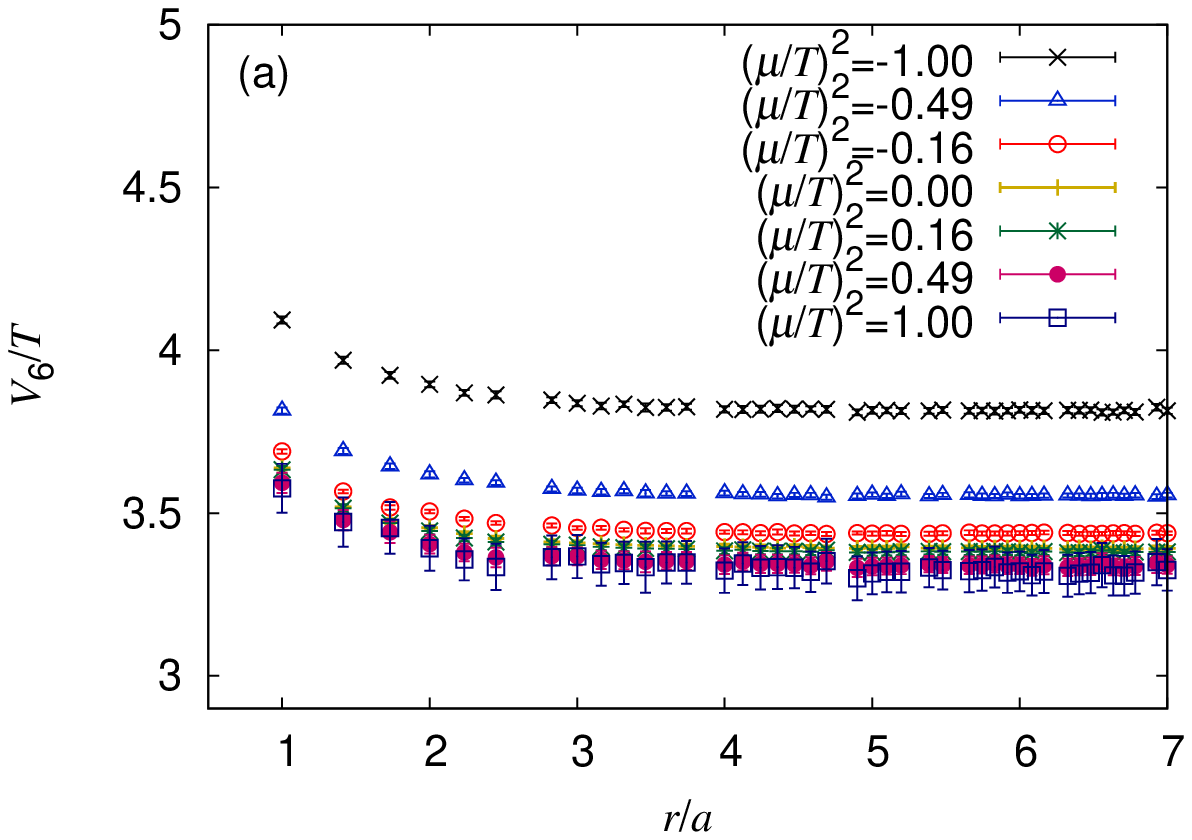}
 \includegraphics[width=0.445\textwidth]{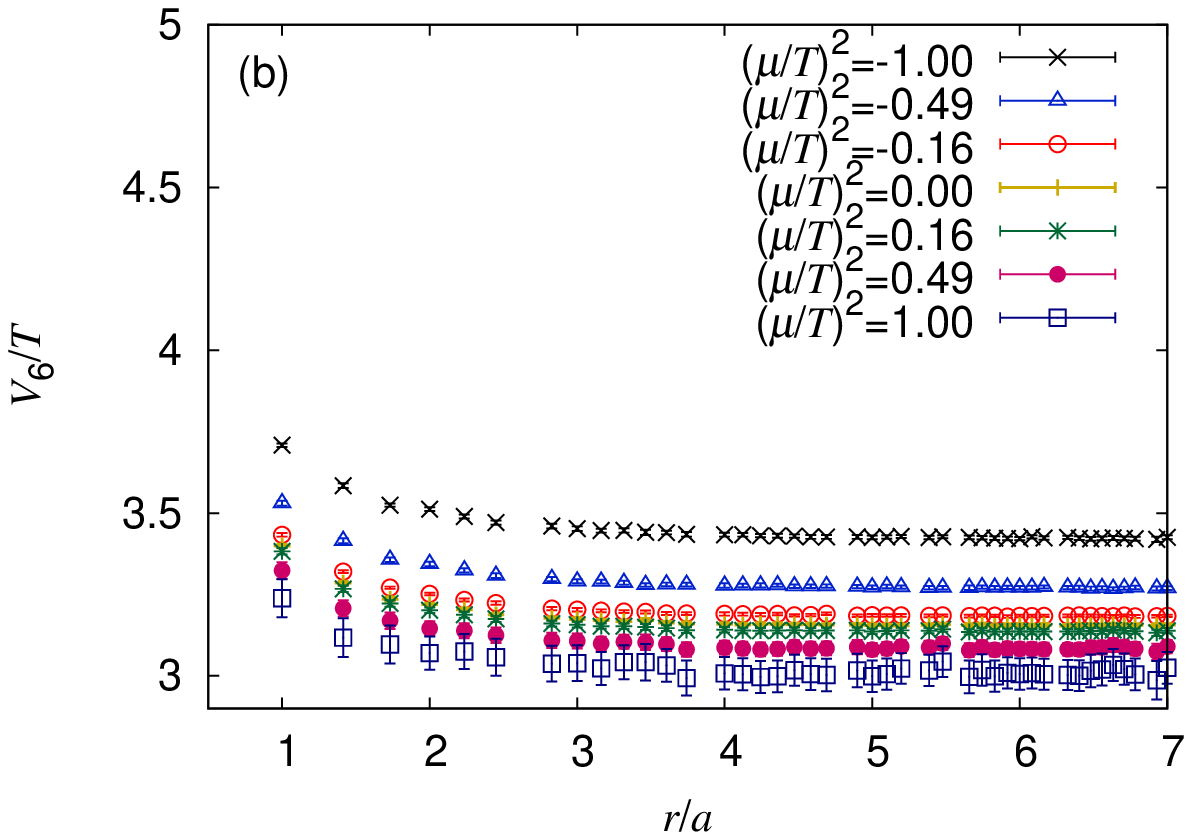}
\end{center}
\vspace{10pt}
\caption{
$\mu/T$ dependence of the real part of the color-sextet $qq$ potential
 for (a) $T/T_{\mathrm{pc}}=1.20$ and (b) $T/T_{\mathrm{pc}}=1.35$ .
}
\label{sextet-pot_C}
\end{figure}
%%%%%%%%%%%%%%

%%%%%%%%%%%%%%%
%    Fig. 10
%%%%%%%%%%%%%%%
\begin{figure}[htbp]
\begin{center}
\hspace{10pt}
 \includegraphics[width=0.445\textwidth]{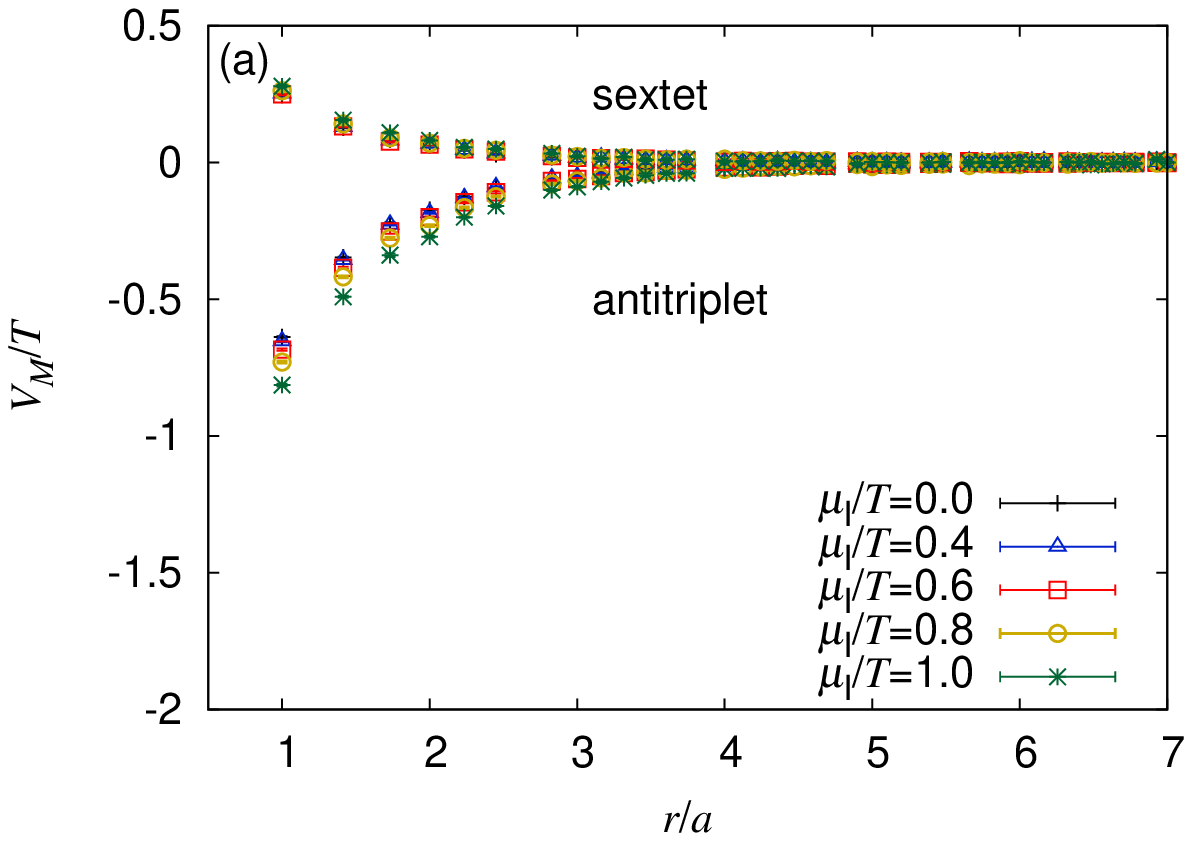}
 \includegraphics[width=0.445\textwidth]{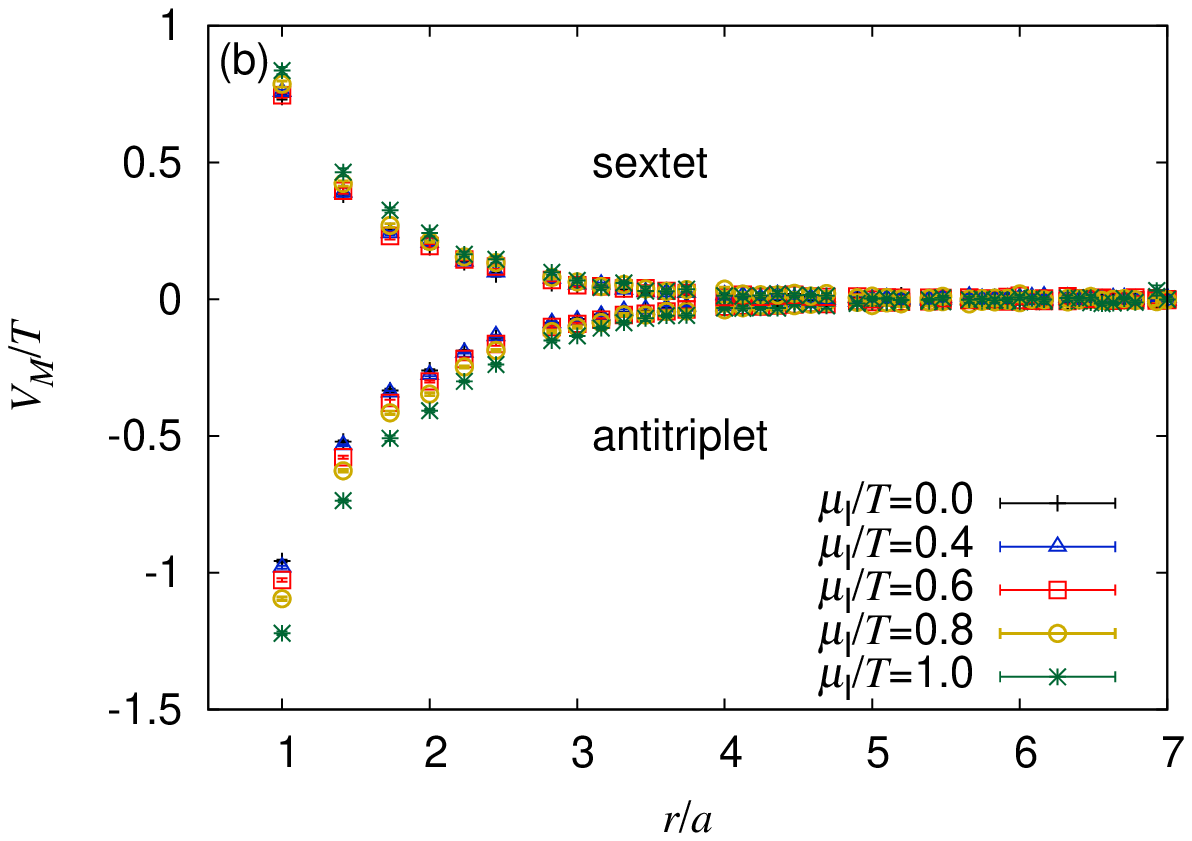}
\end{center}
\vspace{10pt}
\caption{
$\mu_{\rm I}/T$ dependence of the subtracted $qq$ potentials in the sextet and anti-triplet channel at $T/T_{\mathrm{pc}}=1.20$.
The potentials are divided by the absolute values of the corresponding 
Casimir factors in panel (b). Such a normalization is not taken in panel (a).
}
\label{3-6pot_mu-dep_b1950_re}
\end{figure}
%%%%%%%%%%%%%%

\twocolumngrid

%%%%%%%%%%%%%%%
%    Fig. 11
%%%%%%%%%%%%%%%
\begin{figure}[htbp]
\begin{center}
\hspace{10pt}
 \includegraphics[width=0.445\textwidth]{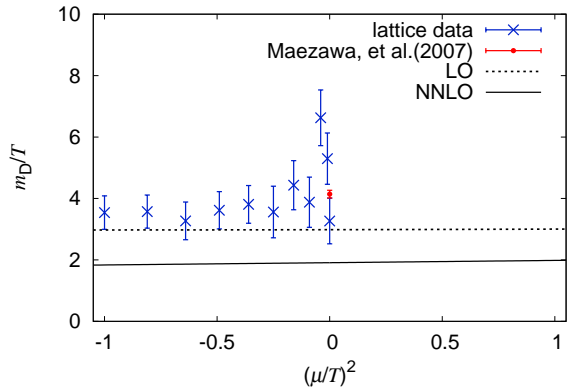}
\end{center}
\vspace{10pt}
\caption{
Comparison of lattice results with HTLpt results for 
$m_{\rm D}$ as a function of $(\mu/T)^{2}$ at $T/T_{\mathrm{pc}}=1.20$. 
The crosses denote the results of the present lattice simulations 
at imaginary $\mu$, whereas the circle is the result of 
the previous lattice simulations at $\mu=0$~\cite{Maezawa2}.
The dotted and solid lines are the HTLpt results 
in LO and NNLO 
at $\nu= 2\pi \sqrt{T^2+\mu^{2}/\pi^{2}}$, respectively.
}
\label{mD_mu2dep-HTL}
\end{figure}
%%%%%%%%%%%%%%

\onecolumngrid

%%%%%%%%%%%%%%%
%    Fig. 12
%%%%%%%%%%%%%%%
\begin{figure}[h]
\begin{center}
\hspace{10pt}
 \includegraphics[width=0.445\textwidth]{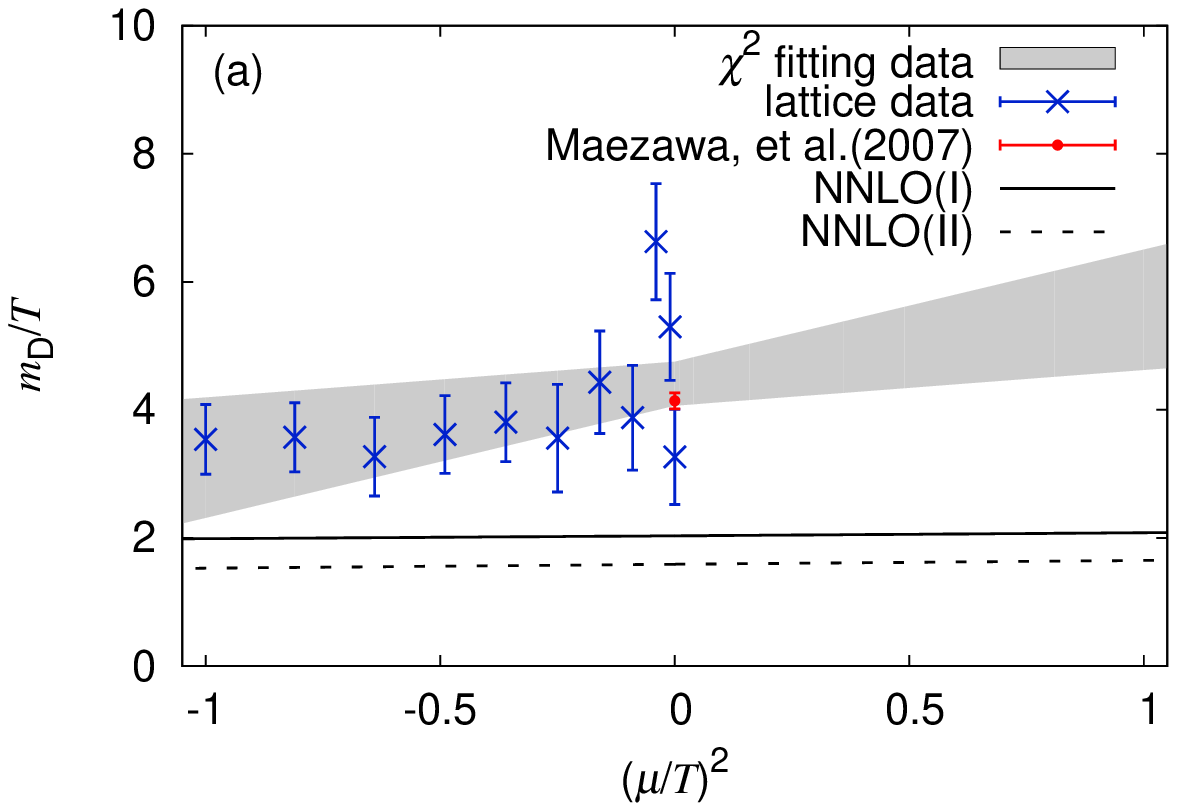}
 \includegraphics[width=0.445\textwidth]{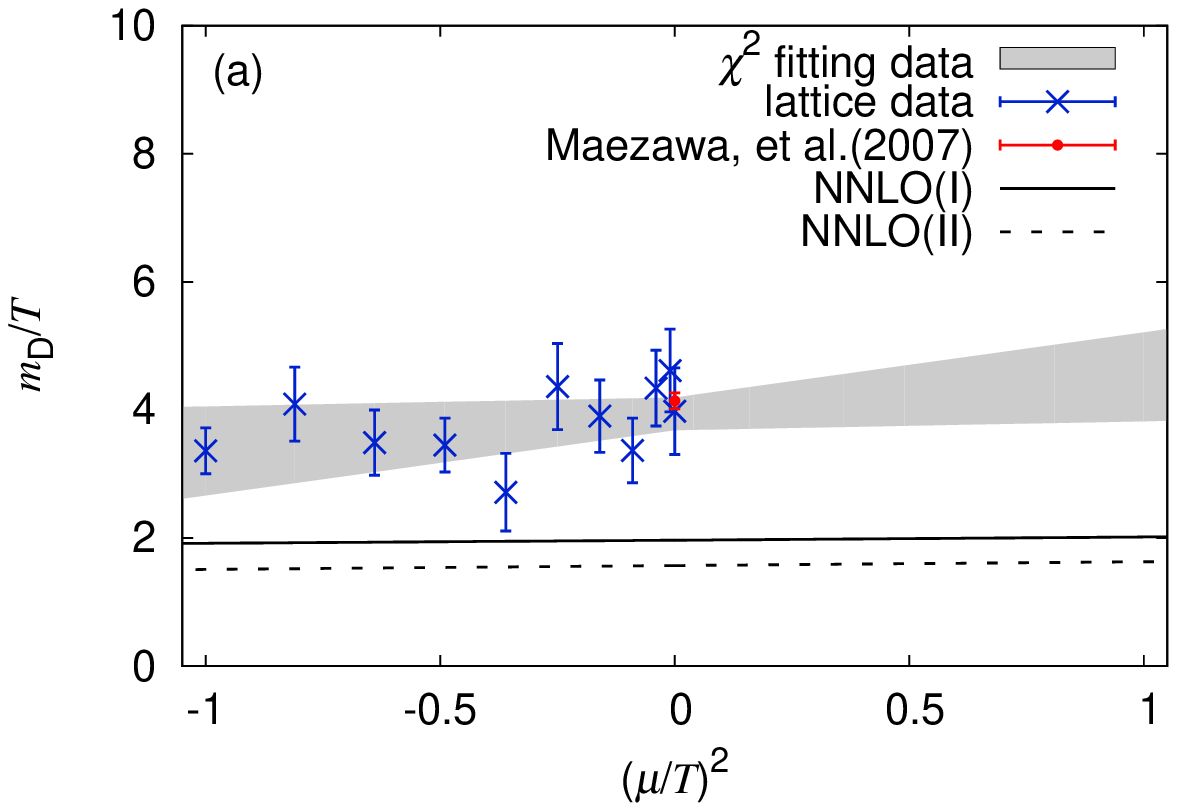}
\end{center}
\vspace{10pt}
\caption{
$(\mu/T)^{2}$ dependence of the color-Debye screening mass for (a) $T/T_{\mathrm{pc}}=1.20$ and (b) $1.35$.
The screening mass is determined from the singlet potential.
The crosses with error bars denote the results of the present lattice simulations at imaginary $\mu$, while the circle with an error bar is the result of the previous lattice simulations at $\mu=0$~\cite{Maezawa2}.
The solid and dashed lines are the NNLO HTLpt results at $\nu= \pi \sqrt{T^2+\mu^{2}/\pi^{2}}$ and $4 \pi \sqrt{T^2+\mu^{2}/\pi^{2}}$, respectively.
See Table~\ref{table-mDb1950} and \ref{table-mDb2000} in Appendix A for the numerical data at imaginary $\mu$.
}
\label{mD_mu2dep}
\end{figure}
%%%%%%%%%%%%%%

%%%%%%%%%%%%%%%
%    Fig. 13
%%%%%%%%%%%%%%%
\begin{figure}[htbp]
\begin{center}
\hspace{10pt}
 \includegraphics[width=0.445\textwidth]{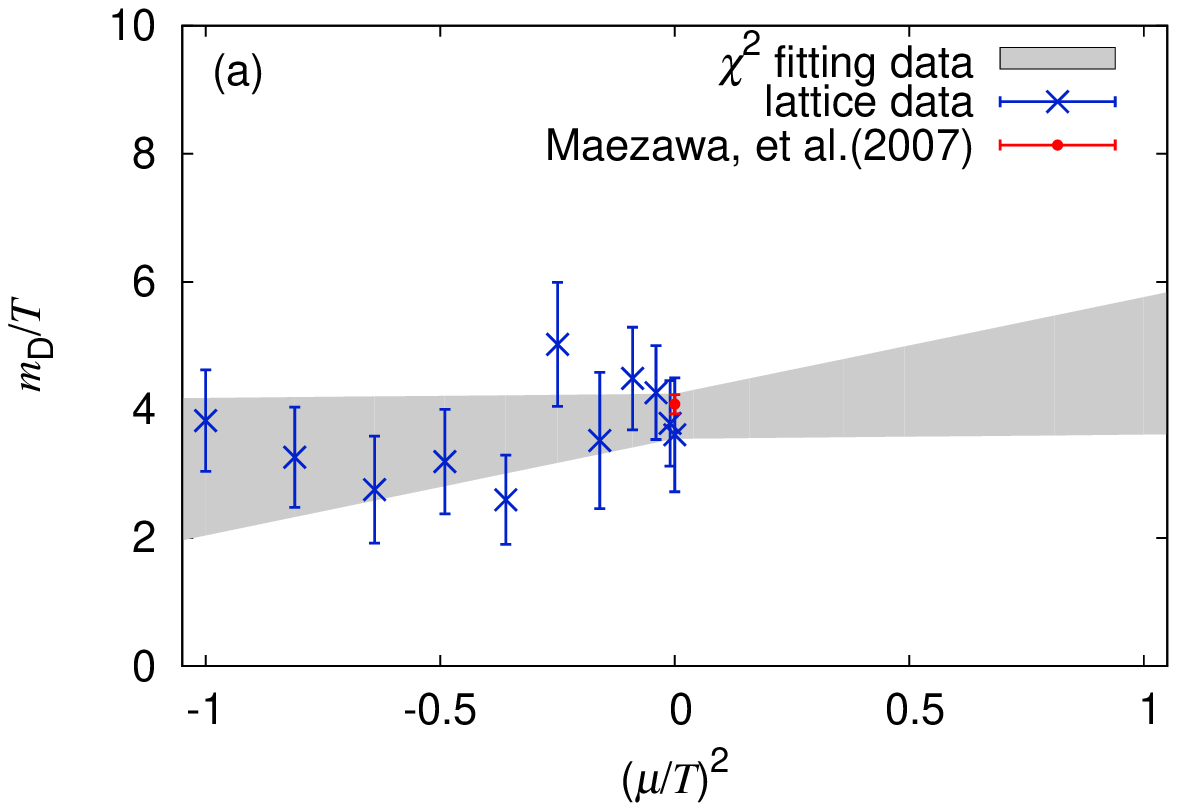}
 \includegraphics[width=0.445\textwidth]{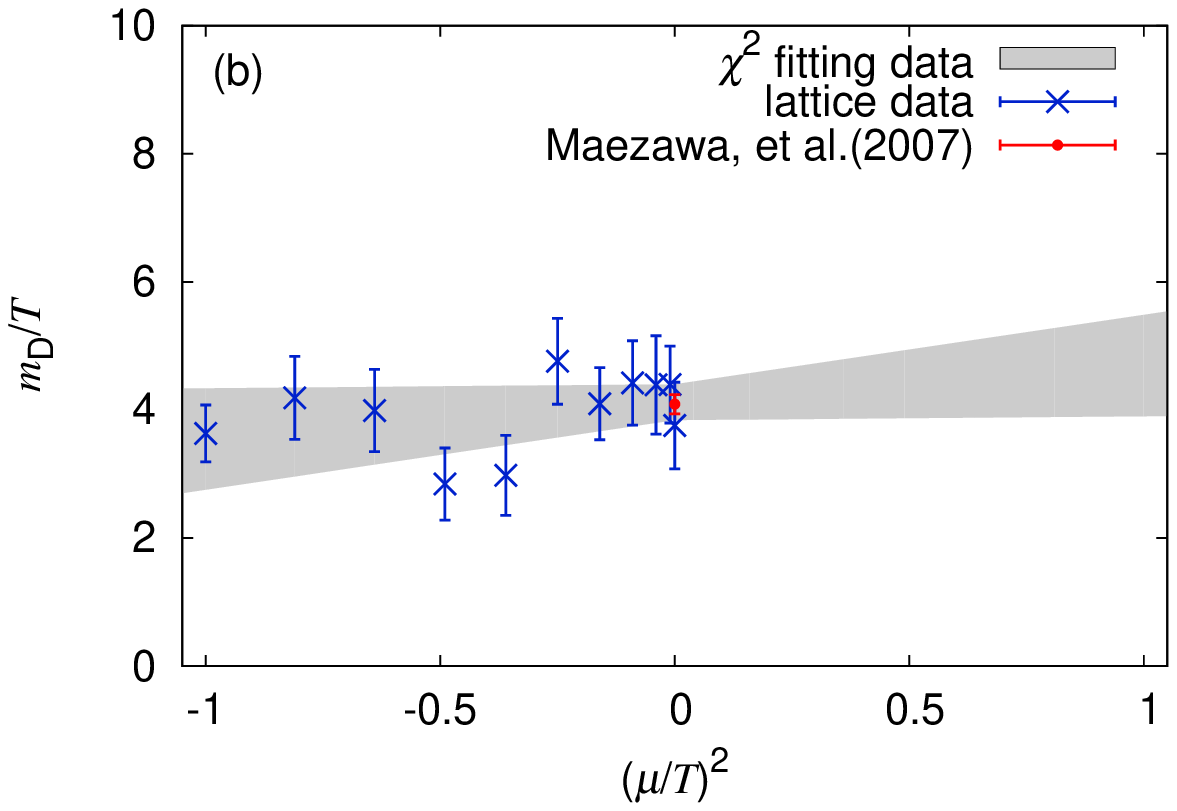}
\end{center}
\vspace{10pt}
\caption{
$(\mu/T)^{2}$ dependence of the color-Debye screening mass for (a) $T/T_{\mathrm{pc}}=1.20$ and (b) $1.35$. 
The screening mass is determined from the antitriplet potential.
The crosses with error bars denote the results of the present lattice simulations at imaginary $\mu$, while the circle with an error bar is the result of the previous lattice simulations at $\mu=0$~\cite{Maezawa2}.
See Tables~\ref{table-mDb1950} and \ref{table-mDb2000} in appendix A for the numerical data at imaginary $\mu$.
}
\label{mD_mu2dep-3*}
\end{figure}
%%%%%%%%%%%%%%

\twocolumngrid

%%%%%%%%%%%%%%%%%%%%%%%%%%%%%%%%%%%%%%%%%%%%%%
%%%%%%%%  q-qbar potential 
%%%%%%%%%%%%%%%%%%%%%%%%%%%%%%%%%%%%%%%%%%%%%%

\subsection{$q\bar{q}$ potential}
At imaginary $\mu$, the $q\bar{q}$ potentials, $V_{M}$ ($M=1, 8$), are ${\cal C}$-even and hence real. 
Meanwhile, the $qq$ potentials, $V_{M}$  ($M=3^{\ast}, 6$), are not ${\cal C}$-even and consequently becomes complex. 
We then consider the real part of $V_{M}$ for all the color channels.\\
\indent
Figure~\ref{oRWt-1pot_CL} shows the color-singlet potential $V_{1}(r)$ at $T/T_{\mathrm{pc}}=1.20$ based on (a) the Coulomb and (b) the Landau gauge fixing.
The potential has a small imaginary component coming from numerical errors in actual calculations, but it is neglected here.
Here imaginary chemical potential is varied from $\mu_{\rm I}/T=0$ to $1.2$.  As mentioned in Sec.~\ref{Sec:Polyakov loop}, ${\cal C}$-even quantities such as  $V_{M}$ are mirror symmetric with respect to the line $\mu_{\rm I}/T=\pi/3$. 
This property is satisfied for the Coulomb gauge fixing, but not for the Landau gauge fixing; compare results of $\mu_{\rm I}/T=0.9$ ($1.0$) 
with those of $\mu_{\rm I}/T=1.2$ ($1.1$). 
This result is natural, because the Coulomb gauge condition is invariant under the ${\mathbb Z}_3$ transformation, but the Landau 
gauge condition is not. 
For this reason, hereafter, we take the Coulomb gauge fixing.\\
\indent
Now we evaluate $v_2(r)$ and $v_4(r)$ from $V_{1}(r)$ at imaginary $\mu$ by expanding it as in \eqref{Eq:expansion-1}, and compare the original value of $V_{1}(r)$ with the value of the right-hand side of \eqref{Eq:expansion-1} in order to investigate the accuracy of the expansion. 
The relative error between the two values is less than 0.5~\%, indicating that the expansion is highly accurate. 
Those coefficients are shown in Fig.~\ref{Taylor-expansion-coefficient}; see Tables~\ref{table-V1V8b1950} and \ref{table-V1V8b2000} in Appendix A for the numerical data. 
The ratio $v_4(r)/v_2(r)$ is about 3/4 for $T/T_{\mathrm{pc}}=1.20$ and about 1/4 for $T/T_{\mathrm{pc}}=1.35$.
The contribution of $v_4(r)$ to $V_{1}(r)$ is thus significant near $T_{\mathrm{pc}}$ such as $T/T_{\mathrm{pc}}=1.20$. 
Even at higher $T$ such as $T/T_{\mathrm{pc}}=1.35$, the contribution is not negligible. 
Our data and the previous ones estimated by the Taylor-expansion method in Ref.~\cite{Ejiri} are plotted on the panel (b) in Fig. 4, for a comparison.
It is found that both data are consistent within error bars.\\
\indent
Figure~\ref{1pot_mu2-dep} shows the color-singlet potential at imaginary and real $\mu$ for  (a) $T/T_{\mathrm{pc}}=1.20$ and (b) $T/T_{\mathrm{pc}}=1.35$. 
The potential $V_{1}(r)$ at real $\mu$ is extracted by replacing $i\mu_{\rm I}/T$ by $\mu_{\rm R}/T$ in the Taylor-expansion 
series up to 4th order. 
The chemical potential is varied from $(\mu/T)^{2}=-1.0$ to 1.0. 
In this study, the lattice spacing $a$ is common to all $\mu$ for each $T$, so one can compare $V_M$ between different values of $\mu/T$ 
without additional adjustments to the short distance~\cite{Kaczmarek2,Maezawa1}.
The potential $V_{1}(r)$ tends to a common value independent of $\mu/T$ as $r$ decreases. Similar behavior is also seen 
for temperature dependence in Ref.~\cite{Maezawa1}.\\
\indent
The potential $V_1$ is ${\cal C}$-even, so that  $v_{1}(r)=v_{3}(r)=0$. 
Furthermore, if $v_{4}(r)=0$, the potential $V_1/T$ will linearly depend on $(\mu/T)^{2}$. 
For $T/T_{\mathrm{pc}}=1.20$, $v_{4}(r)$ is comparable to $v_{2}(r)$. For this property, in panel (a) of Fig.~\ref{1pot_mu2-dep}, $\mu/T$ dependence of  $V_1/T$ is much weaker at real $\mu$ than at imaginary $\mu$. 
In panel (b) of  $T/T_{\mathrm{pc}}=1.35$, the expansion coefficient $v_{4}(r)$ is still non-negligible compared with $v_{2}(r)$, so that $V_1/T$ has still weaker $\mu/T$ dependence at real $\mu$ than at imaginary $\mu$.\\
\indent
As an estimate of the accuracy of the Taylor-expansion series up to 4th order, we assume that the expansion series is reliable when the magnitude of the 4th-order term is less than 10\% of the total potential $V_1$.
We have numerically confirmed that this condition is satisfied at $\mu_{\rm R}/T \lsim 1.2$ for $T/T_{\mathrm{pc}}=1.20$ and 
at $\mu_{\rm R}/T \lsim 1.4$ for $T/T_{\mathrm{pc}}=1.35$. 
The analytic continuation of $V_1$ from imaginary $\mu$ to real $\mu$ may thus be valid at least up to $\mu/T =1.2$.\\
\indent
The same analysis is made in Fig.~\ref{octet-pot_C} for the color-octet potential $V_{8}(r)/T$; see Tables~\ref{table-V1V8b1950} and \ref{table-V1V8b2000} in Appendix A for numerical data on the Taylor-expansion coefficients, $v_2(r)$ and $v_4(r)$, of $V_{8}(r)$.
The potential has a small imaginary component coming from numerical errors in actual calculations, but it is neglected here. 
When $\mu/T$ is varied, the potentials are different in magnitude, but similar in $r$ dependence.
Therefore, the interaction $-d(V_{8}/T)/d(r/a)$ has weak $\mu/T$ dependence.
Unlike the color-singlet potential $V_1(r)$, the color-octet potential $V_{8}(r)$ does not tend to a common value independent of $\mu/T$, as $r$ decreases. 
Meanwhile, the magnitude decreases monotonically as $(\mu/T)^2$ increases from -1 to 1. 
Again, $\mu/T$ dependence of the magnitude is weaker at real $\mu$ than at imaginary $\mu$, because $v_4$ is non-negligible.\\
\indent
For the case of $T > T_{\rm pc}$ and $\mu=0$, the potentials $V_{M}(r)$ are known to tend to twice the single-quark free energy $2F_{q}(T,\mu_{\rm I})$ in the limit of large $r$~\cite{Maezawa1}. 
This behavior persists also for finite $\mu$.
The interactions between static quarks are thus color screened also for finite $\mu$.
Following the previous works~\cite{Maezawa1,Saito1,Saito2,Maezawa2,Ejiri}, we then subtract $2F_{q}(T,\mu_{\rm I})$ from $V_{M}(r)$.
This subtracted static-quark potentials are convenient to see properties of the corresponding interactions $-d(V_{M}/T)/d(r/a)$. 
The subtracted static-quark potentials are shown in Fig.~\ref{1-8pot_mu-dep_b1950_re}(a) for the color-singlet and -octet channels.
The singlet interaction is attractive, while the octet interaction is repulsive. In perturbation, the color-channel dependence comes from the Casimir factor $C_{M}\equiv \langle \sum^{8}_{a=1}t^{a}_{1}\cdot t^{a}_{2}\rangle_{M}$ for channel $M$, i.e., 
\bea
C_{1}=-\frac{4}{3}, \; C_{8}=\frac{1}{6}, \; C_{6}=\frac{1}{3}, \; C_{3^{\ast}}=-\frac{2}{3}.
\eea
The subtracted potentials are then divided by the absolute values of the corresponding Casimir factors in order to extract non-perturbative properties from them. As shown in Fig.~\ref{1-8pot_mu-dep_b1950_re}(b), even after the normalization, the singlet interaction is stronger than the octet one, and the former has larger $\mu_{\rm I}/T$ dependence than the latter.

%%%%%%%%%%%%%%%%%%%%%%%%%%%%%%%%%%%%%%%%%%%%%%
%%%%%%%%  q-q potential 
%%%%%%%%%%%%%%%%%%%%%%%%%%%%%%%%%%%%%%%%%%%%%%
\subsection{$qq$ potential} 
Figures \ref{antitriplet-pot_C} and \ref{sextet-pot_C} show the real parts of the color-sextet and -antitriplet $qq$ potentials, respectively, for (a) $T/T_{\mathrm{pc}}=1.20$ and (b) $T/T_{\mathrm{pc}}=1.35$.
We tabulate numerical data on the Taylor-expansion coefficients in Tables~\ref{table-V3V6b1950} and \ref{table-V3V6b2000} of Appendix A.
Again, the potentials tend to twice  the single-quark free energy at large distance, indicating  that the color screening takes place also for these non-singlet channels, even if $\mu$ is finite. 
Now twice the single-quark free energy are subtracted from the potentials.
The results are shown in Fig.~\ref{3-6pot_mu-dep_b1950_re}(a). 
The antitriplet interaction is attractive, while the sextet interaction is repulsive, as expected from perturbation. 
In Fig.~\ref{3-6pot_mu-dep_b1950_re}(b), the potentials are divided by the absolute values of the corresponding Casimir factors. 
Even after the normalization, the attractive antitriplet interaction is stronger than the repulsive sextet interaction, and the former has stronger $\mu_{\rm I}/T$ dependence than the latter.

%%%%%%%%%%%%%%%%%%%%%%%%%%%%%%%%%%%%%%%%%%%%%%
%%%%%%%%  color-Debye screening mass 
%%%%%%%%%%%%%%%%%%%%%%%%%%%%%%%%%%%%%%%%%%%%%%

\subsection{Color-Debye screening mass}
In order to analyze the color screening effect, we fit the static-quark potential to the screened Coulomb form
\bea
V_{M}(r,T,\mu)=C_{M}\frac{\alpha_{\mathrm{eff}}(T,\mu)}{r}e^{-m_{\rm D}(T,\mu)r},
\label{eq:Yukawa-form}
\eea
where $\alpha_{\mathrm{eff}}$ and $m_{\rm D}(T,\mu)$ are the effective running coupling and the Debye screening mass, respectively. 
We first focus our discussion on the color-singlet channel that is most important in the real world. 
Since $V_{1}=0$ in the limit of large $r$ in (\ref{eq:Yukawa-form}), we extract the screening mass from the subtracted static-quark potential. Following the previous work~\cite{Maezawa2}, we choose a fit range of $\sqrt{11} \le r/a \le 6.0$.\\
\indent
The color-Debye screening mass is calculable with the hard thermal loop perturbation theory (HTLpt). In leading-order (LO), 
$m_{\rm D}$ is obtained \cite{LeBellac} by
\bea
\frac{m_{\rm D}(T,\mu)}{T}=g_{\mathrm{2l}}(\nu)\sqrt{\left(1+\frac{N_{f}}{6}\right)+\frac{N_{f}}{2\pi^{2}}\left(\frac{\mu}{T}\right)^{2}}
\eea
with the 2-loop running coupling $g_{\mathrm{2l}}$ given by
\bea
g^{-2}_{\mathrm{2l}}(\nu)=\beta_{0}\ln\left(\frac{\nu}{\Lambda}\right)^{2}+\frac{\beta_{1}}{\beta_{0}}\ln \ln\left(\frac{\nu}{\Lambda}\right)^{2}, 
\eea
where the argument in the logarithms is rewritten into $\nu/\Lambda=(\nu/T)(T/T_{\mathrm{pc}})(T_{\mathrm{pc}}/\Lambda)$ with $\Lambda=\Lambda^{N_{f}=2}_{\bar{\mathrm{MS}}}\simeq 261$ MeV~\cite{Goeckeler} and $T_{\mathrm{pc}}\simeq 171$ MeV~\cite{Khan1}, and the renormalization point $\nu$ is assumed to be $\nu=2 \pi \sqrt{T^2+\mu^{2}/\pi^{2}}$~\cite{Ipp}. 
Beyond LO, $m_{\rm D}$ was evaluated in next-to next-to leading-order (NNLO)~\cite{Najmul} by the Braaten-Nieto prescription~\cite{Braaten-Nieto} that is the effective-field-theory approach based on ``dimensional reduction".\\
\indent
In Fig. \ref{mD_mu2dep-HTL}, the screening mass is plotted as a function of $(\mu/T)^2$ at $T/T_{\mathrm{pc}}=1.20$. 
The lattice results (crosses) are compared with the LO and NNLO HTLpt results of Refs. \cite{LeBellac,Najmul} at $\nu=2 \pi \sqrt{T^2+\mu^{2}/\pi^{2}}$. 
In both LO and NNLO, $m_{\rm D}$ has weak $\mu$ dependence.\\
\indent
Figure~\ref{mD_mu2dep} shows the $(\mu/T)^{2}$ dependence of $m_{\rm D}$ for (a) $T/T_{\mathrm{pc}}=1.20$ and (b) $T/T_{\mathrm{pc}}=1.35$. 
The lattice-simulation results are plotted by the cross symbols. 
It is not easy to determine $m_{\rm D}$ from the static-quark potential at large $r$, since the potential has weak $r$ dependence there. 
As a result of this problem, the resultant $m_{\rm D}$ is a bit scattered as a function of $\mu$. 
The screening mass is then expanded up to 2nd order of $\mu/T$: 
\bea
 \frac{m_{\rm D}}{T}=a_0(T) + a_2(T)\left(\frac{\mu}{T}\right)^{2},
\label{m_D-expansion}  
\eea
where note that $m_{\rm D}$ is ${\cal C}$-even and hence it has no linear term of $\mu/T$. 
The coefficients, $a_0(T)$ and $a_2(T)$, are determined from the $m_{\rm D}$ at imaginary $\mu$ with the $\chi^2$ fitting:
\bea
\frac{m_{\rm D}}{T}=(4.41 \pm 0.34) 
+  (1.15 \pm 0.60)\left(\frac{\mu}{T}\right)^{2}  
\eea 
for $T/T_{\mathrm{pc}}=1.20$ and 
\bea
\frac{m_{\rm D}}{T}=(3.94 \pm 0.25) 
+ (0.58 \pm  0.44)\left(\frac{\mu}{T}\right)^{2}  
\eea 
for $T/T_{\mathrm{pc}}=1.35$. The screening mass at real $\mu$ is extrapolated from that at imaginary $\mu$ by using \eqref{m_D-expansion}. 
The results of the extrapolation, denoted by the hatching area, are consistent with the previous LQCD result, shown by the circle symbol, at $\mu=0$~\cite{Maezawa2} for both $T/T_{\mathrm{pc}}=1.20$ and  $1.35$.\\
\indent
In Fig.~\ref{mD_mu2dep}, the dashed and solid lines denote the results of NNLO HTLpt calculations at $\nu= \pi \sqrt{T^2+\mu^{2}/\pi^{2}}$ and $4 \pi \sqrt{T^2+\mu^{2}/\pi^{2}}$, respectively. 
Comparing the hatching area with the dashed and solid lines, one may see that the LQCD results have stronger $\mu/T$ dependence than the prediction of the HTLpt.\\
\indent
In principle the same analysis is possible for the non-singlet channels, but in practice the color-Debye screening masses derived from the octet and sextet potentials have large errors since the magnitudes of the potentials are small. 
We then consider only the antitriplet channel and draw the lattice-simulation results with cross symbols in Fig.~\ref{mD_mu2dep-3*}. 
The $\chi^2$ fitting to the lattice results at imaginary $\mu$ yields 
\bea
\frac{m_{\rm D}}{T}=(3.90 \pm 0.35) 
+  (0.79 \pm 0.72)\left(\frac{\mu}{T}\right)^{2}  
\eea 
for $T/T_{\mathrm{pc}}=1.20$ and 
\bea
\frac{m_{\rm D}}{T}=(4.12 \pm 0.28) 
+ (0.58 \pm  0.51)\left(\frac{\mu}{T}\right)^{2}  
\eea 
for $T/T_{\mathrm{pc}}=1.35$.
Results of the extrapolation (the hatching area) are consistent with the previous LQCD result (the circle symbol) at $\mu=0$~\cite{Maezawa2} for both cases of $T/T_{\mathrm{pc}}=1.20$ and  $1.35$. 
Magnitudes and $\mu/T$ dependences of the screening masses thus obtained are similar to each other between the singlet and antitriplet channels. 
The screening masses increase as $\mu$ increases and the $\mu$ dependence is stronger than the prediction of the perturbation theory.
The numerical values of the color-Debye screening masses are tabulated in Tables~\ref{table-mDb1950} and \ref{table-mDb1950} of Appendix A.

%%%%%%%%%%%%%%%%%%%%%%%%%%%%%%%%%%%%%%%%%%%%%%%%%%
%%%%%%%%  Summary 
%%%%%%%%%%%%%%%%%%%%%%%%%%%%%%%%%%%%%%%%%%%%%%%%%%
\section{Summary}
\label{Summary}
We have investigated $\mu$ dependence of the static-quark free energies (potentials) and the color-Debye screening mass in both the imaginary and real $\mu$ regions, performing LQCD simulations at imaginary $\mu$ and extrapolating the result to the real $\mu$ region with analytic continuation. LQCD calculations are done on a $16^{3}\times 4$ lattice with the clover-improved two-flavor Wilson fermion action and the renormalization-group improved Iwasaki gauge action. 
We took an intermediate quark mass and considered two cases of $T/T_{\mathrm{pc}}=1.20$ and $1.35$.\\
\indent
The static-quark potential at real $\mu$ was obtained by expanding the potential at imaginary $\mu$ into a Taylor-expansion series of $i\mu_{\rm I}/T$ up to 4th order and replacing $i\mu_{\rm I}$ to $\mu_{\rm R}$. 
Since the expansion series was taken only up to 2nd order in the previous analysis~\cite{Ejiri}, this is the first analysis that investigates contributions of the 4th-order term to the potential. 
We found that at real $\mu$ the 4th-order term weakens $\mu$ dependence of the potential sizably. This effect becomes more 
significant as $T$ decreases toward $T_{\mathrm{pc}}$.\\
\indent
We have also investigated color-channel dependence of the static-quark potentials. At large distance, all the potentials tend to twice the single-quark free energy, indicating that the interactions are fully color screened. Although this property is known for finite $T$ and zero $\mu$~\cite{Maezawa1}, the present analysis shows that the property persists also for finite $\mu$.
For both real and imaginary $\mu$, the color-singlet $q{\bar q}$ and the color-antitriplet $qq$ interaction are attractive, whereas the color-octet $q{\bar q}$ and the color-sextet $qq$ interaction are repulsive. 
These interactions are divided by the absolute values of the corresponding Casimir factors in order to extract non-perturbative properties from them.  
Even after the normalization, the attractive interactions are stronger than the repulsive interactions, and the former interactions have stronger $\mu/T$ dependence than the latter ones.\\
\indent
The color-Debye screening mass is evaluated from the color-singlet potential at imaginary $\mu$. The screening mass thus obtained at imaginary $\mu$  is extrapolated to real $\mu$ by expanding the mass at imaginary $\mu$ into a power series of $i\mu_{\rm I}/T$ up to 2nd order and replacing $i\mu_{\rm I}/T$ by $\mu_{\rm R}/T$. 
The resulting mass has stronger $\mu$ dependence at both imaginary and  real $\mu$ than the HTLpt prediction. 

%%%%%%%%%%%%%%%%%%%%%%%%%%%%%%%%%%%%%%%%%%%%%%%%%%%%%%%%%%%%%%%%%%%%%%%%%%%%%%%%
%%%%% Acknowledgments 
%%%%%%%%%%%%%%%%%%%%%%%%%%%%%%%%%%%%%%%%%%%%%%%%%%%%%%%%%%%%%%%%%%%%%%%%%%%%%%%%
\noindent
\begin{acknowledgments}
Junichi Takahashi is supported by JSPS KAKENHI (No. 25-3944), Takahiro Sasaki by JSPS KAKENHI (No. 23-2790),Atsushi Nakamura by JSPS KAKENHI (Nos. 23654092, 24340054) and Takuya Saito by JSPS KAKENHI (No. 23740194). Keitaro Nagata is supported in part by Strategic Programs for Innovative Research (SPIRE) Field 5. The numerical calculations were performed on NEC SX-9 and SX-8R at CMC, Osaka University.\\
\end{acknowledgments}

%%%%%%%%%%%%%%%%%%%%%%%%%%%%%%%%%%%%%%%%%%%%%%%%%%%%%%%%%%%%%%%%%%%%%%%%%%%%%%%%
%%%%% Appendix
%%%%%%%%%%%%%%%%%%%%%%%%%%%%%%%%%%%%%%%%%%%%%%%%%%%%%%%%%%%%%%%%%%%%%%%%%%%%%%%%
\noindent
\appendix
\section{Data lists of the Taylor-expansion coefficients of the static-quark potential and the color-Debye screening mass}

Tables \ref{table-V1V8b1950}-\ref{table-V3V6b2000} show data lists of the Taylor-expansion coefficients in all the color-channel potentials for the cases of $T=1.20T_{\rm pc}$ and $1.35T_{\rm pc}$. The color-singlet and -octet potentials, $V_{M}(M=1,8)$, are ${\cal C}$-even and hence real. In actual calculations, these potentials have small imaginary components $(v_{1}(r), \, v_{3}(r))$ coming from numerical errors, but they are not written here.\\
\indent Tables \ref{table-mDb1950} and \ref{table-mDb2000} show data lists of the color-Debye screening masses as a function of $\mu_{\rm I}/T$ in all the color channels at $T=1.20T_{\rm pc}$ and $1.35T_{\rm pc}$. 
Symbols "$ \cdots $" mean that the fitting is unstable because the statistical errors of $V_{8}$ at this parameter point are quite big in the fit range.

\onecolumngrid

%%%%%%%%%%%%%%%%
%%% Table II
%%%%%%%%%%%%%%%%
\begin{table}[h]
\begin{center}
\begin{tabular}{c|ccc|ccc}
\hline 
\hline 
	&			& $V_{1}$ & 		&			& $V_{8}$ & \\
$r/a$ & $v_{0}(r/a)$ & $v_{2}(r/a)$ & $v_{4}(r/a)$ & $v_{0}(r/a)$ & $v_{2}(r/a)$ & $v_{4}(r/a)$\\
\hline
$ 1.000 $ & $ 2.0857 ( 10 ) $ & $ 0.0492 (   62 ) $ & $ 0.0115 (   64 ) $ & $ 3.4983 ( 34 ) $ & $ 0.2481 ( 245 ) $ & $ 0.1731 ( 271 ) $ \\
$ 1.414 $ & $ 2.6807 ( 20 ) $ & $ 0.0933 ( 115 ) $ & $ 0.0329 ( 112 ) $ & $ 3.4432 ( 37 ) $ & $ 0.2562 ( 246 ) $ & $ 0.1675 ( 269 ) $ \\
$ 1.732 $ & $ 2.9364 ( 41 ) $ & $ 0.1431 ( 231 ) $ & $ 0.0415 ( 247 ) $ & $ 3.4281 ( 38 ) $ & $ 0.2231 ( 246 ) $ & $ 0.1976 ( 269 ) $ \\
$ 2.000 $ & $ 3.0268 ( 23 ) $ & $ 0.1661 ( 156 ) $ & $ 0.0443 ( 174 ) $ & $ 3.4150 ( 34 ) $ & $ 0.2546 ( 243 ) $ & $ 0.1662 ( 270 ) $ \\
$ 2.236 $ & $ 3.1277 ( 27 ) $ & $ 0.2160 ( 169 ) $ & $ 0.0401 ( 183 ) $ & $ 3.4082 ( 36 ) $ & $ 0.2574 ( 245 ) $ & $ 0.1604 ( 271 ) $ \\
$ 2.449 $ & $ 3.1980 ( 47 ) $ & $ 0.2138 ( 248 ) $ & $ 0.0705 ( 248 ) $ & $ 3.4040 ( 39 ) $ & $ 0.2638 ( 247 ) $ & $ 0.1568 ( 273 ) $ \\
$ 2.828 $ & $ 3.2771 ( 39 ) $ & $ 0.1983 ( 219 ) $ & $ 0.1264 ( 233 ) $ & $ 3.3980 ( 36 ) $ & $ 0.2564 ( 244 ) $ & $ 0.1659 ( 264 ) $ \\
$ 3.000 $ & $ 3.2869 ( 35 ) $ & $ 0.2659 ( 227 ) $ & $ 0.0711 ( 234 ) $ & $ 3.3961 ( 34 ) $ & $ 0.2563 ( 230 ) $ & $ 0.1669 ( 252 ) $ \\
$ 3.162 $ & $ 3.3101 ( 39 ) $ & $ 0.2323 ( 241 ) $ & $ 0.1236 ( 266 ) $ & $ 3.3960 ( 36 ) $ & $ 0.2529 ( 244 ) $ & $ 0.1671 ( 271 ) $ \\
$ 3.317 $ & $ 3.3201 ( 39 ) $ & $ 0.2567 ( 242 ) $ & $ 0.1088 ( 256 ) $ & $ 3.3948 ( 35 ) $ & $ 0.2518 ( 230 ) $ & $ 0.1729 ( 259 ) $ \\
$ 3.464 $ & $ 3.3415 ( 64 ) $ & $ 0.2123 ( 363 ) $ & $ 0.1612 ( 334 ) $ & $ 3.3958 ( 36 ) $ & $ 0.2409 ( 247 ) $ & $ 0.1811 ( 273 ) $ \\
$ 3.606 $ & $ 3.3485 ( 36 ) $ & $ 0.2344 ( 233 ) $ & $ 0.1431 ( 249 ) $ & $ 3.3940 ( 35 ) $ & $ 0.2434 ( 226 ) $ & $ 0.1810 ( 249 ) $ \\
$ 3.742 $ & $ 3.3549 ( 42 ) $ & $ 0.2397 ( 259 ) $ & $ 0.1310 ( 273 ) $ & $ 3.3941 ( 34 ) $ & $ 0.2426 ( 225 ) $ & $ 0.1829 ( 247 ) $ \\
$ 4.000 $ & $ 3.3616 ( 32 ) $ & $ 0.2313 ( 237 ) $ & $ 0.1664 ( 290 ) $ & $ 3.3905 ( 36 ) $ & $ 0.2658 ( 242 ) $ & $ 0.1555 ( 267 ) $ \\
$ 4.123 $ & $ 3.3656 ( 37 ) $ & $ 0.2468 ( 246 ) $ & $ 0.1489 ( 281 ) $ & $ 3.3925 ( 36 ) $ & $ 0.2516 ( 234 ) $ & $ 0.1707 ( 262 ) $ \\
$ 4.243 $ & $ 3.3715 ( 37 ) $ & $ 0.2512 ( 247 ) $ & $ 0.1419 ( 274 ) $ & $ 3.3907 ( 35 ) $ & $ 0.2545 ( 225 ) $ & $ 0.1703 ( 251 ) $ \\
$ 4.359 $ & $ 3.3752 ( 40 ) $ & $ 0.2522 ( 215 ) $ & $ 0.1384 ( 246 ) $ & $ 3.3903 ( 37 ) $ & $ 0.2564 ( 239 ) $ & $ 0.1680 ( 262 ) $ \\
$ 4.472 $ & $ 3.3748 ( 38 ) $ & $ 0.2298 ( 242 ) $ & $ 0.1793 ( 259 ) $ & $ 3.3900 ( 32 ) $ & $ 0.2559 ( 224 ) $ & $ 0.1696 ( 257 ) $ \\
$ 4.583 $ & $ 3.3792 ( 39 ) $ & $ 0.2361 ( 265 ) $ & $ 0.1693 ( 300 ) $ & $ 3.3907 ( 35 ) $ & $ 0.2598 ( 227 ) $ & $ 0.1633 ( 246 ) $ \\
$ 4.690 $ & $ 3.3813 ( 44 ) $ & $ 0.2354 ( 276 ) $ & $ 0.1611 ( 331 ) $ & $ 3.3906 ( 36 ) $ & $ 0.2528 ( 230 ) $ & $ 0.1712 ( 246 ) $ \\
$ 4.899 $ & $ 3.3781 ( 46 ) $ & $ 0.2542 ( 282 ) $ & $ 0.1604 ( 314 ) $ & $ 3.3904 ( 36 ) $ & $ 0.2597 ( 235 ) $ & $ 0.1622 ( 260 ) $ \\
$ 5.000 $ & $ 3.3816 ( 29 ) $ & $ 0.2511 ( 194 ) $ & $ 0.1646 ( 223 ) $ & $ 3.3904 ( 35 ) $ & $ 0.2515 ( 227 ) $ & $ 0.1716 ( 250 ) $ \\
$ 5.099 $ & $ 3.3845 ( 33 ) $ & $ 0.2374 ( 215 ) $ & $ 0.1709 ( 235 ) $ & $ 3.3903 ( 36 ) $ & $ 0.2520 ( 227 ) $ & $ 0.1725 ( 247 ) $ \\
$ 5.196 $ & $ 3.3870 ( 40 ) $ & $ 0.2320 ( 255 ) $ & $ 0.1795 ( 278 ) $ & $ 3.3889 ( 35 ) $ & $ 0.2598 ( 227 ) $ & $ 0.1616 ( 249 ) $ \\
$ 5.385 $ & $ 3.3839 ( 36 ) $ & $ 0.2517 ( 247 ) $ & $ 0.1628 ( 275 ) $ & $ 3.3905 ( 35 ) $ & $ 0.2479 ( 227 ) $ & $ 0.1784 ( 248 ) $ \\
$ 5.477 $ & $ 3.3872 ( 40 ) $ & $ 0.2423 ( 276 ) $ & $ 0.1698 ( 310 ) $ & $ 3.3910 ( 33 ) $ & $ 0.2541 ( 230 ) $ & $ 0.1698 ( 258 ) $ \\
$ 5.657 $ & $ 3.3836 ( 43 ) $ & $ 0.2916 ( 266 ) $ & $ 0.1200 ( 307 ) $ & $ 3.3922 ( 34 ) $ & $ 0.2421 ( 231 ) $ & $ 0.1795 ( 257 ) $ \\
$ 5.745 $ & $ 3.3904 ( 30 ) $ & $ 0.2319 ( 229 ) $ & $ 0.1777 ( 278 ) $ & $ 3.3903 ( 35 ) $ & $ 0.2543 ( 225 ) $ & $ 0.1691 ( 246 ) $ \\
$ 5.831 $ & $ 3.3850 ( 37 ) $ & $ 0.2568 ( 243 ) $ & $ 0.1651 ( 263 ) $ & $ 3.3910 ( 34 ) $ & $ 0.2464 ( 221 ) $ & $ 0.1761 ( 245 ) $ \\
$ 5.916 $ & $ 3.3853 ( 40 ) $ & $ 0.2559 ( 230 ) $ & $ 0.1642 ( 234 ) $ & $ 3.3903 ( 34 ) $ & $ 0.2574 ( 228 ) $ & $ 0.1675 ( 254 ) $ \\
$ 6.000 $ & $ 3.3901 ( 37 ) $ & $ 0.2481 ( 257 ) $ & $ 0.1624 ( 296 ) $ & $ 3.3917 ( 35 ) $ & $ 0.2479 ( 217 ) $ & $ 0.1754 ( 236 ) $ \\
$ 6.083 $ & $ 3.3840 ( 36 ) $ & $ 0.2593 ( 242 ) $ & $ 0.1648 ( 276 ) $ & $ 3.3911 ( 37 ) $ & $ 0.2531 ( 232 ) $ & $ 0.1677 ( 255 ) $ \\
$ 6.164 $ & $ 3.3881 ( 38 ) $ & $ 0.2584 ( 225 ) $ & $ 0.1572 ( 234 ) $ & $ 3.3911 ( 35 ) $ & $ 0.2582 ( 232 ) $ & $ 0.1643 ( 251 ) $ \\
$ 6.325 $ & $ 3.3829 ( 36 ) $ & $ 0.2690 ( 247 ) $ & $ 0.1562 ( 264 ) $ & $ 3.3908 ( 35 ) $ & $ 0.2600 ( 231 ) $ & $ 0.1620 ( 254 ) $ \\
$ 6.403 $ & $ 3.3871 ( 40 ) $ & $ 0.2571 ( 255 ) $ & $ 0.1661 ( 281 ) $ & $ 3.3901 ( 34 ) $ & $ 0.2572 ( 226 ) $ & $ 0.1649 ( 251 ) $ \\
$ 6.481 $ & $ 3.3839 ( 38 ) $ & $ 0.2748 ( 255 ) $ & $ 0.1461 ( 293 ) $ & $ 3.3884 ( 34 ) $ & $ 0.2653 ( 220 ) $ & $ 0.1588 ( 238 ) $ \\
$ 6.557 $ & $ 3.3883 ( 43 ) $ & $ 0.2481 ( 266 ) $ & $ 0.1767 ( 274 ) $ & $ 3.3923 ( 35 ) $ & $ 0.2473 ( 233 ) $ & $ 0.1724 ( 263 ) $ \\
$ 6.633 $ & $ 3.3918 ( 48 ) $ & $ 0.2363 ( 320 ) $ & $ 0.1997 ( 359 ) $ & $ 3.3908 ( 34 ) $ & $ 0.2556 ( 229 ) $ & $ 0.1629 ( 252 ) $ \\
$ 6.708 $ & $ 3.3858 ( 39 ) $ & $ 0.2675 ( 244 ) $ & $ 0.1583 ( 266 ) $ & $ 3.3893 ( 34 ) $ & $ 0.2593 ( 225 ) $ & $ 0.1645 ( 251 ) $ \\
$ 6.782 $ & $ 3.3889 ( 31 ) $ & $ 0.2608 ( 233 ) $ & $ 0.1570 ( 267 ) $ & $ 3.3902 ( 34 ) $ & $ 0.2533 ( 227 ) $ & $ 0.1686 ( 249 ) $ \\
$ 6.928 $ & $ 3.3928 ( 66 ) $ & $ 0.2079 ( 424 ) $ & $ 0.2015 ( 470 ) $ & $ 3.3888 ( 36 ) $ & $ 0.2523 ( 242 ) $ & $ 0.1772 ( 267 ) $ \\
$ 7.000 $ & $ 3.3887 ( 37 ) $ & $ 0.2425 ( 249 ) $ & $ 0.1852 ( 284 ) $ & $ 3.3905 ( 35 ) $ & $ 0.2525 ( 231 ) $ & $ 0.1720 ( 254 ) $ \\
$ 7.071 $ & $ 3.3892 ( 36 ) $ & $ 0.2474 ( 241 ) $ & $ 0.1779 ( 274 ) $ & $ 3.3912 ( 34 ) $ & $ 0.2486 ( 221 ) $ & $ 0.1730 ( 242 ) $ \\
$ 7.141 $ & $ 3.3862 ( 38 ) $ & $ 0.2798 ( 267 ) $ & $ 0.1477 ( 293 ) $ & $ 3.3909 ( 35 ) $ & $ 0.2599 ( 226 ) $ & $ 0.1610 ( 246 ) $ \\
$ 7.211 $ & $ 3.3901 ( 36 ) $ & $ 0.2530 ( 242 ) $ & $ 0.1736 ( 279 ) $ & $ 3.3897 ( 33 ) $ & $ 0.2522 ( 223 ) $ & $ 0.1730 ( 256 ) $ \\
$ 7.280 $ & $ 3.3878 ( 35 ) $ & $ 0.2642 ( 234 ) $ & $ 0.1603 ( 263 ) $ & $ 3.3903 ( 35 ) $ & $ 0.2539 ( 231 ) $ & $ 0.1693 ( 258 ) $ \\
\hline 
\hline 
\end{tabular}
\caption{
Data list of the Taylor-expansion coefficients of the color-singlet and -octet potential at $T=1.20T_{\rm pc}$ as a function of $r/a$.
}
\label{table-V1V8b1950}
\end{center}
\end{table}
%%%%%%%%%%%

%%%%%%%%%%%%%%%%
%%% Table III
%%%%%%%%%%%%%%%%
\begin{table}[h]
\begin{center}
\begin{tabular}{c|ccccc|ccccc}
\hline 
\hline 
	&			&		& $V_{3^{\ast}}$ &		& 		&			&		& $V_{6}$ &		& \\
$r/a$ & $v_{0}(r/a)$ & $v_{1}(r/a)$ & $v_{2}(r/a)$ & $v_{3}(r/a)$ & $v_{4}(r/a)$ & $v_{0}(r/a)$ & $v_{1}(r/a)$ & $v_{2}(r/a)$ & $v_{3}(r/a)$ & $v_{4}(r/a)$
\\
\hline
$ 1.000 $ & $ 2.7543 ( 22 ) $ & $ 0.0063 ( 18 ) $ & $ 0.1521 ( 130 ) $ & $ -0.0126 ( 34 ) $ & $ 0.0932 ( 138 ) $ & $ 3.6357 ( 36 ) $ & $ 0.1021 ( 63 ) $ & $ 0.2706 ( 264 ) $ & $ 0.0758 ( 96 ) $ & $ 0.1848 ( 292 ) $ \\
$ 1.414 $ & $ 3.0433 ( 28 ) $ & $ 0.0239 ( 29 ) $ & $ 0.1817 ( 176 ) $ & $ 0.0070 ( 49 ) $ & $ 0.0992 ( 195 ) $ & $ 3.5187 ( 40 ) $ & $ 0.0949 ( 54 ) $ & $ 0.2622 ( 274 ) $ & $ 0.0676 ( 89 ) $ & $ 0.1888 ( 299 ) $ \\
$ 1.732 $ & $ 3.1706 ( 32 ) $ & $ 0.0510 ( 46 ) $ & $ 0.1694 ( 170 ) $ & $ 0.0013 ( 71 ) $ & $ 0.1380 ( 174 ) $ & $ 3.4720 ( 40 ) $ & $ 0.0880 ( 66 ) $ & $ 0.2412 ( 274 ) $ & $ 0.0657 ( 114 ) $ & $ 0.2013 ( 300 ) $ \\
$ 2.000 $ & $ 3.2127 ( 28 ) $ & $ 0.0583 ( 35 ) $ & $ 0.2005 ( 185 ) $ & $ 0.0025 ( 54 ) $ & $ 0.1262 ( 210 ) $ & $ 3.4536 ( 36 ) $ & $ 0.0821 ( 52 ) $ & $ 0.2618 ( 248 ) $ & $ 0.0610 ( 84 ) $ & $ 0.1792 ( 269 ) $ \\
$ 2.236 $ & $ 3.2626 ( 29 ) $ & $ 0.0618 ( 34 ) $ & $ 0.2219 ( 193 ) $ & $ 0.0151 ( 54 ) $ & $ 0.1272 ( 217 ) $ & $ 3.4340 ( 37 ) $ & $ 0.0779 ( 44 ) $ & $ 0.2626 ( 239 ) $ & $ 0.0651 ( 65 ) $ & $ 0.1733 ( 257 ) $ \\
$ 2.449 $ & $ 3.2954 ( 35 ) $ & $ 0.0626 ( 48 ) $ & $ 0.2336 ( 216 ) $ & $ 0.0272 ( 71 ) $ & $ 0.1271 ( 231 ) $ & $ 3.4237 ( 40 ) $ & $ 0.0752 ( 49 ) $ & $ 0.2657 ( 266 ) $ & $ 0.0655 ( 69 ) $ & $ 0.1671 ( 290 ) $ \\
$ 2.828 $ & $ 3.3316 ( 32 ) $ & $ 0.0670 ( 42 ) $ & $ 0.2457 ( 221 ) $ & $ 0.0408 ( 67 ) $ & $ 0.1320 ( 238 ) $ & $ 3.4131 ( 39 ) $ & $ 0.0751 ( 50 ) $ & $ 0.2463 ( 247 ) $ & $ 0.0614 ( 69 ) $ & $ 0.1837 ( 266 ) $ \\
$ 3.000 $ & $ 3.3378 ( 31 ) $ & $ 0.0657 ( 41 ) $ & $ 0.2575 ( 203 ) $ & $ 0.0450 ( 53 ) $ & $ 0.1316 ( 224 ) $ & $ 3.4086 ( 36 ) $ & $ 0.0788 ( 43 ) $ & $ 0.2470 ( 247 ) $ & $ 0.0557 ( 68 ) $ & $ 0.1822 ( 270 ) $ \\
$ 3.162 $ & $ 3.3511 ( 36 ) $ & $ 0.0668 ( 43 ) $ & $ 0.2340 ( 221 ) $ & $ 0.0425 ( 59 ) $ & $ 0.1580 ( 252 ) $ & $ 3.4059 ( 38 ) $ & $ 0.0746 ( 44 ) $ & $ 0.2487 ( 246 ) $ & $ 0.0631 ( 62 ) $ & $ 0.1743 ( 272 ) $ \\
$ 3.317 $ & $ 3.3553 ( 36 ) $ & $ 0.0731 ( 48 ) $ & $ 0.2573 ( 231 ) $ & $ 0.0384 ( 80 ) $ & $ 0.1409 ( 254 ) $ & $ 3.4038 ( 37 ) $ & $ 0.0717 ( 51 ) $ & $ 0.2452 ( 233 ) $ & $ 0.0672 ( 75 ) $ & $ 0.1839 ( 252 ) $ \\
$ 3.464 $ & $ 3.3624 ( 50 ) $ & $ 0.0585 ( 75 ) $ & $ 0.2538 ( 312 ) $ & $ 0.0629 ( 113 ) $ & $ 0.1598 ( 326 ) $ & $ 3.4009 ( 36 ) $ & $ 0.0766 ( 60 ) $ & $ 0.2526 ( 273 ) $ & $ 0.0613 ( 95 ) $ & $ 0.1702 ( 316 ) $ \\
$ 3.606 $ & $ 3.3687 ( 33 ) $ & $ 0.0725 ( 44 ) $ & $ 0.2387 ( 221 ) $ & $ 0.0449 ( 65 ) $ & $ 0.1655 ( 245 ) $ & $ 3.3994 ( 37 ) $ & $ 0.0742 ( 45 ) $ & $ 0.2461 ( 234 ) $ & $ 0.0593 ( 66 ) $ & $ 0.1801 ( 262 ) $ \\
$ 3.742 $ & $ 3.3709 ( 34 ) $ & $ 0.0695 ( 44 ) $ & $ 0.2500 ( 222 ) $ & $ 0.0482 ( 68 ) $ & $ 0.1553 ( 238 ) $ & $ 3.3986 ( 35 ) $ & $ 0.0744 ( 49 ) $ & $ 0.2484 ( 238 ) $ & $ 0.0574 ( 74 ) $ & $ 0.1815 ( 265 ) $ \\
$ 4.000 $ & $ 3.3756 ( 35 ) $ & $ 0.0718 ( 51 ) $ & $ 0.2469 ( 235 ) $ & $ 0.0539 ( 76 ) $ & $ 0.1620 ( 270 ) $ & $ 3.3955 ( 37 ) $ & $ 0.0729 ( 54 ) $ & $ 0.2568 ( 240 ) $ & $ 0.0552 ( 86 ) $ & $ 0.1666 ( 262 ) $ \\
$ 4.123 $ & $ 3.3768 ( 34 ) $ & $ 0.0710 ( 44 ) $ & $ 0.2496 ( 229 ) $ & $ 0.0518 ( 67 ) $ & $ 0.1619 ( 262 ) $ & $ 3.3956 ( 36 ) $ & $ 0.0767 ( 46 ) $ & $ 0.2498 ( 238 ) $ & $ 0.0535 ( 69 ) $ & $ 0.1761 ( 267 ) $ \\
$ 4.243 $ & $ 3.3818 ( 36 ) $ & $ 0.0745 ( 45 ) $ & $ 0.2352 ( 231 ) $ & $ 0.0452 ( 65 ) $ & $ 0.1756 ( 253 ) $ & $ 3.3929 ( 36 ) $ & $ 0.0756 ( 46 ) $ & $ 0.2536 ( 243 ) $ & $ 0.0557 ( 69 ) $ & $ 0.1730 ( 274 ) $ \\
$ 4.359 $ & $ 3.3787 ( 39 ) $ & $ 0.0645 ( 54 ) $ & $ 0.2586 ( 255 ) $ & $ 0.0570 ( 83 ) $ & $ 0.1577 ( 282 ) $ & $ 3.3926 ( 35 ) $ & $ 0.0741 ( 43 ) $ & $ 0.2513 ( 245 ) $ & $ 0.0547 ( 74 ) $ & $ 0.1732 ( 276 ) $ \\
$ 4.472 $ & $ 3.3829 ( 35 ) $ & $ 0.0736 ( 50 ) $ & $ 0.2385 ( 232 ) $ & $ 0.0533 ( 67 ) $ & $ 0.1780 ( 262 ) $ & $ 3.3937 ( 34 ) $ & $ 0.0714 ( 48 ) $ & $ 0.2484 ( 229 ) $ & $ 0.0643 ( 67 ) $ & $ 0.1801 ( 260 ) $ \\
$ 4.583 $ & $ 3.3819 ( 34 ) $ & $ 0.0650 ( 48 ) $ & $ 0.2482 ( 231 ) $ & $ 0.0639 ( 67 ) $ & $ 0.1702 ( 255 ) $ & $ 3.3929 ( 36 ) $ & $ 0.0710 ( 45 ) $ & $ 0.2537 ( 229 ) $ & $ 0.0610 ( 62 ) $ & $ 0.1707 ( 248 ) $ \\
$ 4.690 $ & $ 3.3864 ( 39 ) $ & $ 0.0751 ( 52 ) $ & $ 0.2304 ( 249 ) $ & $ 0.0474 ( 85 ) $ & $ 0.1855 ( 272 ) $ & $ 3.3938 ( 37 ) $ & $ 0.0752 ( 50 ) $ & $ 0.2412 ( 248 ) $ & $ 0.0594 ( 75 ) $ & $ 0.1842 ( 277 ) $ \\
$ 4.899 $ & $ 3.3836 ( 42 ) $ & $ 0.0787 ( 49 ) $ & $ 0.2524 ( 259 ) $ & $ 0.0465 ( 77 ) $ & $ 0.1690 ( 288 ) $ & $ 3.3918 ( 36 ) $ & $ 0.0738 ( 51 ) $ & $ 0.2627 ( 242 ) $ & $ 0.0581 ( 78 ) $ & $ 0.1550 ( 263 ) $ \\
$ 5.000 $ & $ 3.3859 ( 34 ) $ & $ 0.0710 ( 47 ) $ & $ 0.2415 ( 226 ) $ & $ 0.0579 ( 71 ) $ & $ 0.1753 ( 256 ) $ & $ 3.3902 ( 35 ) $ & $ 0.0745 ( 46 ) $ & $ 0.2595 ( 233 ) $ & $ 0.0552 ( 68 ) $ & $ 0.1649 ( 261 ) $ \\
$ 5.099 $ & $ 3.3857 ( 33 ) $ & $ 0.0716 ( 48 ) $ & $ 0.2483 ( 218 ) $ & $ 0.0575 ( 71 ) $ & $ 0.1692 ( 245 ) $ & $ 3.3909 ( 36 ) $ & $ 0.0722 ( 40 ) $ & $ 0.2553 ( 231 ) $ & $ 0.0572 ( 64 ) $ & $ 0.1691 ( 253 ) $ \\
$ 5.196 $ & $ 3.3872 ( 32 ) $ & $ 0.0769 ( 54 ) $ & $ 0.2389 ( 205 ) $ & $ 0.0529 ( 77 ) $ & $ 0.1822 ( 233 ) $ & $ 3.3914 ( 35 ) $ & $ 0.0753 ( 46 ) $ & $ 0.2565 ( 230 ) $ & $ 0.0547 ( 73 ) $ & $ 0.1644 ( 253 ) $ \\
$ 5.385 $ & $ 3.3862 ( 33 ) $ & $ 0.0718 ( 43 ) $ & $ 0.2496 ( 225 ) $ & $ 0.0583 ( 64 ) $ & $ 0.1713 ( 251 ) $ & $ 3.3914 ( 34 ) $ & $ 0.0749 ( 38 ) $ & $ 0.2497 ( 223 ) $ & $ 0.0573 ( 59 ) $ & $ 0.1735 ( 248 ) $ \\
$ 5.477 $ & $ 3.3890 ( 35 ) $ & $ 0.0723 ( 52 ) $ & $ 0.2472 ( 223 ) $ & $ 0.0595 ( 78 ) $ & $ 0.1720 ( 243 ) $ & $ 3.3918 ( 34 ) $ & $ 0.0746 ( 36 ) $ & $ 0.2548 ( 223 ) $ & $ 0.0557 ( 56 ) $ & $ 0.1688 ( 247 ) $ \\
$ 5.657 $ & $ 3.3889 ( 35 ) $ & $ 0.0731 ( 61 ) $ & $ 0.2504 ( 230 ) $ & $ 0.0534 ( 89 ) $ & $ 0.1679 ( 252 ) $ & $ 3.3920 ( 36 ) $ & $ 0.0744 ( 44 ) $ & $ 0.2526 ( 239 ) $ & $ 0.0603 ( 71 ) $ & $ 0.1681 ( 260 ) $ \\
$ 5.745 $ & $ 3.3890 ( 35 ) $ & $ 0.0718 ( 54 ) $ & $ 0.2449 ( 228 ) $ & $ 0.0571 ( 82 ) $ & $ 0.1756 ( 255 ) $ & $ 3.3910 ( 36 ) $ & $ 0.0753 ( 44 ) $ & $ 0.2540 ( 233 ) $ & $ 0.0563 ( 67 ) $ & $ 0.1688 ( 253 ) $ \\
$ 5.831 $ & $ 3.3889 ( 33 ) $ & $ 0.0722 ( 50 ) $ & $ 0.2410 ( 226 ) $ & $ 0.0584 ( 72 ) $ & $ 0.1832 ( 253 ) $ & $ 3.3919 ( 35 ) $ & $ 0.0737 ( 43 ) $ & $ 0.2474 ( 227 ) $ & $ 0.0574 ( 68 ) $ & $ 0.1737 ( 257 ) $ \\
$ 5.916 $ & $ 3.3876 ( 37 ) $ & $ 0.0708 ( 49 ) $ & $ 0.2458 ( 217 ) $ & $ 0.0620 ( 75 ) $ & $ 0.1815 ( 238 ) $ & $ 3.3920 ( 35 ) $ & $ 0.0749 ( 41 ) $ & $ 0.2581 ( 229 ) $ & $ 0.0551 ( 64 ) $ & $ 0.1636 ( 251 ) $ \\
$ 6.000 $ & $ 3.3883 ( 33 ) $ & $ 0.0723 ( 48 ) $ & $ 0.2465 ( 223 ) $ & $ 0.0579 ( 64 ) $ & $ 0.1697 ( 250 ) $ & $ 3.3910 ( 35 ) $ & $ 0.0739 ( 41 ) $ & $ 0.2565 ( 224 ) $ & $ 0.0546 ( 65 ) $ & $ 0.1675 ( 243 ) $ \\
$ 6.083 $ & $ 3.3899 ( 39 ) $ & $ 0.0741 ( 51 ) $ & $ 0.2444 ( 237 ) $ & $ 0.0535 ( 75 ) $ & $ 0.1739 ( 260 ) $ & $ 3.3886 ( 35 ) $ & $ 0.0743 ( 44 ) $ & $ 0.2609 ( 225 ) $ & $ 0.0543 ( 65 ) $ & $ 0.1618 ( 253 ) $ \\
$ 6.164 $ & $ 3.3883 ( 36 ) $ & $ 0.0688 ( 46 ) $ & $ 0.2556 ( 215 ) $ & $ 0.0651 ( 63 ) $ & $ 0.1671 ( 232 ) $ & $ 3.3916 ( 36 ) $ & $ 0.0762 ( 48 ) $ & $ 0.2560 ( 239 ) $ & $ 0.0558 ( 73 ) $ & $ 0.1645 ( 264 ) $ \\
$ 6.325 $ & $ 3.3882 ( 38 ) $ & $ 0.0718 ( 49 ) $ & $ 0.2517 ( 238 ) $ & $ 0.0628 ( 71 ) $ & $ 0.1688 ( 259 ) $ & $ 3.3906 ( 36 ) $ & $ 0.0733 ( 47 ) $ & $ 0.2603 ( 232 ) $ & $ 0.0568 ( 71 ) $ & $ 0.1606 ( 258 ) $ \\
$ 6.403 $ & $ 3.3896 ( 35 ) $ & $ 0.0742 ( 46 ) $ & $ 0.2508 ( 229 ) $ & $ 0.0551 ( 66 ) $ & $ 0.1693 ( 251 ) $ & $ 3.3897 ( 34 ) $ & $ 0.0743 ( 45 ) $ & $ 0.2578 ( 231 ) $ & $ 0.0568 ( 66 ) $ & $ 0.1647 ( 258 ) $ \\
$ 6.481 $ & $ 3.3866 ( 32 ) $ & $ 0.0780 ( 51 ) $ & $ 0.2672 ( 214 ) $ & $ 0.0532 ( 79 ) $ & $ 0.1553 ( 242 ) $ & $ 3.3907 ( 35 ) $ & $ 0.0732 ( 47 ) $ & $ 0.2569 ( 231 ) $ & $ 0.0571 ( 69 ) $ & $ 0.1673 ( 251 ) $ \\
$ 6.557 $ & $ 3.3893 ( 35 ) $ & $ 0.0751 ( 49 ) $ & $ 0.2502 ( 223 ) $ & $ 0.0597 ( 76 ) $ & $ 0.1753 ( 238 ) $ & $ 3.3932 ( 38 ) $ & $ 0.0789 ( 45 ) $ & $ 0.2496 ( 246 ) $ & $ 0.0507 ( 74 ) $ & $ 0.1686 ( 273 ) $ \\
$ 6.633 $ & $ 3.3931 ( 40 ) $ & $ 0.0700 ( 56 ) $ & $ 0.2392 ( 239 ) $ & $ 0.0612 ( 79 ) $ & $ 0.1802 ( 257 ) $ & $ 3.3911 ( 32 ) $ & $ 0.0746 ( 44 ) $ & $ 0.2584 ( 229 ) $ & $ 0.0586 ( 62 ) $ & $ 0.1600 ( 250 ) $ \\
$ 6.708 $ & $ 3.3881 ( 36 ) $ & $ 0.0753 ( 46 ) $ & $ 0.2549 ( 235 ) $ & $ 0.0541 ( 72 ) $ & $ 0.1697 ( 259 ) $ & $ 3.3900 ( 33 ) $ & $ 0.0713 ( 43 ) $ & $ 0.2572 ( 221 ) $ & $ 0.0622 ( 64 ) $ & $ 0.1662 ( 247 ) $ \\
$ 6.782 $ & $ 3.3902 ( 34 ) $ & $ 0.0752 ( 45 ) $ & $ 0.2470 ( 212 ) $ & $ 0.0546 ( 66 ) $ & $ 0.1731 ( 224 ) $ & $ 3.3912 ( 37 ) $ & $ 0.0734 ( 48 ) $ & $ 0.2553 ( 239 ) $ & $ 0.0581 ( 65 ) $ & $ 0.1661 ( 262 ) $ \\
$ 6.928 $ & $ 3.3894 ( 41 ) $ & $ 0.0573 ( 89 ) $ & $ 0.2474 ( 273 ) $ & $ 0.0741 ( 128 ) $ & $ 0.1728 ( 315 ) $ & $ 3.3887 ( 39 ) $ & $ 0.0738 ( 59 ) $ & $ 0.2437 ( 250 ) $ & $ 0.0583 ( 83 ) $ & $ 0.1889 ( 279 ) $ \\
$ 7.000 $ & $ 3.3898 ( 35 ) $ & $ 0.0744 ( 46 ) $ & $ 0.2461 ( 219 ) $ & $ 0.0524 ( 73 ) $ & $ 0.1795 ( 243 ) $ & $ 3.3911 ( 37 ) $ & $ 0.0741 ( 41 ) $ & $ 0.2534 ( 235 ) $ & $ 0.0568 ( 64 ) $ & $ 0.1703 ( 258 ) $ \\
$ 7.071 $ & $ 3.3905 ( 35 ) $ & $ 0.0726 ( 42 ) $ & $ 0.2405 ( 220 ) $ & $ 0.0574 ( 65 ) $ & $ 0.1857 ( 242 ) $ & $ 3.3912 ( 35 ) $ & $ 0.0739 ( 44 ) $ & $ 0.2509 ( 230 ) $ & $ 0.0542 ( 64 ) $ & $ 0.1709 ( 252 ) $ \\
$ 7.141 $ & $ 3.3888 ( 36 ) $ & $ 0.0718 ( 46 ) $ & $ 0.2571 ( 224 ) $ & $ 0.0576 ( 75 ) $ & $ 0.1667 ( 237 ) $ & $ 3.3906 ( 34 ) $ & $ 0.0680 ( 48 ) $ & $ 0.2596 ( 229 ) $ & $ 0.0630 ( 69 ) $ & $ 0.1614 ( 256 ) $ \\
$ 7.211 $ & $ 3.3914 ( 37 ) $ & $ 0.0712 ( 52 ) $ & $ 0.2439 ( 236 ) $ & $ 0.0552 ( 77 ) $ & $ 0.1774 ( 262 ) $ & $ 3.3898 ( 33 ) $ & $ 0.0740 ( 45 ) $ & $ 0.2515 ( 221 ) $ & $ 0.0607 ( 63 ) $ & $ 0.1732 ( 253 ) $ \\
$ 7.280 $ & $ 3.3913 ( 34 ) $ & $ 0.0727 ( 45 ) $ & $ 0.2501 ( 226 ) $ & $ 0.0588 ( 65 ) $ & $ 0.1724 ( 249 ) $ & $ 3.3895 ( 35 ) $ & $ 0.0710 ( 47 ) $ & $ 0.2580 ( 236 ) $ & $ 0.0600 ( 68 ) $ & $ 0.1665 ( 263 ) $ \\
\hline 
\hline 
\end{tabular}
\caption{
Data list of the Taylor-expansion coefficients of the color-antitriplet and -sextet potential at $T=1.20T_{\rm pc}$ as a function of $r/a$.
}
\label{table-V3V6b1950}
\end{center}
\end{table}
%%%%%%%%%%%

%%%%%%%%%%%%%%%%
%%% Table IV
%%%%%%%%%%%%%%%%
\begin{table}[h]
\begin{center}
\begin{tabular}{c|ccc|ccc}
\hline 
\hline 
	&			& $V_{1}$ & 		&			& $V_{8}$ & \\
$r/a$ & $v_{0}(r/a)$ & $v_{2}(r/a)$ & $v_{4}(r/a)$ & $v_{0}(r/a)$ & $v_{2}(r/a)$ & $v_{4}(r/a)$
\\
\hline
$ 1.000 $ & $ 1.9616 ( 10 ) $ & $ 0.0394 (   60 ) $ & $ 0.0006 (   56 ) $ & $ 3.2617 ( 32 ) $ & $ 0.2334 ( 197 ) $ & $ 0.0574 ( 204 ) $ \\
$ 1.414 $ & $ 2.5047 ( 16 ) $ & $ 0.0753 (   99 ) $ & $ 0.0051 (   97 ) $ & $ 3.2081 ( 31 ) $ & $ 0.2301 ( 201 ) $ & $ 0.0521 ( 219 ) $ \\
$ 1.732 $ & $ 2.7321 ( 31 ) $ & $ 0.1206 ( 180 ) $ & $ -0.0060 ( 177 ) $ & $ 3.1879 ( 33 ) $ & $ 0.2229 ( 203 ) $ & $ 0.0580 ( 207 ) $ \\
$ 2.000 $ & $ 2.8153 ( 21 ) $ & $ 0.1170 ( 125 ) $ & $ 0.0183 ( 134 ) $ & $ 3.1816 ( 29 ) $ & $ 0.2127 ( 189 ) $ & $ 0.0658 ( 195 ) $ \\
$ 2.236 $ & $ 2.9136 ( 27 ) $ & $ 0.1409 ( 151 ) $ & $ 0.0146 ( 148 ) $ & $ 3.1731 ( 29 ) $ & $ 0.2133 ( 188 ) $ & $ 0.0639 ( 195 ) $ \\
$ 2.449 $ & $ 2.9716 ( 30 ) $ & $ 0.1838 ( 169 ) $ & $ -0.0079 ( 166 ) $ & $ 3.1666 ( 31 ) $ & $ 0.2170 ( 194 ) $ & $ 0.0587 ( 203 ) $ \\
$ 2.828 $ & $ 3.0342 ( 33 ) $ & $ 0.2115 ( 183 ) $ & $ -0.0110 ( 181 ) $ & $ 3.1595 ( 33 ) $ & $ 0.2228 ( 208 ) $ & $ 0.0572 ( 224 ) $ \\
$ 3.000 $ & $ 3.0569 ( 27 ) $ & $ 0.2013 ( 148 ) $ & $ 0.0089 ( 139 ) $ & $ 3.1593 ( 32 ) $ & $ 0.2151 ( 200 ) $ & $ 0.0615 ( 210 ) $ \\
$ 3.162 $ & $ 3.0732 ( 33 ) $ & $ 0.1918 ( 187 ) $ & $ 0.0291 ( 178 ) $ & $ 3.1577 ( 32 ) $ & $ 0.2196 ( 194 ) $ & $ 0.0559 ( 203 ) $ \\
$ 3.317 $ & $ 3.0883 ( 36 ) $ & $ 0.1976 ( 212 ) $ & $ 0.0271 ( 216 ) $ & $ 3.1571 ( 29 ) $ & $ 0.2161 ( 187 ) $ & $ 0.0564 ( 202 ) $ \\
$ 3.464 $ & $ 3.1026 ( 41 ) $ & $ 0.2342 ( 275 ) $ & $ -0.0253 ( 276 ) $ & $ 3.1552 ( 35 ) $ & $ 0.2053 ( 213 ) $ & $ 0.0744 ( 224 ) $ \\
$ 3.606 $ & $ 3.1109 ( 29 ) $ & $ 0.2004 ( 174 ) $ & $ 0.0284 ( 167 ) $ & $ 3.1537 ( 31 ) $ & $ 0.2235 ( 182 ) $ & $ 0.0541 ( 188 ) $ \\
$ 3.742 $ & $ 3.1173 ( 37 ) $ & $ 0.2072 ( 222 ) $ & $ 0.0283 ( 216 ) $ & $ 3.1532 ( 32 ) $ & $ 0.2192 ( 193 ) $ & $ 0.0562 ( 203 ) $ \\
$ 4.000 $ & $ 3.1221 ( 35 ) $ & $ 0.2079 ( 222 ) $ & $ 0.0373 ( 237 ) $ & $ 3.1521 ( 31 ) $ & $ 0.2238 ( 189 ) $ & $ 0.0518 ( 192 ) $ \\
$ 4.123 $ & $ 3.1300 ( 30 ) $ & $ 0.1989 ( 184 ) $ & $ 0.0508 ( 185 ) $ & $ 3.1515 ( 32 ) $ & $ 0.2237 ( 190 ) $ & $ 0.0522 ( 197 ) $ \\
$ 4.243 $ & $ 3.1340 ( 34 ) $ & $ 0.1926 ( 217 ) $ & $ 0.0637 ( 214 ) $ & $ 3.1516 ( 32 ) $ & $ 0.2312 ( 189 ) $ & $ 0.0423 ( 192 ) $ \\
$ 4.359 $ & $ 3.1337 ( 43 ) $ & $ 0.2151 ( 252 ) $ & $ 0.0368 ( 244 ) $ & $ 3.1518 ( 31 ) $ & $ 0.2289 ( 184 ) $ & $ 0.0433 ( 187 ) $ \\
$ 4.472 $ & $ 3.1388 ( 30 ) $ & $ 0.2008 ( 198 ) $ & $ 0.0551 ( 211 ) $ & $ 3.1522 ( 32 ) $ & $ 0.2174 ( 189 ) $ & $ 0.0586 ( 190 ) $ \\
$ 4.583 $ & $ 3.1367 ( 36 ) $ & $ 0.2123 ( 217 ) $ & $ 0.0500 ( 216 ) $ & $ 3.1511 ( 32 ) $ & $ 0.2231 ( 196 ) $ & $ 0.0509 ( 199 ) $ \\
$ 4.690 $ & $ 3.1423 ( 37 ) $ & $ 0.2012 ( 231 ) $ & $ 0.0592 ( 238 ) $ & $ 3.1519 ( 31 ) $ & $ 0.2250 ( 189 ) $ & $ 0.0474 ( 196 ) $ \\
$ 4.899 $ & $ 3.1425 ( 33 ) $ & $ 0.2328 ( 213 ) $ & $ 0.0209 ( 223 ) $ & $ 3.1512 ( 30 ) $ & $ 0.2198 ( 185 ) $ & $ 0.0539 ( 195 ) $ \\
$ 5.000 $ & $ 3.1436 ( 32 ) $ & $ 0.2183 ( 195 ) $ & $ 0.0468 ( 197 ) $ & $ 3.1507 ( 31 ) $ & $ 0.2237 ( 187 ) $ & $ 0.0498 ( 193 ) $ \\
$ 5.099 $ & $ 3.1442 ( 32 ) $ & $ 0.2219 ( 203 ) $ & $ 0.0436 ( 211 ) $ & $ 3.1505 ( 31 ) $ & $ 0.2236 ( 185 ) $ & $ 0.0511 ( 190 ) $ \\
$ 5.196 $ & $ 3.1482 ( 31 ) $ & $ 0.2050 ( 201 ) $ & $ 0.0547 ( 209 ) $ & $ 3.1517 ( 29 ) $ & $ 0.2171 ( 180 ) $ & $ 0.0555 ( 188 ) $ \\
$ 5.385 $ & $ 3.1483 ( 35 ) $ & $ 0.2179 ( 219 ) $ & $ 0.0497 ( 223 ) $ & $ 3.1504 ( 30 ) $ & $ 0.2221 ( 188 ) $ & $ 0.0521 ( 194 ) $ \\
$ 5.477 $ & $ 3.1449 ( 37 ) $ & $ 0.2126 ( 222 ) $ & $ 0.0606 ( 225 ) $ & $ 3.1517 ( 30 ) $ & $ 0.2196 ( 186 ) $ & $ 0.0536 ( 194 ) $ \\
$ 5.657 $ & $ 3.1499 ( 45 ) $ & $ 0.2032 ( 262 ) $ & $ 0.0708 ( 260 ) $ & $ 3.1498 ( 32 ) $ & $ 0.2237 ( 190 ) $ & $ 0.0524 ( 200 ) $ \\
$ 5.745 $ & $ 3.1469 ( 35 ) $ & $ 0.2163 ( 206 ) $ & $ 0.0536 ( 207 ) $ & $ 3.1513 ( 30 ) $ & $ 0.2198 ( 179 ) $ & $ 0.0509 ( 186 ) $ \\
$ 5.831 $ & $ 3.1494 ( 37 ) $ & $ 0.2058 ( 229 ) $ & $ 0.0630 ( 233 ) $ & $ 3.1502 ( 30 ) $ & $ 0.2229 ( 182 ) $ & $ 0.0507 ( 189 ) $ \\
$ 5.916 $ & $ 3.1502 ( 34 ) $ & $ 0.2039 ( 225 ) $ & $ 0.0668 ( 229 ) $ & $ 3.1503 ( 30 ) $ & $ 0.2228 ( 183 ) $ & $ 0.0500 ( 197 ) $ \\
$ 6.000 $ & $ 3.1490 ( 30 ) $ & $ 0.2245 ( 203 ) $ & $ 0.0429 ( 213 ) $ & $ 3.1503 ( 30 ) $ & $ 0.2179 ( 183 ) $ & $ 0.0554 ( 180 ) $ \\
$ 6.083 $ & $ 3.1497 ( 36 ) $ & $ 0.2124 ( 220 ) $ & $ 0.0547 ( 221 ) $ & $ 3.1506 ( 30 ) $ & $ 0.2169 ( 185 ) $ & $ 0.0587 ( 188 ) $ \\
$ 6.164 $ & $ 3.1512 ( 35 ) $ & $ 0.1994 ( 210 ) $ & $ 0.0731 ( 216 ) $ & $ 3.1503 ( 30 ) $ & $ 0.2239 ( 177 ) $ & $ 0.0488 ( 180 ) $ \\
$ 6.325 $ & $ 3.1493 ( 34 ) $ & $ 0.2156 ( 230 ) $ & $ 0.0571 ( 228 ) $ & $ 3.1510 ( 29 ) $ & $ 0.2205 ( 175 ) $ & $ 0.0534 ( 176 ) $ \\
$ 6.403 $ & $ 3.1509 ( 34 ) $ & $ 0.2122 ( 217 ) $ & $ 0.0621 ( 226 ) $ & $ 3.1514 ( 29 ) $ & $ 0.2188 ( 181 ) $ & $ 0.0535 ( 188 ) $ \\
$ 6.481 $ & $ 3.1462 ( 29 ) $ & $ 0.2456 ( 212 ) $ & $ 0.0254 ( 232 ) $ & $ 3.1519 ( 30 ) $ & $ 0.2102 ( 191 ) $ & $ 0.0612 ( 203 ) $ \\
$ 6.557 $ & $ 3.1509 ( 30 ) $ & $ 0.2149 ( 211 ) $ & $ 0.0567 ( 219 ) $ & $ 3.1509 ( 30 ) $ & $ 0.2098 ( 193 ) $ & $ 0.0620 ( 199 ) $ \\
$ 6.633 $ & $ 3.1495 ( 33 ) $ & $ 0.2240 ( 225 ) $ & $ 0.0508 ( 237 ) $ & $ 3.1521 ( 30 ) $ & $ 0.2152 ( 186 ) $ & $ 0.0544 ( 193 ) $ \\
$ 6.708 $ & $ 3.1512 ( 31 ) $ & $ 0.2120 ( 203 ) $ & $ 0.0561 ( 218 ) $ & $ 3.1509 ( 31 ) $ & $ 0.2172 ( 187 ) $ & $ 0.0549 ( 194 ) $ \\
$ 6.782 $ & $ 3.1502 ( 37 ) $ & $ 0.2148 ( 221 ) $ & $ 0.0610 ( 233 ) $ & $ 3.1505 ( 30 ) $ & $ 0.2249 ( 185 ) $ & $ 0.0472 ( 194 ) $ \\
$ 6.928 $ & $ 3.1542 ( 44 ) $ & $ 0.1972 ( 294 ) $ & $ 0.0681 ( 317 ) $ & $ 3.1496 ( 35 ) $ & $ 0.2294 ( 204 ) $ & $ 0.0403 ( 206 ) $ \\
$ 7.000 $ & $ 3.1524 ( 40 ) $ & $ 0.2003 ( 223 ) $ & $ 0.0753 ( 227 ) $ & $ 3.1506 ( 30 ) $ & $ 0.2180 ( 183 ) $ & $ 0.0561 ( 190 ) $ \\
$ 7.071 $ & $ 3.1514 ( 31 ) $ & $ 0.2189 ( 193 ) $ & $ 0.0516 ( 197 ) $ & $ 3.1505 ( 31 ) $ & $ 0.2174 ( 183 ) $ & $ 0.0558 ( 184 ) $ \\
$ 7.141 $ & $ 3.1528 ( 32 ) $ & $ 0.2201 ( 203 ) $ & $ 0.0447 ( 202 ) $ & $ 3.1513 ( 30 ) $ & $ 0.2140 ( 178 ) $ & $ 0.0596 ( 181 ) $ \\
$ 7.211 $ & $ 3.1493 ( 35 ) $ & $ 0.2137 ( 222 ) $ & $ 0.0593 ( 235 ) $ & $ 3.1520 ( 30 ) $ & $ 0.2136 ( 185 ) $ & $ 0.0588 ( 192 ) $ \\
$ 7.280 $ & $ 3.1495 ( 34 ) $ & $ 0.2204 ( 214 ) $ & $ 0.0519 ( 223 ) $ & $ 3.1503 ( 30 ) $ & $ 0.2205 ( 182 ) $ & $ 0.0518 ( 188 ) $ \\
\hline 
\hline 
\end{tabular}
\caption{
Data list of the Taylor-expansion coefficients of the color-singlet and -octet potential at $T=1.35T_{\rm pc}$ as a function of $r/a$.
}
\label{table-V1V8b2000}
\end{center}
\end{table}
%%%%%%%%%%%

%%%%%%%%%%%%%%%%
%%% Table V
%%%%%%%%%%%%%%%%
\begin{table}[h]
\begin{center}
\begin{tabular}{c|ccccc|ccccc}
\hline 
\hline 
	&			&		& $V_{3^{\ast}}$ &		& 		&			&		& $V_{6}$ &		& \\
$r/a$ & $v_{0}(r/a)$ & $v_{1}(r/a)$ & $v_{2}(r/a)$ & $v_{3}(r/a)$ & $v_{4}(r/a)$ & $v_{0}(r/a)$ & $v_{1}(r/a)$ & $v_{2}(r/a)$ & $v_{3}(r/a)$ & $v_{4}(r/a)$
\\
\hline
$ 1.000 $ & $ 2.5677 ( 17 ) $ & $ 0.0054 ( 15 ) $ & $ 0.1310 ( 106 ) $ & $ -0.0057 ( 24 ) $ & $ 0.0270 ( 109 ) $ & $ 3.3938 ( 33 ) $ & $ 0.0752 ( 55 ) $ & $ 0.2486 ( 212 ) $ & $ 0.0469 ( 71 ) $ & $ 0.0651 ( 215 ) $ \\
$ 1.414 $ & $ 2.8337 ( 21 ) $ & $ 0.0220 ( 25 ) $ & $ 0.1482 ( 123 ) $ & $ 0.0011 ( 36 ) $ & $ 0.0318 ( 123 ) $ & $ 3.2800 ( 35 ) $ & $ 0.0706 ( 48 ) $ & $ 0.2498 ( 216 ) $ & $ 0.0342 ( 64 ) $ & $ 0.0511 ( 230 ) $ \\
$ 1.732 $ & $ 2.9464 ( 26 ) $ & $ 0.0352 ( 43 ) $ & $ 0.1645 ( 176 ) $ & $ 0.0037 ( 63 ) $ & $ 0.0320 ( 182 ) $ & $ 3.2346 ( 38 ) $ & $ 0.0618 ( 43 ) $ & $ 0.2255 ( 227 ) $ & $ 0.0386 ( 58 ) $ & $ 0.0641 ( 218 ) $ \\
$ 2.000 $ & $ 2.9867 ( 23 ) $ & $ 0.0343 ( 33 ) $ & $ 0.1707 ( 137 ) $ & $ 0.0113 ( 45 ) $ & $ 0.0352 ( 143 ) $ & $ 3.2171 ( 29 ) $ & $ 0.0546 ( 36 ) $ & $ 0.2279 ( 196 ) $ & $ 0.0405 ( 48 ) $ & $ 0.0668 ( 203 ) $ \\
$ 2.236 $ & $ 3.0339 ( 27 ) $ & $ 0.0397 ( 30 ) $ & $ 0.1801 ( 153 ) $ & $ 0.0129 ( 42 ) $ & $ 0.0373 ( 152 ) $ & $ 3.1995 ( 32 ) $ & $ 0.0602 ( 40 ) $ & $ 0.2213 ( 205 ) $ & $ 0.0315 ( 46 ) $ & $ 0.0680 ( 217 ) $ \\
$ 2.449 $ & $ 3.0631 ( 26 ) $ & $ 0.0454 ( 35 ) $ & $ 0.1942 ( 157 ) $ & $ 0.0138 ( 52 ) $ & $ 0.0322 ( 162 ) $ & $ 3.1869 ( 33 ) $ & $ 0.0596 ( 41 ) $ & $ 0.2209 ( 215 ) $ & $ 0.0304 ( 57 ) $ & $ 0.0622 ( 224 ) $ \\
$ 2.828 $ & $ 3.0935 ( 31 ) $ & $ 0.0474 ( 33 ) $ & $ 0.2054 ( 178 ) $ & $ 0.0194 ( 46 ) $ & $ 0.0325 ( 176 ) $ & $ 3.1723 ( 33 ) $ & $ 0.0636 ( 43 ) $ & $ 0.2299 ( 201 ) $ & $ 0.0248 ( 55 ) $ & $ 0.0562 ( 217 ) $ \\
$ 3.000 $ & $ 3.1054 ( 29 ) $ & $ 0.0519 ( 32 ) $ & $ 0.2005 ( 169 ) $ & $ 0.0177 ( 46 ) $ & $ 0.0384 ( 161 ) $ & $ 3.1691 ( 33 ) $ & $ 0.0606 ( 40 ) $ & $ 0.2228 ( 204 ) $ & $ 0.0283 ( 52 ) $ & $ 0.0606 ( 220 ) $ \\
$ 3.162 $ & $ 3.1134 ( 29 ) $ & $ 0.0491 ( 34 ) $ & $ 0.2040 ( 175 ) $ & $ 0.0216 ( 44 ) $ & $ 0.0425 ( 177 ) $ & $ 3.1659 ( 31 ) $ & $ 0.0593 ( 36 ) $ & $ 0.2269 ( 190 ) $ & $ 0.0283 ( 46 ) $ & $ 0.0532 ( 200 ) $ \\
$ 3.317 $ & $ 3.1205 ( 29 ) $ & $ 0.0530 ( 39 ) $ & $ 0.2001 ( 180 ) $ & $ 0.0166 ( 51 ) $ & $ 0.0464 ( 189 ) $ & $ 3.1637 ( 30 ) $ & $ 0.0604 ( 38 ) $ & $ 0.2190 ( 191 ) $ & $ 0.0251 ( 53 ) $ & $ 0.0621 ( 207 ) $ \\
$ 3.464 $ & $ 3.1264 ( 35 ) $ & $ 0.0619 ( 45 ) $ & $ 0.2109 ( 218 ) $ & $ 0.0088 ( 66 ) $ & $ 0.0353 ( 211 ) $ & $ 3.1600 ( 33 ) $ & $ 0.0634 ( 47 ) $ & $ 0.2212 ( 202 ) $ & $ 0.0217 ( 58 ) $ & $ 0.0618 ( 221 ) $ \\
$ 3.606 $ & $ 3.1305 ( 30 ) $ & $ 0.0543 ( 35 ) $ & $ 0.2106 ( 174 ) $ & $ 0.0196 ( 44 ) $ & $ 0.0443 ( 172 ) $ & $ 3.1586 ( 32 ) $ & $ 0.0591 ( 36 ) $ & $ 0.2193 ( 193 ) $ & $ 0.0280 ( 46 ) $ & $ 0.0620 ( 197 ) $ \\
$ 3.742 $ & $ 3.1329 ( 31 ) $ & $ 0.0572 ( 33 ) $ & $ 0.2055 ( 186 ) $ & $ 0.0167 ( 46 ) $ & $ 0.0516 ( 181 ) $ & $ 3.1565 ( 32 ) $ & $ 0.0552 ( 38 ) $ & $ 0.2318 ( 200 ) $ & $ 0.0331 ( 48 ) $ & $ 0.0468 ( 217 ) $ \\
$ 4.000 $ & $ 3.1370 ( 29 ) $ & $ 0.0541 ( 41 ) $ & $ 0.2117 ( 180 ) $ & $ 0.0252 ( 60 ) $ & $ 0.0487 ( 193 ) $ & $ 3.1554 ( 31 ) $ & $ 0.0582 ( 30 ) $ & $ 0.2279 ( 192 ) $ & $ 0.0275 ( 42 ) $ & $ 0.0496 ( 201 ) $ \\
$ 4.123 $ & $ 3.1398 ( 31 ) $ & $ 0.0532 ( 32 ) $ & $ 0.2092 ( 182 ) $ & $ 0.0259 ( 46 ) $ & $ 0.0529 ( 182 ) $ & $ 3.1535 ( 31 ) $ & $ 0.0581 ( 33 ) $ & $ 0.2294 ( 189 ) $ & $ 0.0269 ( 43 ) $ & $ 0.0499 ( 195 ) $ \\
$ 4.243 $ & $ 3.1427 ( 33 ) $ & $ 0.0545 ( 32 ) $ & $ 0.1998 ( 196 ) $ & $ 0.0227 ( 48 ) $ & $ 0.0654 ( 193 ) $ & $ 3.1541 ( 32 ) $ & $ 0.0560 ( 34 ) $ & $ 0.2305 ( 191 ) $ & $ 0.0288 ( 44 ) $ & $ 0.0458 ( 192 ) $ \\
$ 4.359 $ & $ 3.1431 ( 31 ) $ & $ 0.0596 ( 33 ) $ & $ 0.2169 ( 185 ) $ & $ 0.0191 ( 50 ) $ & $ 0.0433 ( 183 ) $ & $ 3.1542 ( 32 ) $ & $ 0.0569 ( 35 ) $ & $ 0.2295 ( 183 ) $ & $ 0.0283 ( 46 ) $ & $ 0.0456 ( 192 ) $ \\
$ 4.472 $ & $ 3.1441 ( 30 ) $ & $ 0.0510 ( 36 ) $ & $ 0.2080 ( 178 ) $ & $ 0.0309 ( 50 ) $ & $ 0.0594 ( 180 ) $ & $ 3.1531 ( 29 ) $ & $ 0.0581 ( 36 ) $ & $ 0.2227 ( 181 ) $ & $ 0.0252 ( 49 ) $ & $ 0.0540 ( 182 ) $ \\
$ 4.583 $ & $ 3.1437 ( 31 ) $ & $ 0.0549 ( 34 ) $ & $ 0.2176 ( 192 ) $ & $ 0.0241 ( 54 ) $ & $ 0.0483 ( 194 ) $ & $ 3.1529 ( 31 ) $ & $ 0.0558 ( 31 ) $ & $ 0.2250 ( 187 ) $ & $ 0.0275 ( 44 ) $ & $ 0.0494 ( 193 ) $ \\
$ 4.690 $ & $ 3.1467 ( 30 ) $ & $ 0.0589 ( 33 ) $ & $ 0.2230 ( 182 ) $ & $ 0.0213 ( 49 ) $ & $ 0.0409 ( 187 ) $ & $ 3.1521 ( 32 ) $ & $ 0.0624 ( 32 ) $ & $ 0.2332 ( 188 ) $ & $ 0.0196 ( 51 ) $ & $ 0.0412 ( 196 ) $ \\
$ 4.899 $ & $ 3.1468 ( 33 ) $ & $ 0.0575 ( 37 ) $ & $ 0.2276 ( 194 ) $ & $ 0.0199 ( 57 ) $ & $ 0.0349 ( 194 ) $ & $ 3.1521 ( 31 ) $ & $ 0.0589 ( 40 ) $ & $ 0.2223 ( 188 ) $ & $ 0.0243 ( 55 ) $ & $ 0.0526 ( 198 ) $ \\
$ 5.000 $ & $ 3.1489 ( 31 ) $ & $ 0.0599 ( 37 ) $ & $ 0.2112 ( 188 ) $ & $ 0.0208 ( 52 ) $ & $ 0.0563 ( 188 ) $ & $ 3.1511 ( 32 ) $ & $ 0.0564 ( 31 ) $ & $ 0.2270 ( 189 ) $ & $ 0.0281 ( 39 ) $ & $ 0.0468 ( 193 ) $ \\
$ 5.099 $ & $ 3.1483 ( 31 ) $ & $ 0.0580 ( 35 ) $ & $ 0.2157 ( 188 ) $ & $ 0.0224 ( 51 ) $ & $ 0.0531 ( 187 ) $ & $ 3.1509 ( 31 ) $ & $ 0.0591 ( 31 ) $ & $ 0.2285 ( 184 ) $ & $ 0.0225 ( 44 ) $ & $ 0.0467 ( 189 ) $ \\
$ 5.196 $ & $ 3.1484 ( 30 ) $ & $ 0.0556 ( 35 ) $ & $ 0.2138 ( 194 ) $ & $ 0.0267 ( 51 ) $ & $ 0.0558 ( 206 ) $ & $ 3.1513 ( 30 ) $ & $ 0.0598 ( 35 ) $ & $ 0.2222 ( 178 ) $ & $ 0.0201 ( 49 ) $ & $ 0.0540 ( 181 ) $ \\
$ 5.385 $ & $ 3.1485 ( 31 ) $ & $ 0.0553 ( 36 ) $ & $ 0.2201 ( 197 ) $ & $ 0.0264 ( 50 ) $ & $ 0.0504 ( 199 ) $ & $ 3.1514 ( 30 ) $ & $ 0.0571 ( 33 ) $ & $ 0.2202 ( 182 ) $ & $ 0.0248 ( 48 ) $ & $ 0.0534 ( 187 ) $ \\
$ 5.477 $ & $ 3.1486 ( 30 ) $ & $ 0.0563 ( 34 ) $ & $ 0.2208 ( 193 ) $ & $ 0.0242 ( 47 ) $ & $ 0.0506 ( 211 ) $ & $ 3.1526 ( 29 ) $ & $ 0.0616 ( 34 ) $ & $ 0.2136 ( 180 ) $ & $ 0.0181 ( 50 ) $ & $ 0.0614 ( 181 ) $ \\
$ 5.657 $ & $ 3.1493 ( 37 ) $ & $ 0.0682 ( 43 ) $ & $ 0.2074 ( 222 ) $ & $ 0.0126 ( 63 ) $ & $ 0.0673 ( 218 ) $ & $ 3.1500 ( 32 ) $ & $ 0.0562 ( 30 ) $ & $ 0.2292 ( 191 ) $ & $ 0.0267 ( 47 ) $ & $ 0.0469 ( 200 ) $ \\
$ 5.745 $ & $ 3.1486 ( 31 ) $ & $ 0.0590 ( 32 ) $ & $ 0.2297 ( 189 ) $ & $ 0.0250 ( 41 ) $ & $ 0.0401 ( 196 ) $ & $ 3.1516 ( 31 ) $ & $ 0.0570 ( 34 ) $ & $ 0.2187 ( 184 ) $ & $ 0.0255 ( 44 ) $ & $ 0.0539 ( 190 ) $ \\
$ 5.831 $ & $ 3.1501 ( 32 ) $ & $ 0.0582 ( 37 ) $ & $ 0.2158 ( 200 ) $ & $ 0.0247 ( 46 ) $ & $ 0.0552 ( 203 ) $ & $ 3.1500 ( 30 ) $ & $ 0.0557 ( 31 ) $ & $ 0.2260 ( 183 ) $ & $ 0.0282 ( 45 ) $ & $ 0.0478 ( 188 ) $ \\
$ 5.916 $ & $ 3.1492 ( 30 ) $ & $ 0.0610 ( 37 ) $ & $ 0.2154 ( 191 ) $ & $ 0.0197 ( 48 ) $ & $ 0.0592 ( 203 ) $ & $ 3.1497 ( 29 ) $ & $ 0.0584 ( 34 ) $ & $ 0.2239 ( 179 ) $ & $ 0.0240 ( 50 ) $ & $ 0.0492 ( 188 ) $ \\
$ 6.000 $ & $ 3.1499 ( 28 ) $ & $ 0.0598 ( 38 ) $ & $ 0.2176 ( 184 ) $ & $ 0.0206 ( 52 ) $ & $ 0.0524 ( 189 ) $ & $ 3.1502 ( 32 ) $ & $ 0.0565 ( 33 ) $ & $ 0.2224 ( 189 ) $ & $ 0.0272 ( 44 ) $ & $ 0.0490 ( 188 ) $ \\
$ 6.083 $ & $ 3.1507 ( 29 ) $ & $ 0.0612 ( 39 ) $ & $ 0.2080 ( 193 ) $ & $ 0.0193 ( 58 ) $ & $ 0.0643 ( 204 ) $ & $ 3.1500 ( 31 ) $ & $ 0.0551 ( 31 ) $ & $ 0.2225 ( 184 ) $ & $ 0.0273 ( 43 ) $ & $ 0.0535 ( 185 ) $ \\
$ 6.164 $ & $ 3.1503 ( 30 ) $ & $ 0.0610 ( 38 ) $ & $ 0.2177 ( 183 ) $ & $ 0.0212 ( 51 ) $ & $ 0.0550 ( 187 ) $ & $ 3.1501 ( 30 ) $ & $ 0.0551 ( 34 ) $ & $ 0.2237 ( 180 ) $ & $ 0.0262 ( 46 ) $ & $ 0.0498 ( 184 ) $ \\
$ 6.325 $ & $ 3.1510 ( 31 ) $ & $ 0.0532 ( 41 ) $ & $ 0.2116 ( 197 ) $ & $ 0.0237 ( 56 ) $ & $ 0.0623 ( 203 ) $ & $ 3.1507 ( 29 ) $ & $ 0.0556 ( 34 ) $ & $ 0.2245 ( 175 ) $ & $ 0.0290 ( 46 ) $ & $ 0.0500 ( 179 ) $ \\
$ 6.403 $ & $ 3.1511 ( 30 ) $ & $ 0.0586 ( 36 ) $ & $ 0.2171 ( 190 ) $ & $ 0.0235 ( 47 ) $ & $ 0.0548 ( 196 ) $ & $ 3.1509 ( 29 ) $ & $ 0.0559 ( 31 ) $ & $ 0.2255 ( 181 ) $ & $ 0.0273 ( 42 ) $ & $ 0.0469 ( 190 ) $ \\
$ 6.481 $ & $ 3.1498 ( 31 ) $ & $ 0.0554 ( 37 ) $ & $ 0.2251 ( 193 ) $ & $ 0.0291 ( 54 ) $ & $ 0.0456 ( 203 ) $ & $ 3.1509 ( 31 ) $ & $ 0.0559 ( 33 ) $ & $ 0.2169 ( 194 ) $ & $ 0.0273 ( 45 ) $ & $ 0.0532 ( 200 ) $ \\
$ 6.557 $ & $ 3.1509 ( 25 ) $ & $ 0.0601 ( 41 ) $ & $ 0.2105 ( 175 ) $ & $ 0.0200 ( 57 ) $ & $ 0.0619 ( 176 ) $ & $ 3.1512 ( 30 ) $ & $ 0.0520 ( 36 ) $ & $ 0.2112 ( 185 ) $ & $ 0.0320 ( 49 ) $ & $ 0.0600 ( 191 ) $ \\
$ 6.633 $ & $ 3.1509 ( 29 ) $ & $ 0.0610 ( 48 ) $ & $ 0.2158 ( 186 ) $ & $ 0.0169 ( 67 ) $ & $ 0.0576 ( 188 ) $ & $ 3.1518 ( 31 ) $ & $ 0.0575 ( 33 ) $ & $ 0.2117 ( 189 ) $ & $ 0.0235 ( 38 ) $ & $ 0.0598 ( 198 ) $ \\
$ 6.708 $ & $ 3.1515 ( 30 ) $ & $ 0.0579 ( 37 ) $ & $ 0.2140 ( 186 ) $ & $ 0.0223 ( 48 ) $ & $ 0.0568 ( 198 ) $ & $ 3.1512 ( 30 ) $ & $ 0.0566 ( 31 ) $ & $ 0.2161 ( 183 ) $ & $ 0.0261 ( 42 ) $ & $ 0.0554 ( 192 ) $ \\
$ 6.782 $ & $ 3.1500 ( 31 ) $ & $ 0.0580 ( 33 ) $ & $ 0.2198 ( 191 ) $ & $ 0.0238 ( 43 ) $ & $ 0.0553 ( 206 ) $ & $ 3.1503 ( 31 ) $ & $ 0.0589 ( 34 ) $ & $ 0.2265 ( 185 ) $ & $ 0.0227 ( 42 ) $ & $ 0.0447 ( 194 ) $ \\
$ 6.928 $ & $ 3.1532 ( 37 ) $ & $ 0.0642 ( 58 ) $ & $ 0.2153 ( 225 ) $ & $ 0.0115 ( 87 ) $ & $ 0.0555 ( 212 ) $ & $ 3.1475 ( 34 ) $ & $ 0.0560 ( 47 ) $ & $ 0.2324 ( 212 ) $ & $ 0.0226 ( 67 ) $ & $ 0.0382 ( 230 ) $ \\
$ 7.000 $ & $ 3.1505 ( 31 ) $ & $ 0.0590 ( 38 ) $ & $ 0.2192 ( 189 ) $ & $ 0.0215 ( 54 ) $ & $ 0.0558 ( 195 ) $ & $ 3.1517 ( 30 ) $ & $ 0.0539 ( 33 ) $ & $ 0.2119 ( 187 ) $ & $ 0.0302 ( 42 ) $ & $ 0.0615 ( 196 ) $ \\
$ 7.071 $ & $ 3.1509 ( 30 ) $ & $ 0.0592 ( 35 ) $ & $ 0.2172 ( 187 ) $ & $ 0.0222 ( 50 ) $ & $ 0.0542 ( 187 ) $ & $ 3.1504 ( 31 ) $ & $ 0.0550 ( 29 ) $ & $ 0.2205 ( 184 ) $ & $ 0.0283 ( 40 ) $ & $ 0.0521 ( 187 ) $ \\
$ 7.141 $ & $ 3.1512 ( 32 ) $ & $ 0.0615 ( 41 ) $ & $ 0.2216 ( 200 ) $ & $ 0.0199 ( 54 ) $ & $ 0.0484 ( 204 ) $ & $ 3.1511 ( 29 ) $ & $ 0.0572 ( 34 ) $ & $ 0.2165 ( 175 ) $ & $ 0.0261 ( 45 ) $ & $ 0.0559 ( 181 ) $ \\
$ 7.211 $ & $ 3.1502 ( 31 ) $ & $ 0.0563 ( 39 ) $ & $ 0.2197 ( 194 ) $ & $ 0.0266 ( 49 ) $ & $ 0.0497 ( 204 ) $ & $ 3.1521 ( 30 ) $ & $ 0.0524 ( 29 ) $ & $ 0.2148 ( 184 ) $ & $ 0.0295 ( 40 ) $ & $ 0.0581 ( 192 ) $ \\
$ 7.280 $ & $ 3.1501 ( 30 ) $ & $ 0.0594 ( 36 ) $ & $ 0.2219 ( 187 ) $ & $ 0.0204 ( 49 ) $ & $ 0.0510 ( 198 ) $ & $ 3.1503 ( 30 ) $ & $ 0.0563 ( 29 ) $ & $ 0.2249 ( 182 ) $ & $ 0.0267 ( 39 ) $ & $ 0.0472 ( 186 ) $ \\
\hline 
\hline 
\end{tabular}
\caption{
Data list of the Taylor-expansion coefficients of the color-antitriplet and -sextet potential at $T=1.35T_{\rm pc}$ as a function of $r/a$.
}
\label{table-V3V6b2000}
\end{center}
\end{table}
%%%%%%%%%%%

\clearpage

\begin{table}[t]
\begin{tabular}{cc}
\begin{minipage}{0.4\hsize}
%%%%%%%%%%%%%%%%
%%% Table VI
%%%%%%%%%%%%%%%%
\begin{center}
\begin{tabular}{c|cccc}
\hline 
\hline 
& \multicolumn{4}{|c}{$m_{\rm D}/T$} \\
$\mu_{\rm I}/T$ & $1$ & $8$ & $3^{\ast}$ & $6$ \\
\hline
$ 0.0 $ & $ 3.27 ( 74 ) $ & $ 5.60 ( 359 ) $ & $ 3.61 ( 89 ) $ & $ 3.43 ( 149 ) $ \\
$ 0.1 $ & $ 5.30 ( 84 ) $ & $ \cdots $ & $ 3.79 ( 67 ) $ & $ 7.01 ( 234 ) $ \\
$ 0.2 $ & $ 6.63 ( 91 ) $ & $ 3.02 ( 298 ) $ & $ 4.27 ( 73 ) $ & $ 3.77 ( 151 ) $ \\
$ 0.3 $ & $ 3.88 ( 82 ) $ & $ \cdots $ & $ 4.49 ( 80 ) $ & $ 6.94 ( 303 ) $ \\
$ 0.4 $ & $ 4.43 ( 80 ) $ & $ 4.21 ( 249 ) $ & $ 3.52 ( 106 ) $ & $ 5.86 ( 195 ) $ \\
$ 0.5 $ & $ 3.56 ( 84 ) $ & $ \cdots $ & $ 5.02 ( 97 ) $ & $ 7.21 ( 285 ) $ \\
$ 0.6 $ & $ 3.81 ( 61 ) $ & $ \cdots $ & $ 2.60 ( 70 ) $ & $ 6.06 ( 176 ) $ \\
$ 0.7 $ & $ 3.62 ( 61 ) $ & $ \cdots $ & $ 3.19 ( 82 ) $ & $ 4.90 ( 153 ) $ \\
$ 0.8 $ & $ 3.27 ( 62 ) $ & $ \cdots $ & $ 2.76 ( 84 ) $ & $ 4.50 ( 162 ) $ \\
$ 0.9 $ & $ 3.57 ( 54 ) $ & $ 5.86 ( 448 ) $ & $ 3.26 ( 78 ) $ & $ 2.98 ( 130 ) $ \\
$ 1.0 $ & $ 3.54 ( 55 ) $ & $ \cdots $ & $ 3.83 ( 79 ) $ & $ 4.78 ( 141 ) $ \\
\hline 
\hline 
\end{tabular}
\caption{
Data list of the color-Debye screening masses in all the color channels at $T=1.20T_{\rm pc}$ as a function of $\mu_{\rm I}/T$.
}
\label{table-mDb1950}
\end{center}
%%%%%%%%%%%

\end{minipage}
\hspace{33pt}
\begin{minipage}{0.4\hsize}

%%%%%%%%%%%%%%%%
%%% Table VI
%%%%%%%%%%%%%%%%
\begin{center}
\begin{tabular}{c|cccc}
\hline 
\hline 
& \multicolumn{4}{|c}{$m_{\rm D}/T$} \\
$\mu_{\rm I}/T$ & $1$ & $8$ & $3^{\ast}$ & $6$ \\
\hline
$ 0.0 $ & $ 3.98 ( 68 ) $ & $ 8.74 ( 342 ) $ & $ 3.76 ( 68 ) $ & $ 5.68 ( 137 ) $ \\
$ 0.1 $ & $ 4.61 ( 65 ) $ & $ \cdots $ & $ 4.40 ( 60 ) $ & $ 3.18 ( 120 ) $ \\
$ 0.2 $ & $ 4.34 ( 59 ) $ & $ 3.18 ( 166 ) $ & $ 4.39 ( 77 ) $ & $ 3.24 ( 119 ) $ \\
$ 0.3 $ & $ 3.37 ( 50 ) $ & $ 3.47 ( 176 ) $ & $ 4.42 ( 66 ) $ & $ 3.67 ( 97 ) $ \\
$ 0.4 $ & $ 3.90 ( 57 ) $ & $ 3.89 ( 242 ) $ & $ 4.10 ( 56 ) $ & $ 2.33 ( 112 ) $ \\
$ 0.5 $ & $ 4.36 ( 67 ) $ & $ 2.25 ( 213 ) $ & $ 4.76 ( 67 ) $ & $ 5.07 ( 124 ) $ \\
$ 0.6 $ & $ 2.71 ( 61 ) $ & $ 5.93 ( 279 ) $ & $ 2.98 ( 62 ) $ & $ 4.86 ( 138 ) $ \\
$ 0.7 $ & $ 3.45 ( 42 ) $ & $ \cdots $ & $ 2.84 ( 56 ) $ & $ 1.36 ( 100 ) $ \\
$ 0.8 $ & $ 3.49 ( 51 ) $ & $ \cdots $ & $ 3.99 ( 64 ) $ & $ 1.31 ( 102 ) $ \\
$ 0.9 $ & $ 4.09 ( 58 ) $ & $ 2.19 ( 192 ) $ & $ 4.19 ( 65 ) $ & $ 4.35 ( 91 ) $ \\
$ 1.0 $ & $ 3.36 ( 36 ) $ & $ 3.61 ( 160 ) $ & $ 3.63 ( 45 ) $ & $ 2.58 ( 60 ) $ \\
\hline 
\hline 
\end{tabular}
\caption{
Data list of the color-Debye screening masses in all the color channels at $T=1.35T_{\rm pc}$ as a function of $\mu_{\rm I}/T$.
}
\label{table-mDb2000}
\end{center}
%%%%%%%%%%%

\end{minipage}
\end{tabular}
\end{table}

\twocolumngrid

%%%%%%%%%%%%%%%%%%%%%%%%%%%%%%%%%%%%%%%%%%%%%%%%%%%%%%%%%%%%%%%%%%%%%%%%%%%%%%%%
%%%%% References 
%%%%%%%%%%%%%%%%%%%%%%%%%%%%%%%%%%%%%%%%%%%%%%%%%%%%%%%%%%%%%%%%%%%%%%%%%%%%%%%%

\end{document}